\documentclass[a4paper,12pt]{article}
\usepackage{float}
\usepackage[percent]{overpic}
\usepackage[margin=2.5cm]{geometry}
\usepackage{setspace}
\usepackage{indentfirst}
\usepackage{footmisc}
\usepackage{amsmath,amssymb,mathtools}
\usepackage{mathrsfs}
\usepackage{graphicx}
\usepackage{booktabs}
\usepackage{caption}
\usepackage[nosort]{cite}
\usepackage[svgnames]{xcolor}
\usepackage[colorlinks=true,allcolors=.,bookmarksnumbered=true,pdfpagemode=UseNone,pdfstartview=FitH]{hyperref}

\newcommand{\dif}{\mathop{}\!\mathrm{d}}

\newcommand{\calK}{\mathcal K}
\newcommand{\calR}{\mathcal R}
\newcommand{\calX}{\mathcal X}

\numberwithin{equation}{section}
\makeatletter
\renewcommand\section{\@startsection {section}{1}{\z@}{-3.5ex \@plus -1ex \@minus -.2ex}{2.3ex \@plus.2ex}{\normalfont\large\sffamily\bfseries}}
\renewcommand\subsection{\@startsection{subsection}{2}{\z@}{-3.25ex\@plus -1ex \@minus -.2ex}{1.5ex \@plus .2ex}{\normalfont\normalsize\sffamily\bfseries}}
\renewcommand\subsubsection{\@startsection{subsubsection}{3}{\z@}{-3.25ex\@plus -1ex \@minus -.2ex}{1.0ex \@plus .2ex}{\normalfont\normalsize\sffamily\itshape}}
\makeatother

\begin{document}
\thispagestyle{empty}
\vbox{}\vspace{0.5cm}
\begin{center}
{\sffamily\LARGE The type-D accelerating Taub--NUT spacetime:\\[2mm]
construction, geometry and physical properties}\\[10mm]
{\sffamily\large Kenneth Hong}\\[4mm]
{\sffamily\slshape Department of Physics, National University of Singapore, Singapore}\\[8mm]
\end{center}
\vspace{0.7cm}

\centerline{\sffamily\bfseries Abstract}
\bigskip
\noindent
We investigate the geometry and physical properties of the vacuum type-D accelerating Taub--NUT spacetime in the sector with vanishing Kerr rotation. We recover its normalized metric from the Pleba\'nski--Demia\'nski family and a dressed finite binary, tracking the horizon Killing directions through the large-companion limit. The black-hole and acceleration horizons have distinct projective Killing generators, and their intrinsic horizon geometry yields a relation between polar Gaussian curvatures and cone angles that differs from that of the type-I accelerating Taub--NUT solution without requiring a matching of acceleration parameters. We establish separability of the type-D null Hamilton--Jacobi equation and construct an unwrapped radial conformal atlas with four simple-horizon extensions and five angular conformal diagrams. We also examine axial identifications and chronology-violating regions. The resulting geometrical and dynamical distinctions clarify how the two accelerating NUT families differ despite sharing their nonaccelerating and zero-NUT limits; the radial atlas is not claimed to be a unique maximal four-dimensional extension.
\newpage

\section{Introduction}
\label{intro}

The $C$-metric provides an exact description of accelerated black holes and illustrates how acceleration changes the interpretation of horizons, axes and conformal infinity. In contrast to an isolated asymptotically flat black hole, an accelerating geometry contains acceleration horizons and typically requires conical sources along the symmetry axis. Its global structure therefore cannot be inferred solely from the zeros of a radial metric function: one must also specify the angular domain, the conformal boundary and the identifications of the Killing coordinates. The factorized forms of the static and rotating $C$-metrics given by Hong and Teo, together with subsequent analyses of their global structure, supply the starting point for these questions~\cite{HongTeo2003,HongTeo2005,GP2005,GKP}. The addition of NUT charge makes the problem more subtle still. Even without acceleration, Taub--NUT geometry admits different global treatments of its axial strings and time coordinate, and distinct analytic extensions need not be compatible~\cite{Misner1963,Bonnor1969,MillerKruskalGodfrey1971}. Consequently, local regularity of a metric and global regularity of a chosen spacetime quotient must be carefully distinguished.

The interplay of acceleration and NUT charge has a further algebraic aspect. The Pleba\'nski--Demia\'nski family is a canonical setting for vacuum and electrovacuum solutions of Petrov type D, and its physical parametrizations make the roles of mass, rotation, NUT charge and acceleration accessible~\cite{PlebanskiDemianski1976,GriffithsPodolsky2006PD}. Nevertheless, setting the Kerr-like rotation parameter to zero in familiar forms does not automatically expose a nontrivial accelerating Taub--NUT sector. This apparent absence was historically puzzling because the vanishing of one physical parameter need not commute transparently with the coordinate and Killing transformations that relate different forms of the general metric. The issue is not resolved by finding an arbitrary metric with both acceleration and NUT charge: one must distinguish its Petrov character and establish precisely which branch of the solution family is being considered.

An accelerating NUT solution was obtained by Chng, Mann and Stelea and subsequently studied in a more transparent parametrization by Podolsk\'y and Vr\'atn\'y~\cite{CMS,PVtypeI}. The generic member of this family is of Petrov type I. Its horizon and conformal structure, as well as its axial and chronology-related features, provide an important precedent, but it is not the geometry studied throughout the main part of the present paper. Later Ehlers-based constructions clarified how accelerating NUT solutions may arise and how different NUT parameters can enter the black-hole and background sectors~\cite{BarrientosCisterna,AstorinoBoldi}. These developments show why the phrase ``accelerating Taub--NUT'' alone does not uniquely specify a local spacetime.

The type-D sector has now been identified explicitly. Astorino constructed an accelerating Kerr--NUT family admitting a nontrivial zero-Kerr specialization, and Wu and Wu clarified the embedding of the apparently missing NUT $C$-metric within the Pleba\'nski--Demia\'nski class~\cite{Astorino,WuWu}. Ovcharenko, Podolsk\'y and Astorino subsequently established detailed relationships among modern forms of the type-D family, including the compact $A^+$ representation used here~\cite{OPA}. These works establish the existence of the solution and the relation between its parametrizations. They do not make a separate investigation of every geometric and physical property redundant. In particular, knowing that two metric expressions are locally equivalent does not determine the normalized horizon generators, the intrinsic geometry of their horizon sections, the structure of their distinguished null rays or the conformal truncation of their extended radial blocks.

We study the vacuum type-D accelerating Taub--NUT metric in the nondegenerate four-horizon regime, using the parameters $m$, $l$ and $\alpha$ and a fixed normalization of the Killing coordinates. We first determine its horizon, geodesic and conformal properties. The type-I metric is introduced only in the final comparison, where the common Taub--NUT and $C$-metric limits provide checks but not a criterion for equivalence.

Two complementary constructions establish the origin and normalization of our starting metric. First, we recover the compact $A^+$ form directly from the original Pleba\'nski--Demia\'nski metric. This calculation makes the required coordinate and Killing transformations explicit and checks the full metric, including the orbit-space factor; it is a reconstruction of an established local solution, not a claim to discover a new type-D branch. Second, we revisit the finite-binary inverse-scattering construction. The Belinski--Zakharov dressing method and its subsequent refinements provide a systematic framework for stationary axisymmetric solutions~\cite{BelinskiZakharov1978,Pomeransky2006}; the Weyl and rod descriptions help interpret the resulting Killing directions~\cite{EmparanReall2002,Harmark2004}. The accelerating binary limit itself has an established precedent in Astorino's construction~\cite{Astorino}. Our additional emphasis is on computing the kernels of the dressed finite-binary Killing block on the two open horizon rods \emph{before} the large-companion limit, and then tracing their projective directions into the limiting accelerating geometry. This clarifies which horizon information follows from the finite binary and which normalization or global-identification questions remain separate.

Once the metric has been established, its horizons and axes provide the first intrinsic tests. The two black-hole and two acceleration horizons are roots of the same radial polynomial, but their Killing generators need not coincide. In the type-D geometry their projective directions are generally distinct. This feature motivates the horizon-mechanics calculation, where surface gravities depend on the chosen normalization, and the intrinsic-horizon calculation, where Gaussian curvature can be compared without relying on a coordinate identification between different solutions. In particular, the relation among polar Gaussian curvatures and cone-angle coefficients provides a sharp distinction from the type-I family. The comparison is formulated for horizon sections equipped with their axial circle action; it is not a statement that arbitrary, unmarked two-dimensional surfaces can be compared by their coordinate poles.

Petrov type D suggests a special algebraic structure, but null separability must be established in the physical coordinates. We derive the null Hamilton--Jacobi separation in the conformal Carter geometry and identify the fixed-angle null congruences as a distinguished specialization of the separated equations. These congruences supply the radial null directions used in the subsequent conformal construction. The broader literature on Carter separability and conformal hidden symmetries provides the appropriate context~\cite{Carter1968CMP,Carter1968PR,FrolovKrtousKubiznak2017,Gray2026}. Our claims concern the stated null dynamics and do not imply a coordinate-independent nonintegrability theorem for the type-I solution.

The radial conformal analysis then asks how the four simple horizons are crossed and how the resulting Kruskal blocks terminate at infinity. The $C$-metric construction of Griffiths, Krtou\v{s} and Podolsk\'y and the general type-D framework of Podolsk\'y and Vr\'atn\'y provide essential precedents~\cite{GKP,PV2021,PV2023}. In the present zero-Kerr $A^+$ parametrization, the reciprocal coordinate $y=(\alpha r)^{-1}$ is regular at $y=0$ away from the equatorial conformal endpoint. Its relation to the angular boundary $y=x$ distinguishes the two acceleration sectors and yields five angular classes of radial conformal diagrams. We construct an unwrapped repeated radial atlas and explicitly retain its limits: this is not a proof of a unique maximal four-dimensional analytic extension or a resolution of every possible axial quotient.

Finally, NUT charge makes the interpretation of the axes inseparable from the global choice of Killing identifications. We distinguish conical regularity from the Misner-string transition, and chronology-violating coordinate circles from stronger statements about closed causal geodesics. The interpretation of ordinary Taub--NUT developed by Cl\'ement, Gal'tsov and Guenouche is useful here, but its conclusions cannot simply be transferred to an accelerating spacetime without an additional argument~\cite{CGG}. These issues also explain why the local radial atlas should not be conflated with a complete classification of global causal structure.

In the final comparison we introduce the type-I metric and examine its horizon directions, intrinsic horizon geometry and null Hamilton--Jacobi equation alongside the corresponding type-D results. The shared parameter limits do not imply equivalence for nonzero acceleration and NUT charge.

The paper is organized as follows. Section~\ref{metric} introduces the type-D metric and its principal limits. Sections~\ref{pdorigin} and \ref{binaryorigin} reconstruct its canonical and finite-binary origins. Section~\ref{killing} studies the horizons, axes and projective Killing structure, and Sec.~\ref{horizongeom} develops horizon mechanics and intrinsic geometry. Section~\ref{geodesics} treats null dynamics and hidden symmetry; Sec.~\ref{causal} uses the distinguished null congruences to construct the radial conformal atlas and examines axial identifications and chronology. Section~\ref{comparison} introduces the type-I metric and presents the systematic comparison. Section~\ref{conclusion} summarizes the conclusions and their scope.

\section{Type-D accelerating Taub--NUT geometry}
\label{metric}

\subsection{Canonical form and conventions}

We work with the compact $A^+$ representation of the vacuum type-D accelerating Taub--NUT metric.  We denote the mass, NUT and acceleration parameters by $m$, $l$ and $\alpha$, respectively, and set
\begin{equation}
\sigma=\sqrt{m^2+l^2},
\qquad
r_\pm=m\pm\sigma .
\label{sigmarpm}
\end{equation}
The radial and angular structure functions are
\begin{equation}
\Delta_r=(1-\alpha^2r^2)(r-r_+)(r-r_-),
\label{Dr}
\end{equation}
\begin{equation}
\Delta_x=(1-x^2)(1-2\alpha mx-\alpha^2l^2x^2),
\label{Dx}
\end{equation}
with $x=\cos\theta$, so that $x=+1$ and $x=-1$ denote the north and south axes.  We take $\alpha>0$ unless otherwise stated.

The metric is written as
\begin{equation}
\begin{aligned}
\dif s^2=\frac{1}{\Omega^2}\bigg[&
- \frac{\Delta_r}{\rho^2}\big(A\,\dif t-B\,\dif\phi\big)^2
+ \frac{\Delta_x}{\rho^2}\big(C\,\dif t+D\,\dif\phi\big)^2 +\rho^2\left(\frac{\dif r^2}{\Delta_r}+\frac{\dif x^2}{\Delta_x}\right)
\bigg],
\end{aligned}
\label{Dmetric}
\end{equation}
where
\begin{equation}
A=1+\alpha^2l^2x^2,
\qquad
B=2lx, \qquad C=2\alpha lr,
\qquad
D=r^2+l^2,
\label{ABOmega}
\end{equation}
and
\begin{equation}
\Omega=1-\alpha rx,
\qquad \rho^2=(r^2+l^2)(1+\alpha^2l^2x^2)+4\alpha l^2rx .
\label{rho2}
\end{equation}
We adopt $C_f=1$, the local orbit-space normalization of the physical $A^+$ representation obtained from the two-soliton construction. The global period of $\phi$ is specified separately when discussing conical defects and horizon areas.

\subsection{Principal limits}

At zero acceleration, $\Omega\to1$, $\Delta_x\to1-x^2$, and the metric reduces to ordinary Taub--NUT after the corresponding stationary-axisymmetric coordinate normalization.  At $l=0$, the off-diagonal Killing terms vanish and Eq.~\eqref{Dmetric} reduces to the vacuum $C$-metric in a factorized form~\cite{HongTeo2003}.  These common limits will be important because the type-I accelerating NUT solution reduces to the same two geometries when $\alpha=0$ and $l=0$, respectively~\cite{PVtypeI}.

The existence of a nontrivial $a=0$ accelerating-NUT sector in the type-D family is not obtained by the naive $a\to0$ limit of the older physical Pleba\'nski--Demia\'nski parametrization.  Its recovery requires the nontrivial reparametrizations and Killing transformations discussed in Refs.~\cite{WuWu,OPA}.  We use Eq.~\eqref{Dmetric} as the master physical metric and give a direct canonical map in Sec.~\ref{pdorigin}.

\section{Direct recovery from the original Pleba\'nski--Demia\'nski metric}\label{pdorigin}
We first recover the physical metric from the canonical vacuum PD solution. Both the Killing and orbit-space sectors must agree; matching the structure functions alone is insufficient. The transformations are local and singular when either acceleration or NUT charge vanishes. These limits will instead be taken in the physical coordinates. No global identification of the Killing orbits is implied.

\subsection{Canonical PD form and the two coordinate scales}\label{sec:pd-canonical}
\subsubsection{Original coordinates and integration constants}
We begin with the expanding vacuum Pleba\'nski--Demia\'nski metric~\cite{PlebanskiDemianski1976,GriffithsPodolsky2006PD}, setting the electromagnetic charges and cosmological constant to zero. In coordinates $(\hat p,\hat q,\hat\tau,\hat\sigma)$, write the original metric as
\begin{align}
\dif s^2=\frac{1}{(1-\hat p\hat q)^2}\bigg[&\frac{X(\hat p)}{\hat p^2+\hat q^2}(\dif\hat\tau+\hat q^2\dif\hat\sigma)^2
-\frac{Y(\hat q)}{\hat p^2+\hat q^2}(\dif\hat\tau-\hat p^2\dif\hat\sigma)^2\notag\\
&+(\hat p^2+\hat q^2)\left(\frac{\dif\hat p^2}{X(\hat p)}+\frac{\dif\hat q^2}{Y(\hat q)}\right)\bigg],
\end{align}
with
\begin{align}
X(\hat p)&=k_0+2n_0\hat p-\epsilon_0\hat p^2+2m_0\hat p^3-k_0\hat p^4,\\
Y(\hat q)&=k_0-2m_0\hat q+\epsilon_0\hat q^2-2n_0\hat q^3-k_0\hat q^4.
\end{align}
The constants $(m_0,n_0,\epsilon_0,k_0)$ are canonical coefficients; $n_0$ is not yet the physical NUT charge. Independent rescalings of the orbit coordinates make the acceleration scale in the conformal factor and the relative scale in the quadratic denominator explicit.
\subsubsection{Introducing the auxiliary acceleration and twist scales}
Introduce two nonzero scales $a_s,b_s$, initially taken positive to fix square-root branches, and set
\begin{equation}
\hat p=a_sp,\qquad \hat q=b_sr,\qquad
\hat\tau=\frac{\tau}{b_s},\qquad
\hat\sigma=\frac{\sigma}{a_sb_s^2}.
\end{equation}
The conformal factor and quadratic denominator become
\begin{equation}
1-\hat p\hat q=1-a_sb_spr,\qquad
\hat p^2+\hat q^2=b_s^2\left(r^2+\frac{a_s^2}{b_s^2}p^2\right).
\end{equation}
It is therefore convenient to define
\begin{equation}
\alpha=a_sb_s,\qquad \omega=\frac{a_s}{b_s},\qquad
 a_s=\sqrt{\alpha\omega},\quad b_s=\sqrt{\alpha/\omega}.
\end{equation}
The Killing one-forms transform according to
\begin{align}
\dif\hat\tau+\hat q^2\dif\hat\sigma
&=\frac{1}{b_s\omega}(\omega\dif\tau+r^2\dif\sigma),\\
\dif\hat\tau-\hat p^2\dif\hat\sigma
&=\frac{1}{b_s}(\dif\tau-\omega p^2\dif\sigma).
\end{align}
Writing $\rho^2=r^2+\omega^2p^2$, define
\begin{equation}
P(p)=\frac{X(a_sp)}{a_s^2b_s^2},\qquad
Q(r)=\frac{Y(b_sr)}{b_s^4}.
\end{equation}
Direct substitution then yields the equivalent line element
\begin{align}
\dif s^2=\frac{1}{(1-\alpha pr)^2}\bigg[&\frac{P(p)}{\rho^2}(\omega\dif\tau+r^2\dif\sigma)^2
-\frac{Q(r)}{\rho^2}(\dif\tau-\omega p^2\dif\sigma)^2\notag\\
&+\rho^2\left(\frac{\dif p^2}{P(p)}+\frac{\dif r^2}{Q(r)}\right)\bigg].
\end{align}
Define the rescaled coefficients by
\begin{equation}
k=\frac{k_0}{\alpha^2},\qquad n=\frac{n_0}{b_s^3},\qquad
\epsilon=\frac{\epsilon_0}{b_s^2},\qquad m=\frac{m_0}{b_s^3}.
\end{equation}
The quartics become
\begin{align}
P(p)&=k+2\omega^{-1}np-\epsilon p^2+2\alpha mp^3-\alpha^2\omega^2kp^4,\\
Q(r)&=\omega^2k-2mr+\epsilon r^2-2\alpha\omega^{-1}nr^3-\alpha^2kr^4.
\end{align}
For example, the cubic coefficient of $P$ follows from
\begin{equation}
\frac{2m_0a_s^3}{a_s^2b_s^2}
=2\frac{m_0}{b_s^3}(a_sb_s)=2\alpha m.
\end{equation}
Thus the cubic coefficient is determined by the same coordinate rescaling that fixes the other coefficients.
The result is the vacuum specialization of Eqs.~(2)--(5) of
Ref.~\cite{GP2005}, with the opposite overall signature. In particular,
the same factor \(b_s^3\) relates both \(m\) and \(n\) to their
original canonical coefficients. The product \(\alpha\) and ratio
\(\omega\) describe the coordinate rescaling, not additional constants
of the vacuum solution. In this and the next section, the symbols
\(m,n,\alpha,\omega\) are the rescaled PD quantities; the physical
parameters introduced later require a separate identification.

\subsection{Degeneracy of the affine angular parametrization}
\subsubsection{The angular scale and the two axes}

The rescaled PD metric admits the affine angular parametrization of
Griffiths and Podolsk\'y (GP)~\cite{GP2005}. In our conventions it is
\begin{equation}\label{eq:gp-affine-chart}
 p=\frac{l+a x}{\omega},\qquad
 \tau=t-\frac{(l+a x_0)^2}{a}\phi,\qquad
 \sigma=-\frac{\omega}{a}\phi.
\end{equation}
Here $x_0$ specifies the reference axis for the Killing-coordinate shift,
and $a\omega\ne0$ is assumed. In this section $a$ denotes the scale in
the GP angular map, and $l$ is its shift parameter. The mass-like
coefficient $m$ retains the meaning assigned in the rescaled quartics.
Neither this affine choice nor its degeneracy identifies these symbols
with the physical parameters used later in the $A^+$ chart.

The angular part of the orbit-space metric becomes
$\rho^2\,\dif x^2/\widetilde P(x)$, with
\begin{equation}
 \rho^2=r^2+(l+ax)^2
\end{equation}
and
\begin{equation}
 \widetilde P(x)=\frac{\omega^2}{a^2}
 P\!\left(\frac{l+ax}{\omega}\right).
\end{equation}
The prefactor in $\widetilde P$ comes from $\dif p^2=(a/\omega)^2\dif x^2$.
For axes placed at $x=\pm1$, the corresponding values of $p$ are
\begin{equation}
 p_\pm=\frac{l\pm a}{\omega},\qquad
 p_+-p_-=\frac{2a}{\omega}.
\end{equation}
The Jacobian is
\begin{equation}
 \frac{\dif p}{\dif x}=\frac{a}{\omega}.
\end{equation}
Thus, at fixed $\omega\ne0$, the endpoints coalesce and the angular
map loses rank as $a\to0$. The normalization of $\widetilde P$ and
the Killing transformation in Eq.~\eqref{eq:gp-affine-chart} also
involve inverse powers of $a$. Even if their limiting metric
combinations can be made finite, setting $a=0$ in the transformation
does not define an invertible chart. This is a statement about the GP
angular scale, not the physical zero-Kerr condition.

\subsubsection{Angular-root constraints}

Write
\begin{equation}
 \widetilde P(x)=A_0+A_1x+A_2x^2+A_3x^3+A_4x^4 .
\end{equation}
The conditions $\widetilde P(1)=\widetilde P(-1)=0$ give
\begin{equation}
 A_1+A_3=0,\qquad A_0+A_2+A_4=0,
\end{equation}
so that
\begin{equation}
 \widetilde P(x)=(1-x^2)(A_0-A_3x-A_4x^2).
\end{equation}
These are root conditions on the chosen domain, not a guarantee of
the required sign of $\widetilde P$ throughout the interval. The
vacuum coefficients follow by expanding the rescaled quartic:
\begin{align}
 A_0&=\frac{k\omega^2}{a^2}+\frac{2nl}{a^2}-\frac{\epsilon l^2}{a^2}
 +\frac{2\alpha m l^3}{a^2\omega}-\frac{\alpha^2 k l^4}{a^2},\\
 A_1&=\frac{2}{a}\left(n-\epsilon l+\frac{3\alpha m l^2}{\omega}
                         -2\alpha^2kl^3\right),\\
 A_2&=-\epsilon+\frac{6\alpha ml}{\omega}-6\alpha^2kl^2,\\
 A_3&=\frac{2\alpha a}{\omega}(m-2\alpha kl\omega),\\
 A_4&=-\alpha^2ka^2.
\end{align}
The odd root condition determines the NUT-like PD coefficient,
\begin{equation}\label{eq:gp-odd-root}
 n=\epsilon l-\frac{\alpha m}{\omega}(3l^2+a^2)
 +2\alpha^2kl(l^2+a^2).
\end{equation}
Substituting this expression in the even root condition gives
\begin{equation}\label{eq:gp-even-root}
 k\big[\omega^2+\alpha^2(3l^4-2a^2l^2-a^4)\big]
 +\frac{4\alpha ml(a^2-l^2)}{\omega}
 +\epsilon(l^2-a^2)=0 .
\end{equation}
For $a^2\ne l^2$, one may equivalently write
\begin{equation}
 \epsilon=\frac{k[\omega^2+\alpha^2(3l^4-2a^2l^2-a^4)]
 +4\alpha ml(a^2-l^2)/\omega}{a^2-l^2}.
\end{equation}
At $a^2=l^2$ the undivided equation must be used. The same
distinction arises in the nonaccelerating Kerr--NUT quadratic:
$r^2+(l+ax)^2$ can vanish for real $x\in[-1,1]$ only when
$r=0$ and $|l|\leq |a|$. This observation about the nonaccelerating
quadratic does not by itself determine the curvature of the
accelerating metric.

\subsubsection{Normalization and the domain of the affine description}

The further GP choice $A_0=1$ fixes the scale of the angular polynomial
rather than the locations of its roots. It does not fix the axial
period or impose elementary flatness. Define
\begin{equation}
 D=\omega^2+3\alpha^2l^2(a^2-l^2).
\end{equation}
For $a\omega D\ne0$, eliminating $\epsilon$ between $A_0=1$ and
Eq.~\eqref{eq:gp-even-root} gives
\begin{equation}
 k=\frac{(a^2-l^2)(\omega+2\alpha lm)}{\omega D}.
\end{equation}
The two undivided equations for $k$ and $\epsilon$ are equivalent to
this procedure even at $a^2=l^2$, where Eq.~\eqref{eq:gp-even-root}
gives $k=0$ and $A_0=1$ fixes $\epsilon$. At $D=0$ the displayed
formula for $k$ is not applicable, and the defining equations must
instead be examined. These are algebraic qualifications to the
normalization; neither establishes that the chosen interval is
admissible for every set of parameters. The generic relations agree
with the vacuum specialization of GP's Eqs.~(10)--(13). GP note that
the remaining scaling freedom can be used to choose
$\omega=|a+l|$ where that choice does not vanish~\cite{GP2005}.

A joint limit of $a$ and $\omega$ could keep the ratio $a/\omega$
nonzero, but this alone would not control the shift $l/\omega$,
the Killing-coordinate transformation, or either structure
function. Assessing such a limit requires the full metric. In
particular, the specialization considered in Sec.~8 of
Ref.~\cite{GP2005} reduces to ordinary NUT after a further
coordinate transformation. Its apparent acceleration is removable;
this does not exclude a genuinely accelerating Taub--NUT
spacetime from the original PD family.

The degeneracy at fixed $\omega\ne0$ therefore arises at the
affine angular identification, not at the preceding two-scale
rescaling. We seek a chart for the physical zero-Kerr sector
directly, using the separated Killing directions as input, including
the distinct time shifts of the two axial directions. The following
section derives a projective
angular map whose Jacobian in Eq.~\eqref{eq:nondegenerate-nut-angular-map}
is nonzero for $\alpha l\ne0$ away from its projective pole, with
$l$ now denoting the physical NUT parameter.

\subsection{The projective map to the accelerating Taub--NUT chart}
\label{sec:consolidated-forward}

\subsubsection{Separated Killing directions}

We return to the original vacuum PD chart and, from here on, write
\((p,q,T,\Phi)=(\hat p,\hat q,\hat\tau,\hat\sigma)\). In particular,
\(p\) is now the original angular coordinate, not the rescaled
variable in the preceding affine construction. The canonical
conformal factor is \(1-pq\), and the two separated Killing one-forms
are \(\dif T-p^2\dif\Phi\) and \(\dif T+q^2\dif\Phi\).

Let \((t,\phi,R,x)\) be the physical coordinates of the known
zero-Kerr accelerating Taub--NUT representation. Let \(m,l,\alpha\)
denote its mass, NUT and acceleration parameters, and set
\(b=\alpha l\ne0\). We take the directions of the two separated
Killing one-forms in this representation as input, obtained by setting
\(a=e=g=\Lambda=0\) in the \(A^+\) metric of Ref.~\cite{OPA}:
\begin{equation}
 \mathcal A=(1+b^2x^2)\dif t-2lx\,\dif\phi,\qquad
 \mathcal B=2bR\,\dif t+(R^2+l^2)\dif\phi.
\end{equation}
For example, the two axial directions at \(x=\pm1\) annihilate
\(\mathcal A\), so in the representative
\(\partial_\phi+\lambda_\pm\partial_t\) they have
\begin{equation}
 \lambda_\pm=\pm\frac{2l}{1+b^2}.
\end{equation}
The opposite time components distinguish the two axial
directions. We use the full angular and radial dependence of
\(\mathcal A,\mathcal B\), not just their values on the rods. Neither
the structure functions nor the normalization of their squared
contributions to the metric is used here.

We seek a separated map \(p=p(x)\), \(q=q(R)\) such that the two
canonical PD forms are individually proportional to
\(\mathcal A\) and \(\mathcal B\), respectively. We require \(p(x)\)
to be rational and the transformation of the Killing coordinates
to be constant and invertible. These conditions specify the
class in which the map is constructed.

\subsubsection{Angular map from a rational square}

Write the most general constant Killing change as
\begin{equation}
 \dif T=A\,\dif t+B\,\dif\phi,\qquad
 \dif\Phi=C\,\dif t+D\,\dif\phi,\qquad AD-BC\ne0 .
\end{equation}
The coefficients of the first PD form are \(A-Cp^2\) and
\(B-Dp^2\). Proportionality to \(\mathcal A\) requires
\((B-Dp^2)(1+b^2x^2)+2lx(A-Cp^2)=0\), or
\begin{equation}\label{eq:rational-p-square}
 p(x)^2=\frac{B(1+b^2x^2)+2lAx}{D(1+b^2x^2)+2lCx}.
\end{equation}
The numerator and denominator are coprime over \(\mathbb C\):
a common zero would force both \(1+b^2x^2\) and \(2lx\) to vanish
by \(AD-BC\ne0\). Since \(p(x)\) is rational, the numerator and
denominator in this reduced quotient must each be a constant
times a polynomial square. In particular, neither \(B\) nor \(D\)
vanishes: the corresponding polynomial would otherwise be
nonzero and linear. Vanishing discriminants of the two quadratics
therefore give
\begin{equation}
 l^2A^2=b^2B^2,\qquad l^2C^2=b^2D^2 .
\end{equation}
Putting \(A=\varepsilon_1\alpha B\) and \(C=\varepsilon_2\alpha D\),
with \(\varepsilon_{1,2}=\pm1\), yields
\(AD-BC=\alpha BD(\varepsilon_1-\varepsilon_2)\).
The signs must be opposite. On the branch
\((\varepsilon_1,\varepsilon_2)=(+1,-1)\), the numerator and
denominator of \eqref{eq:rational-p-square} are
\(B(1+bx)^2\) and \(D(1-bx)^2\), respectively. Thus
\begin{equation}
 p=\gamma\frac{1+bx}{1-bx},\qquad \gamma^2=\frac{B}{D}>0.
\end{equation}
The opposite sign assignment gives the reciprocal fractional-linear
dependence on \(x\), with the same free scale. We choose the displayed
branch and the positive sign of \(\gamma\) for \(p\). The angular
matching has not determined \(\gamma\) or the common Killing scale.

\subsubsection{Radial map from the second Killing direction}

On the chosen angular branch, set \(K=D\ne0\), so that
\begin{equation}
 \dif T=K\gamma^2(\alpha\,\dif t+\dif\phi),\qquad
 \dif\Phi=K(-\alpha\,\dif t+\dif\phi).
\end{equation}
The form \(\dif T+q^2\dif\Phi\) then has coefficients
\(K\alpha(\gamma^2-q^2)\) and \(K(\gamma^2+q^2)\).
Their ratio must agree with that of \(\mathcal B\):
\begin{equation}
 \frac{\gamma^2+q^2}{\alpha(\gamma^2-q^2)}
 =\frac{R^2+l^2}{2bR}.
\end{equation}
The cross-multiplied condition is valid also where the displayed
ratios are undefined. Using \(b=\alpha l\) gives
\begin{equation}
 2lR(\gamma^2+q^2)=(R^2+l^2)(\gamma^2-q^2),
 \qquad q^2(R+l)^2=\gamma^2(R-l)^2 .
\end{equation}
Thus the radial dependence follows without an ansatz for \(q(R)\):
\begin{equation}
 q=\varepsilon\gamma\frac{R-l}{R+l},\qquad \varepsilon=\pm1 .
\end{equation}

\subsubsection{Fixing the projective scale}

The two Killing directions leave a scale and a radial sign to be
fixed. We now require the canonical conformal factor to represent
the physical conformal boundary \(\Omega_+=1-\alpha Rx=0\).
This condition concerns the defining curves away from the coordinate
poles; it does not require every local patch to reach the boundary.
Substituting the orbit maps gives
\begin{align}
 (R+l)(1-bx)(1-pq)
 ={}&(1-\varepsilon\gamma^2)R+(1+\varepsilon\gamma^2)l
 \notag\\
 &-bx\big[(1+\varepsilon\gamma^2)R
              +(1-\varepsilon\gamma^2)l\big].
\end{align}
On the target boundary, \(bx=l/R\), and the right-hand side reduces
to \((1-\varepsilon\gamma^2)(R-l^2/R)\). Its vanishing on the
boundary away from the coordinate poles therefore requires
\(\varepsilon\gamma^2=1\). Conversely, this choice reduces the
right-hand side to \(2l(1-\alpha Rx)\), so it matches the defining
function up to the stated coordinate factors. For real
\(\gamma>0\), the choice is \(\varepsilon=+1\) and \(\gamma=1\).

Writing \(H=K\), the resulting transformation on the selected
orientation branch is
\begin{equation}\label{eq:direct-map}
 p=\frac{1+bx}{1-bx},\qquad q=\frac{R-l}{R+l},\qquad
 \dif T=H(\alpha\,\dif t+\dif\phi),\qquad
 \dif\Phi=H(-\alpha\,\dif t+\dif\phi).
\end{equation}
The Jacobians on this branch are
\begin{equation}\label{eq:nondegenerate-nut-angular-map}
 \frac{\dif p}{\dif x}=\frac{2b}{(1-bx)^2},\qquad
 \frac{\dif q}{\dif R}=\frac{2l}{(R+l)^2},\qquad
 \det\frac{\partial(T,\Phi)}{\partial(t,\phi)}=2\alpha H^2.
\end{equation}
Unlike the affine chart, the angular Jacobian does not vanish in the
zero-Kerr sector. The map is locally invertible for \(\alpha lH\ne0\)
away from \(1-bx=0\) and \(R+l=0\).

The coordinate dependence has thus been determined within the
separated rational class, using the two one-form directions and
the target conformal boundary. This is not a uniqueness statement
for general local isometries or a reconstruction from rod data
alone. The normalization \(H\) and the canonical quartics require
a comparison of the full metrics, to which we now turn.

\subsection{Reconstruction of the quartics and local metric equivalence}
\label{sec:metric-equivalence}

\subsubsection{Matching the separated Killing terms}

For the vacuum $A^+$ solution with zero Kerr parameter, define
\begin{align}
 \Omega_+&=1-\alpha Rx,&
 \mathcal H&=(R^2+l^2)(1+b^2x^2)+4blRx,\\
 \Delta_R&=(1-\alpha^2R^2)(R^2-2mR-l^2),&
 \Delta_x&=(1-x^2)(1-2\alpha mx-b^2x^2).
\end{align}
The target metric in the two Killing one-forms
$\mathcal A,\mathcal B$ of the preceding section is
\begin{align}\label{eq:target-metric}
 \dif s^2_{A^+}=\frac1{\Omega_+^2}\bigg[
 &-\frac{\Delta_R}{\mathcal H}\mathcal A^2
 +\frac{\Delta_x}{\mathcal H}\mathcal B^2+C_f\mathcal H
 \left(\frac{\dif R^2}{\Delta_R}+\frac{\dif x^2}{\Delta_x}\right)
 \bigg].
\end{align}
The factors of $\Delta_R$ distinguish the black-hole and
acceleration horizon candidates. The zeros $x=\pm1$ of
$\Delta_x$ specify the candidate axes in the physical angular chart.
Their inclusion in a particular coordinate domain is considered in
Sec.~\ref{sec:axial-direction-check}.

Using the map \eqref{eq:direct-map} in $\Omega=1-pq$ and
$\Sigma=p^2+q^2$ gives
\begin{align}\label{eq:pda-basic-pullbacks}
 \Omega&=\frac{2l\Omega_+}{(R+l)(1-bx)},&
 \Sigma&=\frac{2\mathcal H}{(R+l)^2(1-bx)^2},\\
 \dif T-p^2\dif\Phi&=\frac{2\alpha H}{(1-bx)^2}\mathcal A,&
 \dif T+q^2\dif\Phi&=\frac{2H}{(R+l)^2}\mathcal B .
\end{align}
It follows that the coefficients of $\mathcal B^2$ and
$-\mathcal A^2$ in the transformed PD metric are, respectively,
\[
 \frac{H^2(1-bx)^4P(p(x))}{2l^2\Omega_+^2\mathcal H},
 \qquad
 \frac{\alpha^2H^2(R+l)^4Q(q(R))}{2l^2\Omega_+^2\mathcal H}.
\]
Equating them to their counterparts in \eqref{eq:target-metric}
fixes the two functions:
\begin{align}\label{eq:pda-pq-composition}
 P(p(x))&=\frac{2l^2\Delta_x(x)}{H^2(1-bx)^4},&
 Q(q(R))&=\frac{2l^2\Delta_R(R)}{\alpha^2H^2(R+l)^4}.
\end{align}
At this stage, matching establishes the Killing block. It does not
yet show that $P,Q$ belong to the original canonical PD family
or that the orbit-space block has the required normalization.

\subsubsection{Canonical coefficient relations}

The inverse orbit-coordinate transformations are
\[
 x=\frac{p-1}{b(p+1)},\qquad R=l\frac{1+q}{1-q}.
\]
They convert \eqref{eq:pda-pq-composition} into
\begin{align}\label{eq:pda-pq-factored}
 P(p)&=-\frac{l}{4H^2b^2}(mp^2-2lp-m)
 [(b-1)p+b+1][(b+1)p+b-1],\\
 Q(q)&=-\frac{l}{4H^2b^2}(mq^2+2lq-m)
 [(b-1)q+b+1][(b+1)q+b-1].
\end{align}
For example, the angular-axis factor simplifies as
\[
 b^2(p+1)^2(1-x^2)=
 [(b-1)p+b+1][(b+1)p+b-1].
\]
The product on the right is the same reciprocal quadratic
in each variable:
\[
 S(z)=(b^2-1)(z^2+1)+2(b^2+1)z,\qquad z^2 S(1/z)=S(z).
\]
Moreover, $mq^2+2lq-m=-q^2(m/q^2-2l/q-m)$.
Hence \eqref{eq:pda-pq-factored} gives the polynomial identity
$Q(q)=-q^4P(1/q)$, and the highest and lowest coefficients of
$P$ are opposite. Writing $P=\sum_{j=0}^4p_jp^j$ and
$Q=\sum_{j=0}^4q_jq^j$, the six vacuum PD coefficient relations are
\begin{equation}\label{eq:pda-pq-residuals}
 q_0-p_0=q_1+p_3=q_2+p_2=q_3+p_1=q_4-p_4=p_4+p_0=0.
\end{equation}
Thus the quartics reconstructed separately from the two Killing
terms form a single canonical pair. In particular, they have a
common constant term and opposite coefficients of the quadratic terms.

In the conventions of the original vacuum PD line element, the
coefficients of $p^0$, $p^1$, $p^2$ and $p^3$ in $P$ give
$k_0$, $2n_0$, $-\epsilon_0$ and $2m_0$, respectively. This yields
\begin{align}\label{eq:pda-canonical-dictionary}
 k_0&=\frac{lm(b^2-1)}{4H^2b^2},&
 n_0&=\frac{l[(b^2-1)l+(b^2+1)m]}{4H^2b^2},\\
 m_0&=\frac{l[(b^2-1)l-(b^2+1)m]}{4H^2b^2},&
 \epsilon_0&=-\frac{l^2(1+b^2)}{H^2b^2}.
\end{align}
The canonical coefficients $n_0,m_0$ are not the physical NUT
and mass parameters $l,m$.

\subsubsection{Orbit-space normalization and the metric identity}

For the orbit-space block, differentiation of \eqref{eq:direct-map}
gives $\dif p=2b\,\dif x/(1-bx)^2$ and
$\dif q=2l\,\dif R/(R+l)^2$. Inserting the functions already fixed in
\eqref{eq:pda-pq-composition} gives
\begin{equation}
 \frac{\dif p^2}{P}=2\alpha^2H^2\frac{\dif x^2}{\Delta_x},
 \qquad
 \frac{\dif q^2}{Q}=2\alpha^2H^2\frac{\dif R^2}{\Delta_R},
 \qquad
 \frac{\Sigma}{\Omega^2}=\frac{\mathcal H}{2l^2\Omega_+^2}.
\end{equation}
Let $\mathcal F$ denote the map from the physical chart to the
canonical PD chart. The two blocks just computed give
\begin{align}\label{eq:pda-complete-pullback}
 \mathcal F^*\dif s^2_{\rm PD}=\frac1{\Omega_+^2}\bigg[
 &-\frac{\Delta_R}{\mathcal H}\mathcal A^2
 +\frac{\Delta_x}{\mathcal H}\mathcal B^2+\frac{\alpha^2H^2}{l^2}\mathcal H
 \left(\frac{\dif R^2}{\Delta_R}+\frac{\dif x^2}{\Delta_x}\right)
 \bigg].
\end{align}
The Killing block is unchanged when $H$ is varied together with
the canonical coefficients in \eqref{eq:pda-canonical-dictionary}.
The remaining coefficient therefore fixes $H$ for a given $C_f$:
\begin{equation}\label{eq:pda-killing-normalization}
 H^2=\frac{l^2 C_f}{\alpha^2}.
\end{equation}
For $C_f>0$ and $\alpha l\ne0$, this gives a nonzero real Killing
normalization. The resulting equality with \eqref{eq:target-metric}
holds for the full four-dimensional metric; no overall homothety is
introduced. In these conventions $[H]=L^2$ and $[P]=[Q]=L^{-2}$,
so the Killing and orbit-space terms both have dimensions $L^2$.
The equality is local. Coordinate domains and the normalization
and periodicity of axial generators are separate issues.

\subsection{Coordinate domains and axial Killing directions}
\label{sec:axial-direction-check}

The Jacobians in Eq.~\eqref{eq:nondegenerate-nut-angular-map} show that the transformation is locally invertible for $\alpha lH\ne0$,
$R+l\ne0$ and $1-bx\ne0$, on a regular coordinate patch. On
$-1\le x\le1$, the bound
\[
 1-2\alpha mx-b^2x^2\ \ge\ 1-2|\alpha m|-b^2
\]
shows that
\begin{equation}\label{eq:pda-angular-domain}
 1-b^2>2|\alpha m|
\end{equation}
is sufficient both for $\Delta_x>0$ on the open interval and
for the absence of an angular projective pole on the closed interval.
It is not a classification of Lorentzian coordinate domains and
does not establish global regularity.

For $b=\alpha l\ne0$, $b^2\ne1$, the endpoints $x=\pm1$, when
present in the chosen domain, map to
\begin{equation}
 p_+=\frac{1+b}{1-b},\qquad p_-=\frac{1-b}{1+b},\qquad
 p_+p_-=1.
\end{equation}
At a regular portion of a simple angular root $p_i$ of $P$, the
surviving Killing square is proportional to
$(\dif T-p_i^2\dif\Phi)^2$. Its kernel is generated by
\begin{equation}
 L_i^{\rm PD}=p_i^2\partial_T+\partial_\Phi .
\end{equation}
In the physical chart the surviving square is proportional to
$\mathcal A^2$. Its kernels at the two endpoints are
\begin{equation}
 L_\pm^{A^+}=\partial_\phi\pm\frac{2l}{1+b^2}\partial_t .
\end{equation}
The constant Killing transformation gives
$\partial_t=\alpha H(\partial_T-\partial_\Phi)$ and
$\partial_\phi=H(\partial_T+\partial_\Phi)$. Thus
\begin{align}
 L_+^{A^+}&=\frac{H(1-b)^2}{1+b^2}L_+^{\rm PD},\\
 L_-^{A^+}&=\frac{H(1+b)^2}{1+b^2}L_-^{\rm PD}.
\end{align}
These nonzero factors relate two choices of unnormalized generators.
They are not conical surface gravities, which would require a
near-axis metric calculation and a choice of periodic generator.
In particular, the comparison of axial directions does not fix
conical deficits, common angular identifications or Misner strings.

\section{Finite-binary origin of the accelerating geometry}\label{binaryorigin}
The PD reconstruction establishes the local type-D embedding, but does not show how the acceleration-horizon direction emerges from a finite binary. We therefore dress a Bach--Weyl seed with two Belinski--Zakharov solitons~\cite{BelinskiZakharov1978} and calculate both open-rod kernels before taking the large-companion limit. The accelerating Kerr--NUT family is retained until the zero-Kerr specialization. The binary origin of this family is known; here the finite-rod calculation provides an explicit check of the limiting Killing directions in a common basis.

\subsection{Weyl conventions and the binary seed}
In the canonical stationary-axisymmetric reduction and rod description~\cite{Harmark2004}, related to the generalized Weyl framework~\cite{EmparanReall2002}, write the four-dimensional metric as
\begin{equation}\label{ism:eq:weyl}
 \dif s^2=G_{ab}\dif x^a\dif x^b+f(\dif\rho^2+\dif z^2),\qquad
 (x^1,x^2)=(t,\phi),\qquad \det G=-\rho^2.
\end{equation}
Let $z_1<z_2<z_3<z_4$ be the ordered turning points and define
\begin{equation}\label{ism:eq:mu}
 \mu_i=\sqrt{\rho^2+(z-z_i)^2}-(z-z_i),\qquad
 R_{ij}=\rho^2+\mu_i\mu_j.
\end{equation}
The finite horizon intervals of the static binary seed are $[z_1,z_2]$ and $[z_3,z_4]$. The Bach--Weyl Killing matrix and its orbit-space factor are
\begin{align}\label{ism:eq:BWfull}
 G_{\rm BW}&=\operatorname{diag}\left(-\frac{\mu_1\mu_3}{\mu_2\mu_4},\frac{\rho^2\mu_2\mu_4}{\mu_1\mu_3}\right),\\
 f_{\rm BW}&=k_B^2\frac{\mu_2\mu_4}{\mu_1\mu_3}
 \frac{R_{12}^2R_{14}^2R_{23}^2R_{34}^2}
 {R_{13}^2R_{24}^2R_{11}R_{22}R_{33}R_{44}}.\label{ism:eq:BWf}
\end{align}
Following the remove--and--readd procedure~\cite{Pomeransky2006}, we remove the first pair of trivial poles from the first diagonal component and factor out $\mu_2/\mu_1$ before readdition. We use the intermediate diagonal seed
\begin{equation}\label{ism:eq:g0b}
 G_{0B}=\operatorname{diag}\left(-\frac{\mu_3}{\mu_4},\frac{\rho^2\mu_4}{\mu_3}\right),
\end{equation}
obtained by the trivial removal stage of the two-soliton construction. Its generating matrix is
\begin{equation}\label{ism:eq:psib}
 \Psi_B(\lambda)=\operatorname{diag}\left(
 -\frac{\mu_3-\lambda}{\mu_4-\lambda},
 \left(\lambda+\frac{\rho^2}{\mu_3}\right)(\mu_4-\lambda)\right).
\end{equation}
We choose the poles and constant BZ rows
\begin{equation}\label{ism:eq:poles}
 p_1=-\frac{\rho^2}{\mu_1},\qquad p_2=\mu_2,
 \qquad m_{0B}^{(i)}=(1,C_i^B),\quad i=1,2.
\end{equation}
The dressed rows are $m_B^{(i)}=m_{0B}^{(i)}\Psi_B^{-1}(p_i)$, and the Gram matrix is
\begin{equation}\label{ism:eq:gram}
 \Gamma_{ij}=\frac{m^{(i)}G_0(m^{(j)})^T}{\rho^2+p_ip_j}.
\end{equation}
With $L^{(i)}=m^{(i)}G_0$, the dressing reconstruction reads
\begin{equation}\label{ism:eq:dress}
 \widetilde G_{ab}=(G_0)_{ab}
 -\sum_{i,j=1}^2\frac{(\Gamma^{-1})_{ij}}{p_ip_j}L_a^{(i)}L_b^{(j)},
 \qquad G=\frac{\mu_2}{\mu_1}\widetilde G.
\end{equation}
The final scalar factor restores the canonical determinant. We verify this directly from the finite reconstruction rather than assuming it.  Define the $2\times2$ matrices $P_*=\operatorname{diag}(p_1,p_2)$ and $U$ whose $i$th column is $(L^{(i)})^T/p_i$.  Then $\widetilde G=G_0-U\Gamma^{-1}U^T$.  The matrix determinant lemma gives
\begin{equation}\label{ism:eq:detlemma}
 \frac{\det\widetilde G}{\det G_0}
 =\det\!\left(I-\Gamma^{-1}U^TG_0^{-1}U\right).
\end{equation}
The definition of the Gram matrix implies
\begin{equation}\label{ism:eq:detgramidentity}
 (U^TG_0^{-1}U)_{ij}
 =\frac{m^{(i)}G_0(m^{(j)})^T}{p_ip_j}
 =\Gamma_{ij}\left(1+\frac{\rho^2}{p_ip_j}\right),
\end{equation}
so that $U^TG_0^{-1}U=\Gamma+\rho^2P_*^{-1}\Gamma P_*^{-1}$.  Consequently,
\begin{equation}\label{ism:eq:detdressing}
 \det\widetilde G
 =\det G_0\,\frac{\rho^4}{p_1^2p_2^2}
 =-\rho^2\frac{\mu_1^2}{\mu_2^2},
\end{equation}
where $\det G_0=-\rho^2$, $p_1=-\rho^2/\mu_1$, and $p_2=\mu_2$ have been used.  Multiplying the entire $2\times2$ Killing matrix by $\mu_2/\mu_1$ therefore yields
\begin{equation}\label{ism:eq:canonicalproof}
 \det G=\left(\frac{\mu_2}{\mu_1}\right)^2\det\widetilde G=-\rho^2.
\end{equation}
The proof holds on the generic patch where the matrices and denominators in the determinant lemma are defined.  It applies both to the finite binary and to its limiting two-soliton reconstruction. The orbit-space factor is calculated from the static seed factor and the ratio of Gram determinants,
\begin{equation}\label{ism:eq:conf}
 f_B=f_{0B}\frac{\det\Gamma_B(C_1^B,C_2^B)}{\det\Gamma_B(0,0)}.
\end{equation}
The trivial-polarization limit recovers the original static binary. Throughout, matrix inversion and rational identities are understood on a generic local patch where the relevant denominators do not vanish.

\subsection{Explicit finite two-soliton binary metric}\label{ism:sec:explicit}
We first evaluate the finite stationary binary before taking either limit. The result is expressed explicitly in terms of the four Weyl functions and two BZ parameters.

The dressed BZ rows, with the poles in Eq.~\eqref{ism:eq:poles}, are
\begin{equation}\label{ism:eq:binaryrows}
 m_B^{(i)}=\left(\frac{p_i-\mu_4}{\mu_3-p_i},
 \frac{C_i^B\mu_3}{(\rho^2+\mu_3p_i)(\mu_4-p_i)}\right),\qquad i=1,2.
\end{equation}
The Gram entries are given explicitly by
\begin{align}\label{ism:eq:binarygramentries}
 \Gamma_{ij}^B=\frac{1}{\rho^2+p_ip_j}\bigg[&
 -\frac{\mu_3}{\mu_4}\frac{(p_i-\mu_4)(p_j-\mu_4)}{(\mu_3-p_i)(\mu_3-p_j)}\\
 &+\frac{\rho^2\mu_4}{\mu_3}
 \frac{C_i^BC_j^B\mu_3^2}{(\rho^2+\mu_3p_i)(\rho^2+\mu_3p_j)(\mu_4-p_i)(\mu_4-p_j)}
 \bigg].
\end{align}
The two components of $L^{(i)}=m_B^{(i)}G_{0B}$, which are already fixed by Eqs.~\eqref{ism:eq:g0b} and \eqref{ism:eq:binaryrows}, are
\begin{equation}\label{ism:eq:binaryL}
 L_t^{(i)}=-\frac{\mu_3}{\mu_4}\frac{p_i-\mu_4}{\mu_3-p_i},\qquad
 L_\phi^{(i)}=\frac{\rho^2\mu_4}{\mu_3}
 \frac{C_i^B\mu_3}{(\rho^2+\mu_3p_i)(\mu_4-p_i)}.
\end{equation}
The finite binary metric follows without an unevaluated matrix inverse:
\begin{align}\label{ism:eq:binaryexplicit}
 g_{tt}^B=\frac{\mu_2}{\mu_1}\bigg[-\frac{\mu_3}{\mu_4}
 &-\frac{1}{\det\Gamma_B}\left(\frac{\Gamma_{22}^B(L_t^{(1)})^2}{p_1^2}
 -\frac{2\Gamma_{12}^BL_t^{(1)}L_t^{(2)}}{p_1p_2}
 +\frac{\Gamma_{11}^B(L_t^{(2)})^2}{p_2^2}\right)\bigg],\\
 g_{t\phi}^B=-\frac{\mu_2}{\mu_1\det\Gamma_B}&\left(
 \frac{\Gamma_{22}^BL_t^{(1)}L_\phi^{(1)}}{p_1^2}
 -\frac{\Gamma_{12}^B(L_t^{(1)}L_\phi^{(2)}+L_t^{(2)}L_\phi^{(1)})}{p_1p_2}
 +\frac{\Gamma_{11}^BL_t^{(2)}L_\phi^{(2)}}{p_2^2}\right),\label{ism:eq:binaryexplicit12}\\
 g_{\phi\phi}^B=\frac{\mu_2}{\mu_1}\bigg[\frac{\rho^2\mu_4}{\mu_3}
 &-\frac{1}{\det\Gamma_B}\left(\frac{\Gamma_{22}^B(L_\phi^{(1)})^2}{p_1^2}
 -\frac{2\Gamma_{12}^BL_\phi^{(1)}L_\phi^{(2)}}{p_1p_2}
 +\frac{\Gamma_{11}^B(L_\phi^{(2)})^2}{p_2^2}\right)\bigg].\label{ism:eq:binaryexplicit22}
\end{align}
The orbit-space coefficient is
\begin{equation}\label{ism:eq:binaryexplicitf}
 f_B=k_B^2\frac{\mu_2\mu_4}{\mu_1\mu_3}
 \frac{R_{12}^2R_{14}^2R_{23}^2R_{34}^2}{R_{13}^2R_{24}^2R_{11}R_{22}R_{33}R_{44}}
 \frac{\det\Gamma_B(C_1^B,C_2^B)}{\det\Gamma_B(0,0)}.
\end{equation}
The determinant in this formula is the ordinary $2\times2$ determinant of the entries in Eq.~\eqref{ism:eq:binarygramentries}. Thus Eqs.~\eqref{ism:eq:binaryrows}--\eqref{ism:eq:binaryexplicitf} specify the complete finite binary by rational operations in the original Weyl and BZ variables. The determinant and zero-polarization identities are independent checks of the reconstruction.

\subsection{Direct large-companion limit of the dressed binary}\label{ism:sec:direct}
\subsubsection{The finite binary is constructed first}
The relevant object is the full stationary metric $g_B$, given explicitly by Eqs.~\eqref{ism:eq:binaryexplicit}--\eqref{ism:eq:binaryexplicitf} at \emph{finite} $z_4$. Its Killing block is a rational function of the four soliton functions and the two BZ polarizations. We do not first replace the double-Schwarzschild seed by a C-metric seed and then dress the latter. The limiting spectral data will be used below to evaluate this direct limit, not to define a second dressing operation.

For the large companion we hold the three finite endpoints fixed and set
\begin{equation}\label{ism:eq:directlimit}
 z_4=L\to+\infty,\qquad z_1,z_2,z_3\ \text{fixed},\qquad \varepsilon=L^{-1}.
\end{equation}
At fixed $\rho,z$ away from the rods, $\mu_4=2L-2z+O(L^{-1})$. To determine the required Killing-coordinate rescaling, consider the trivial-polarization member of the finite binary. In the original coordinates its Killing components behave as
\begin{align}
 (G_{\rm BW})_{tt}&=-\frac{\mu_1\mu_3}{2L\mu_2}+O(L^{-2}),\\
 (G_{\rm BW})_{\phi\phi}&=2L\frac{\rho^2\mu_2}{\mu_1\mu_3}+O(1).
\end{align}
Thus the untransformed components have no finite nondegenerate limit. This failure is not, by itself, a curvature singularity: a parameter-dependent identification of the Killing coordinates can remove it. Consider the rescaling $t=L^pT$, $\phi=L^q\Phi$. The two displayed components then scale as $L^{2p-1}$ and $L^{2q+1}$. Requiring finite nonzero values fixes $p=1/2$ and $q=-1/2$. The factors of $2$ below choose the conventional Weyl normalization. This argument determines the diagonal scaling in the static reference branch; it does not establish uniqueness among general $L$-dependent mixtures of Killing fields.

The polarization scaling is a separate question. It is determined by requiring the nontrivial stationary deformation to survive in this limiting basis. The finite dressed rows of Eq.~\eqref{ism:eq:binaryrows} have first components of order $L$ and second components of order $C_i^B/L$. After the contragredient Killing-basis change and an allowed overall normalization of each BZ row, the representative $(m_{B,1}^{(i)}/(2L),m_{B,2}^{(i)})$ has a finite nonzero second entry precisely when $C_i^B$ is of order $L$. We write the finite limiting coefficients as $C_i^C$, fixing the numerical factor by the convention of the limiting generating matrix. The nondegenerate limit therefore uses
\begin{equation}\label{ism:eq:directscaling}
 S_L=\operatorname{diag}\bigl(\sqrt{2L},(2L)^{-1/2}\bigr),\qquad
 C_i^B=2LC_i^C\quad(i=1,2).
\end{equation}
Define the transformed \emph{finite} Killing matrix by
\begin{equation}\label{ism:eq:finiteGB}
 G_B^{\rm tr}(L)=S_L^T G_B(L,C_1^B,C_2^B)S_L.
\end{equation}
The transformation is an ordinary constant change of Killing basis for each finite $L$; its singularity occurs only as $L\to\infty$. In particular $\det S_L=1$, so the canonical Weyl determinant remains $-\rho^2$.

\subsubsection{Covariance of the finite dressing and normalization of BZ rows}
The limiting calculation can be performed on the finite reconstruction formula using its covariance under constant Killing-coordinate transformations and the freedom to rescale individual Belinski--Zakharov rows.

For any invertible constant matrix $S$, let $x=Sx'$ and $G_0'=S^TG_0S$. With our row-vector convention the dressed polarizations transform as $m'^{(i)}=m^{(i)}S^{-T}$. Their Gram matrix is unchanged, since
\begin{equation}\label{ism:eq:covgram}
 \Gamma'_{ij}=\frac{m^{(i)}S^{-T}S^TG_0S S^{-1}(m^{(j)})^T}{\rho^2+p_ip_j}=\Gamma_{ij},
 \qquad L'^{(i)}=L^{(i)}S.
\end{equation}
Substituting into the reconstruction formula, including the scalar remove--readd factor, gives the exact finite-parameter identity
\begin{equation}\label{ism:eq:covmetric}
 \frac{\mu_2}{\mu_1}\left[G'_0-\sum_{i,j=1}^2\frac{(\Gamma'^{-1})_{ij}}{p_ip_j}(L'^{(i)})^TL'^{(j)}\right]=S^TG_BS.
\end{equation}
No limit has been taken in Eq.~\eqref{ism:eq:covmetric}.

There is also an independent freedom to multiply each BZ row by a nonzero factor $d_i$. For $D=\operatorname{diag}(d_1,d_2)$ this changes $\Gamma$ into $D\Gamma D$ and the matrix of dressing rows $L$ into $DL$. Because $(D\Gamma D)^{-1}=D^{-1}\Gamma^{-1}D^{-1}$, the dressed Killing matrix is unaffected. This observation permits a normalization of the transformed BZ rows that remains finite as $L\to\infty$.

For the particular $S_L$ in Eq.~\eqref{ism:eq:directscaling}, take $d_i=(2L)^{-1/2}$ for both rows. Then
\begin{equation}\label{ism:eq:rowexact}
 \widehat m_B^{(i)}=\frac{1}{\sqrt{2L}}m_B^{(i)}S_L^{-T}
 =\left(\frac{m_{B,1}^{(i)}}{2L},m_{B,2}^{(i)}\right),
 \qquad \widehat\Gamma_B=\frac{\Gamma_B}{2L}.
\end{equation}
The last equality is exact at finite $L$, not merely asymptotic. In particular the normalized Gram determinant acquires a factor $(2L)^{-2}$, which cancels from the polarized-to-trivial determinant ratio used in the conformal factor. The scalar factor $\mu_2/\mu_1$ is unaffected by $S_L$ and is independent of the distant endpoint $z_4$.

Equation~\eqref{ism:eq:covmetric} identifies the normalized reconstruction with the original finite binary metric in a different Killing basis. The limit can therefore be taken in the normalized coordinates.

\subsubsection{Explicit componentwise limiting prescription}
The direct limit is taken on Eqs.~\eqref{ism:eq:binaryexplicit}--\eqref{ism:eq:binaryexplicitf} using the single substitution
\begin{equation}\label{ism:eq:substitutionlimit}
 \mu_4=\frac2\varepsilon-2z+\frac{\rho^2\varepsilon}{2}+O(\varepsilon^2),\qquad
 C_i^B=\frac{2C_i^C}{\varepsilon},\qquad \varepsilon\to0^+.
\end{equation}
For clarity, the three entries to be limited are $2g_{tt}^B/\varepsilon$, $g_{t\phi}^B$, and $\varepsilon g_{\phi\phi}^B/2$, respectively, because the transformation $S_L$ acts on the two Killing coordinates. These are precisely the entries of $S_L^TG_BS_L$. In particular, \emph{one must not} take a naive entrywise limit of $G_B$ before the Killing-basis rescaling.

The scaling of the three Killing components is fixed by the diagonal entries of $S_L$. The factors $2L$, $1$, and $(2L)^{-1}$ multiplying $g^B_{tt}$, $g^B_{t\phi}$, and $g^B_{\phi\phi}$ respectively are therefore not independent normalization choices. To see explicitly how the singular factors cancel, consider the dressed rows of Eq.~\eqref{ism:eq:binaryrows} before taking the limit. In the normalized representatives of Eq.~\eqref{ism:eq:rowexact}, set $q_i=(\mu_4-p_i)/(2L)$. The first entry is $q_i/(p_i-\mu_3)$ and the second is $q_i^{-1}C_i^C\mu_3/(\rho^2+\mu_3p_i)$. At fixed interior Weyl coordinates $q_i\to1$. Also the two transformed seed entries acquire the factors $2L/\mu_4$ and $\mu_4/(2L)$ relative to their limits. These exact relations give the limiting data directly from the finite reconstruction.

In the limit the intermediate matrix becomes
$G_{0C}=\operatorname{diag}(-\mu_3,\rho^2/\mu_3)$, while the dressed rows reduce to
\begin{equation}\label{ism:eq:Cexplicitrows}
 m_C^{(i)}=\left(\frac{1}{p_i-\mu_3},
 \frac{C_i^C\mu_3}{\rho^2+\mu_3p_i}\right),\qquad i=1,2.
\end{equation}
The Gram entries are consequently
\begin{equation}\label{ism:eq:CexplicitGram}
 \Gamma_{ij}^C=\frac{1}{\rho^2+p_ip_j}
 \left[-\frac{\mu_3}{(p_i-\mu_3)(p_j-\mu_3)}
 +\frac{\rho^2\mu_3C_i^CC_j^C}{(\rho^2+\mu_3p_i)(\rho^2+\mu_3p_j)}\right].
\end{equation}
The limiting components can be written without an implicit matrix inversion. In the following equations, the entries of $\Gamma_C$ are those in Eq.~\eqref{ism:eq:CexplicitGram}, and $L^{(i)}=m_C^{(i)}G_{0C}$, with $G_{0C}=\operatorname{diag}(-\mu_3,\rho^2/\mu_3)$. Thus
\begin{align}
 L_t^{(i)}&=-\frac{\mu_3}{p_i-\mu_3},&
 L_\phi^{(i)}&=\frac{\rho^2 C_i^C}{\rho^2+\mu_3p_i}.
\end{align}
Taking the limit of the three entries of the \emph{finite} metric in Eq.~\eqref{ism:eq:binaryexplicit}, with the rescaling in Eq.~\eqref{ism:eq:directscaling}, gives
\begin{align}
 g_{tt}^C&=\frac{\mu_2}{\mu_1}\left[-\mu_3
 -\frac{1}{\det\Gamma_C}\left(
 \frac{\Gamma_{22}^C(L_t^{(1)})^2}{p_1^2}
 -\frac{2\Gamma_{12}^C L_t^{(1)}L_t^{(2)}}{p_1p_2}
 +\frac{\Gamma_{11}^C(L_t^{(2)})^2}{p_2^2}\right)\right],\\
 g_{t\phi}^C&=-\frac{\mu_2}{\mu_1\det\Gamma_C}\left(
 \frac{\Gamma_{22}^C L_t^{(1)}L_\phi^{(1)}}{p_1^2}
 -\frac{\Gamma_{12}^C(L_t^{(1)}L_\phi^{(2)}+L_t^{(2)}L_\phi^{(1)})}{p_1p_2}
 +\frac{\Gamma_{11}^C L_t^{(2)}L_\phi^{(2)}}{p_2^2}\right),\\
 g_{\phi\phi}^C&=\frac{\mu_2}{\mu_1}\left[\frac{\rho^2}{\mu_3}
 -\frac{1}{\det\Gamma_C}\left(
 \frac{\Gamma_{22}^C(L_\phi^{(1)})^2}{p_1^2}
 -\frac{2\Gamma_{12}^C L_\phi^{(1)}L_\phi^{(2)}}{p_1p_2}
 +\frac{\Gamma_{11}^C(L_\phi^{(2)})^2}{p_2^2}\right)\right].\label{ism:eq:limitcomponents}
\end{align}
These are explicit rational expressions in $\rho,\mu_1,\mu_2,\mu_3,C_1^C,C_2^C$ after insertion of Eqs.~\eqref{ism:eq:poles} and \eqref{ism:eq:CexplicitGram}. They are obtained as the componentwise limit of the finite binary, not by applying a new soliton transformation to a limiting seed.

\subsubsection{Direct limiting Killing block}
Taking $L\to\infty$ in the already reconstructed $G_B^{\rm tr}$ gives
\begin{equation}\label{ism:eq:directG}
 \lim_{L\to\infty}G_B^{\rm tr}(L)=G_C,
\end{equation}
where $G_C$ denotes the \emph{dressed} two-soliton C-metric Killing block, not its static seed. Indeed, the normalized rows and the transformed seed converge with the poles fixed. Their Gram matrix and the projected dressing rows therefore converge entry by entry. Wherever the limiting Gram determinant is nonzero, inversion is continuous, and the finite reconstruction identity \eqref{ism:eq:covmetric} gives \eqref{ism:eq:directG}. This is a limit off the rods; it does not in itself give the limit of a finite rod kernel, which is checked in Sec.~\ref{ism:sec:finite-rod-directions}.

\subsection{Direct limit of the orbit-space conformal factor}\label{ism:sec:factor}
The Killing-block calculation does not determine the coefficient of the two-dimensional orbit space. In particular, a constant change of Killing basis leaves $f_B$ unchanged. We must therefore take the large-companion limit of the complete finite-binary conformal factor in Eq.~\eqref{ism:eq:binaryexplicitf} independently. The argument has two parts: the asymptotics of the Bach--Weyl seed and the behavior of the polarization-dependent Gram determinant.

\subsubsection{Asymptotics of the original binary seed}
Write $f_{0B}=k_B^2f_{0B}^{\rm unit}$. In the limit $z_4=L\to\infty$, with $\rho,z,z_1,z_2,z_3$ fixed, $\mu_4=2L-2z+O(L^{-1})$. Consequently, for $i=1,2,3$,
\begin{equation}
 R_{i4}=\mu_i\mu_4[1+O(L^{-1})],\qquad R_{44}=\mu_4^2[1+O(L^{-2})].
\end{equation}
Separating the factors involving the remote endpoint in the exact Bach--Weyl expression gives
\begin{align}
 \frac{f_{0B}}{k_B^2}
 &=\frac{\mu_2\mu_4}{\mu_1\mu_3}
 \frac{R_{12}^2R_{23}^2}{R_{13}^2R_{11}R_{22}R_{33}}
 \frac{R_{14}^2R_{34}^2}{R_{24}^2R_{44}}=\mu_4\frac{\mu_1\mu_3}{\mu_2}
 \frac{R_{12}^2R_{23}^2}{R_{13}^2R_{11}R_{22}R_{33}}
 [1+O(L^{-1})].
\end{align}
The unit-normalized C-metric seed coefficient is thus naturally identified as
\begin{equation}\label{ism:eq:f0c}
 F_{0C}=\frac{\mu_1\mu_3}{\mu_2}
 \frac{R_{12}^2R_{23}^2}{R_{13}^2R_{11}R_{22}R_{33}},
 \qquad f_{0C}=k_C^2F_{0C}.
\end{equation}
It follows, without using the stationary dressing, that
\begin{equation}\label{ism:eq:F0}
 \lim_{L\to\infty}\frac{f_{0B}}{Lk_B^2}=2F_{0C}.
\end{equation}
The factor of two follows from $\mu_4\sim 2L$.

\subsubsection{The determinant ratio and the complete direct limit}
The dressing multiplies the static conformal factor by the ratio
\begin{equation}\label{ism:eq:ratios}
 \mathcal R_B=\frac{\det\Gamma_B(C_1^B,C_2^B)}{\det\Gamma_B(0,0)},\qquad
 \mathcal R_C=\frac{\det\Gamma_C(C_1^C,C_2^C)}{\det\Gamma_C(0,0)}.
\end{equation}
The denominator is the same two-pole construction evaluated with trivial polarizations, not the determinant of the static seed. By Eq.~\eqref{ism:eq:rowexact}, the normalized Gram determinant is $(2L)^{-2}\det\Gamma_B$ for both the polarized and trivial-polarization constructions. This factor cancels exactly in their ratio at every finite $L$. Since the normalized Gram entries have finite limits, on the branch where the limiting reference determinant does not vanish,
\begin{equation}\label{ism:eq:ratiolim}
 \lim_{L\to\infty}\mathcal R_B=\mathcal R_C.
\end{equation}
These observations apply directly to Eq.~\eqref{ism:eq:binaryexplicitf}, the conformal factor of the \emph{already dressed finite binary}. Combining them gives
\begin{equation}\label{ism:eq:directf}
 \lim_{L\to\infty}\frac{f_B}{Lk_B^2}=2F_{0C}\mathcal R_C.
\end{equation}
The C-metric-based dressed coefficient is $f_C=k_C^2F_{0C}\mathcal R_C$. Therefore the normalization required for the same finite orbit-space metric is
\begin{equation}\label{ism:eq:knorm}
 k_B^2=\frac{k_C^2}{2L},\qquad
 \lim_{L\to\infty}f_B=f_C.
\end{equation}
The constant $k_C$ is subsequently fixed by the Jacobian of the physical accelerating coordinates in Sec.~\ref{ism:sec:physical}.

\subsection{Accelerating Kerr--NUT in physical coordinates}\label{ism:sec:physical}
We identify the limiting metric by introducing coordinates adapted to its three turning points and reducing the Killing and orbit-space sectors separately. The accelerating Kerr--NUT form will then follow from the resulting parameter relations.
For the coordinate calculation we use
\begin{align}
 \Omega&=1+\alpha rx, &
 \Delta&=(r-r_+)(r-r_-)(1-\alpha^2r^2),\\
 P&=(1-x^2)(1+\alpha r_+x)(1+\alpha r_-x), &
 \Sigma&=[r+\alpha(a^2-n^2)x]^2+[ax-n-\alpha r(a-nx)]^2.
\end{align}
The expression for $\Sigma$ will be recovered from the ISM denominator below.
To adapt the coordinates to the finite black-hole rod and the semi-infinite acceleration-horizon rod, put $z_0=(z_1+z_2)/2$, $\sigma=(z_2-z_1)/2$ and $r_\pm=m\pm\sigma$. We choose $r=r_+$ and $r=1/\alpha$ as the respective horizon locations, and $x=\pm1$ as the axis locations. A canonical Weyl transformation consistent with these boundary assignments is \begin{align}
 \rho^2&=\frac{(r-r_+)(r-r_-)(1-\alpha^2r^2)(1-x^2)(1+\alpha r_+x)(1+\alpha r_-x)}{\Omega^4},\\
 z-z_0&=\frac{(x+\alpha r)[(r-m)(1+\alpha mx)+\alpha\sigma^2x]}{\Omega^2}.
\end{align}
\subsubsection{Accelerating coordinates and horizon rods}
The ordered turning points imply $\sigma>0$ and $c=z_3-z_0>\sigma$.
Using the turning-point relation $2\alpha c=1-\alpha^2r_+r_-$,
\begin{equation}
 2\alpha(c-\sigma)=(1-\alpha r_+)(1+\alpha r_-).
\end{equation}
On the conventional real branch $\alpha>0$, $r_+>r_-$ and
$|\alpha r_\pm|<1$, a natural open exterior patch is
$r_+<r<1/\alpha$, $-1<x<1$.  There $\Delta>0$, $P>0$ and
$\Omega>0$, and hence $\rho^2>0$.
The coordinate map can be checked directly on its horizon boundaries:
\begin{align}
 z(r_+,x)-z_0&=\sigma\frac{x+\alpha r_+}{1+\alpha r_+x},\\
 z(1/\alpha,x)-z_3&=
 \frac{(1-\alpha r_+)(1-\alpha r_-)}{2\alpha}
 \frac{1-x}{1+x}.
\end{align}
The first map runs monotonically from $z_1$ to $z_2$ as $x$
runs from $-1$ to $1$, because
\begin{equation}
 \left.\partial_x z\right|_{r=r_+}=
 \frac{\sigma(1-\alpha^2r_+^2)}{(1+\alpha r_+x)^2}>0.
\end{equation}
The second map runs from $z_3$ at $x=1$ to $+\infty$ as
$x\to-1^+$.  Thus the finite horizon rod and the semi-infinite
acceleration-horizon candidate have the expected coordinate locations.
This boundary correspondence is distinct from verifying their stationary
Killing generators, and it does not establish global axis regularity.

\subsubsection{Gram-determinant structure}
Substitution of the rational Weyl functions into the limiting Gram determinant gives a sum-of-squares expression before any physical parameter identification. Define
\begin{align}
 X={}&(1-\alpha^2r_+^2)(r-r_-)(1+\alpha r_-x)
 +C_1^CC_2^C(1-\alpha^2r_-^2)(r-r_+)(1+\alpha r_+x),\\
 Y={}&-C_1^C(1-x)(1+\alpha r)(1-\alpha r_-)(1-\alpha r_+)-C_2^C(1+x)(1-\alpha r)(1+\alpha r_-)(1+\alpha r_+).
\end{align}
The factorization of the $2\times2$ Gram determinant is
\begin{equation}\label{ism:eq:gramphysical}
 \mathcal R_C=\frac{X^2+\sigma^2Y^2}
 {(1-\alpha^2r_+^2)^2(r-r_-)^2(1+\alpha r_-x)^2}.
\end{equation}
This equality is an algebraic identity in the independent polarization parameters. The identity follows by comparing the coefficients of $(C_1^C)^2$, $(C_2^C)^2$, $C_1^CC_2^C$ and $(C_1^CC_2^C)^2$ after substituting the rational Weyl functions; the unpolarized term agrees separately.

\subsubsection{Factorization of the Killing matrix}
The identification of the physical metric likewise requires a componentwise
comparison, not just recognition of its common denominator. The componentwise polynomial reduction leads to the following factorized
Killing sector:
\begin{equation}
 \dif s_K^2=\frac{-\Delta(A\dif t+B\dif\phi)^2
 +P(C\dif t+D\dif\phi)^2}{\alpha\Omega^2W},
\end{equation}
Its three independent components consequently satisfy
\begin{align}
 \alpha\Omega^2Wg_{tt}&=-\Delta A^2+PC^2,\\
 \alpha\Omega^2Wg_{t\phi}&=-\Delta AB+PCD,\\
 \alpha\Omega^2Wg_{\phi\phi}&=-\Delta B^2+PD^2.
\end{align}
The determinant of this Killing matrix is
\begin{equation}
 \det G=-\frac{\Delta P}{\alpha^2\Omega^4W^2}(AD-BC)^2.
\end{equation}
Since $\rho^2=\Delta P/\Omega^4$, canonical normalization demands
$(AD-BC)^2=\alpha^2W^2$. This condition is necessary, not sufficient:
the individual component identities must still be established from the
limiting ISM expressions. The explicit coefficientwise mixed-component
comparison is supplied in Appendix~\ref{ism:app:factorwise}.  The time--time
comparison deserves a separate normalization check: factoring the Gram
determinant and factoring the reconstructed numerator are not sufficient
unless they are expressed over the same rational denominator.  Appendix~\ref{ism:app:timepoly}
therefore derives both polynomials with one common multiplier $Q$ and then
checks their equality with the accelerating-coordinate two-square expression.
The common denominator normalization is fixed entirely by the ISM reconstruction. The last component then follows from the first two: the reconstructed block has determinant $-\rho^2$ by Eq.~\eqref{ism:eq:canonicalproof}, and the two-square block has the same determinant by the coefficient identity $AD-BC=\alpha W$. Thus $g_{\phi\phi}=(g_{t\phi}^2-\rho^2)/g_{tt}$ wherever $g_{tt}\ne0$. Equality extends across its zeros wherever both rational expressions remain regular.

The algebraic origin of the sum of squares is apparent from the limiting $tt$ component before the physical parameter dictionary is imposed. Write $v=\rho^2$, $R_{ij}=v+\mu_i\mu_j$, and introduce the polynomial combinations
\begin{align}
 F_0&=R_{12}R_{23}(\mu_1-\mu_3), &
 F_{12}&=R_{12}R_{13}(\mu_2-\mu_3),\\
 G_1&=(\mu_1-\mu_2)R_{23}R_{13}, &
 G_2&=v(\mu_1-\mu_2)(\mu_1-\mu_3)(\mu_2-\mu_3).
\end{align}
The entries of $\Gamma_C$ are the sum of a polarization-independent term and a term proportional to $C_i^CC_j^C$. Hence its determinant has exactly five polarization monomials,\begin{equation*}
1,\quad (C_1^C)^2,\quad (C_2^C)^2,\quad C_1^CC_2^C,\quad (C_1^CC_2^C)^2.
\end{equation*}
Clearing the rational denominators with one common polarization-independent factor $Q$ gives
\begin{equation}\label{ism:eq:mainpolydet}
 Q\det\Gamma_C=\mu_1\mu_2\bigl[(F_0+C_1^CC_2^CF_{12})^2+(C_1^CG_1+C_2^CG_2)^2\bigr].
\end{equation}
The same $Q$ must be used in the time--time numerator. Its reduction is
\begin{equation}\label{ism:eq:mainpolytt}
 g_{tt}^C=\frac{\mu_3}{Q\det\Gamma_C}\left\{v\left(C_1^CG_1-\frac{\mu_1\mu_2}{v}C_2^CG_2\right)^2-(\mu_1F_0+C_1^CC_2^C\mu_2F_{12})^2\right\}.
\end{equation}
At zero polarization Eq.~\eqref{ism:eq:mainpolytt} reduces to $g_{tt}=-\mu_1\mu_3/\mu_2$. This fixes the factor $\mu_3$ in the numerator. The five coefficient identities and the common multiplier $Q$ are stated in Appendix~\ref{ism:app:timepoly}. After substituting the rational Weyl functions, Eq.~\eqref{ism:eq:mainpolytt} agrees with $(-\Delta A^2+PC^2)/(\alpha\Omega^2W)$; the four mixed-component identities are recorded in Appendix~\ref{ism:app:factorwise}. Together with the independently derived canonical determinant, these two component equalities fix the complete Killing block on the generic regular patch. This supplies the algebraic bridge between the original Gram reconstruction and the two-square representation, rather than taking the latter as an ansatz for the desired Kerr--NUT geometry.

The physical parameters are related to the turning points and BZ polarizations as follows. On the generic branch, the turning-point equation is $2\alpha(z_3-z_0)=1-\alpha^2r_+r_-$, while the remaining chart freedom is fixed by eliminating the mixed $rx$ term in the first square of the transformed ISM denominator:
\begin{equation}
 C_1^CC_2^C=-\frac{r_-(1-\alpha^2r_+^2)}{r_+(1-\alpha^2r_-^2)}.
\end{equation}
With $h=(1-\alpha^2r_+^2)/r_+$, the two linear polarization combinations define rotation and NUT parameters through
\begin{align}
 C_1^C(1-\alpha r_-)(1-\alpha r_+)&=h(a+n),\\
 C_2^C(1+\alpha r_-)(1+\alpha r_+)&=h(n-a).
\end{align}
The product relation follows by setting the coefficient of $rx$ in $X$ to zero. This coefficient is
\begin{equation}
 \alpha\bigl[r_-(1-\alpha^2r_+^2)+C_1^CC_2^C r_+(1-\alpha^2r_-^2)\bigr].
\end{equation}
Its cancellation gives the product relation above. Substitution into the full expression for $X$, not just its $rx$ coefficient, then yields the exact linear reduction
\begin{equation}
 X=2\sigma\frac{1-\alpha^2r_+^2}{r_+}
 \bigl(r+\alpha r_+r_-x\bigr).
\end{equation}
The preceding cancellation condition is meaningful only if it can be used to
reparametrize the original Weyl and BZ data. To establish this locally, keep
$c=z_3-(z_1+z_2)/2$, $\sigma=(z_2-z_1)/2$ and
$u=C_1^CC_2^C$ fixed, and regard $m,\alpha$ as the unknowns. The turning-point
relation and the product relation, with denominators cleared, become
\begin{align}
 F(m,\alpha)&=2\alpha c-1+\alpha^2(m^2-\sigma^2)=0,\\
 H(m,\alpha)&=u(m+\sigma)[1-\alpha^2(m-\sigma)^2]
 +(m-\sigma)[1-\alpha^2(m+\sigma)^2]=0.
\end{align}
These are equations for a change of parameters, not additional field equations.
The implicit-function theorem applies wherever the Jacobian
$J=\partial(F,H)/\partial(m,\alpha)$ does not vanish. Direct differentiation gives
\begin{align}
 F_m&=2\alpha^2m, \\
 F_\alpha&=2c+2\alpha(m^2-\sigma^2),\\
 H_m&=u[1-\alpha^2(m-\sigma)^2]+[1-\alpha^2(m+\sigma)^2]-2\alpha^2(m^2-\sigma^2)(u+1),\\
 H_\alpha&=-2\alpha(m^2-\sigma^2)[u(m-\sigma)+(m+\sigma)].
\end{align}
On substituting $F=H=0$ into $J=F_mH_\alpha-F_\alpha H_m$,
\begin{equation}\label{ism:eq:physicalJacobian}
 J\big|_{F=H=0}=-\frac{2\sigma}{\alpha r_+}
 (1-\alpha^2r_+^2)(1-\alpha^2r_+r_-).
\end{equation}
Thus, for a real solution of the two equations, the coordinate parameters
$m,\alpha$ depend locally and uniquely on $(c,u)$ at fixed $\sigma$ when
$\sigma\alpha r_+\ne0$, $1-\alpha^2r_+^2\ne0$ and
$1-\alpha^2r_+r_-\ne0$; the rational form of the product relation also
requires $1-\alpha^2r_-^2\ne0$. This is a local statement, not a proof
that a real solution exists for every choice of original parameters or that
the parametrization is globally one-to-one. The nonaccelerating limit and
the exceptional zero factors must be approached separately. In particular,
$1-\alpha^2r_+r_-=0$ would force $c=0$ through the turning-point relation
at finite nonzero $\alpha$, inconsistent with the strictly ordered exterior
turning points.

The second square retains the two polarizations separately. Substituting their defining relations into $Y$ gives
\begin{align}
 Y={}&-h(a+n)(1-x)(1+\alpha r)-h(n-a)(1+x)(1-\alpha r)=2h\bigl[ax-n-\alpha r(a-nx)\bigr].
\end{align}
Since $\lambda=2\sigma h$, this establishes
\begin{equation}\label{ism:eq:Yreduction}
 \sigma Y=\lambda\bigl[ax-n-\alpha r(a-nx)\bigr].
\end{equation}
The terms proportional to $1$, $x$, $\alpha r$, and $\alpha rx$ combine separately. Thus the two polarizations enter the rotation and NUT parameters through different linear combinations.

Multiplying the two defining polarization relations gives
\begin{equation}
 C_1^CC_2^C(1-\alpha^2r_-^2)(1-\alpha^2r_+^2)
 =h^2(n^2-a^2).
\end{equation}
Using the product relation above and $h=(1-\alpha^2r_+^2)/r_+$ yields $a^2-n^2=r_+r_-=m^2-\sigma^2$, hence $\sigma^2=m^2+n^2-a^2$. Together with the reduction of $X$, Eq.~\eqref{ism:eq:Yreduction} now gives the explicit denominator identity
\begin{align}\label{ism:eq:Wphysical}
 W=X^2+\sigma^2Y^2
 =\lambda^2\bigl\{&[r+\alpha(a^2-n^2)x]^2+[ax-n-\alpha r(a-nx)]^2\bigr\}
 =\lambda^2\Sigma.
\end{align}
The determinant of the physical Killing one-form coefficients provides a further check:
\begin{align}
 &\bigl[1-\alpha^2(a^2-n^2)x^2\bigr](r^2-a^2+n^2)+\bigl[a(1+x^2)-2nx\bigr]
 \bigl[a(1+\alpha^2r^2)+2\alpha nr\bigr]
 =\Sigma.
\end{align}
The equality follows by expansion and collection of powers of $\alpha$; it is an independent algebraic consistency check of the proposed physical one-forms, not a substitute for matching them with the ISM Killing components.

\subsubsection{Physical Killing coordinates}
The denominator identity does not fix the Killing numerators. Collecting the limiting Killing components into terms proportional to $\Delta$ and $P$ gives four coefficients:
\begin{align}
 A(x)={}&(1-\alpha^2r_+^2)(1+\alpha r_-x)^2
 +C_1^CC_2^C(1-\alpha^2r_-^2)(1+\alpha r_+x)^2,\\
 B(x)={}&\alpha\sigma\bigl[C_1^C(1-\alpha r_+)(1-\alpha r_-)(1-x)^2\nonumber\\
 &\hspace{25mm}-C_2^C(1+\alpha r_+)(1+\alpha r_-)(1+x)^2\bigr],\\
 C(r)={}&\sigma\bigl[-C_1^C(1-\alpha r_+)(1-\alpha r_-)(1+\alpha r)^2\nonumber\\
 &\hspace{25mm}+C_2^C(1+\alpha r_+)(1+\alpha r_-)(1-\alpha r)^2\bigr],\\
 D(r)={}&\alpha\bigl[(1-\alpha^2r_+^2)(r-r_-)^2
 +C_1^CC_2^C(1-\alpha^2r_-^2)(r-r_+)^2\bigr].
\end{align}
The componentwise identities for this factorization are checked against the reconstructed Killing matrix, with the mixed-component polynomial calculation given in Appendix~\ref{ism:app:factorwise}.

Substituting the polarization product relation into $A$ gives
\begin{align}
 A(x)&=h\bigl[r_+(1+\alpha r_-x)^2-r_-(1+\alpha r_+x)^2\bigr]=2\sigma h(1-\alpha^2r_+r_-x^2).
\end{align}
The terms linear in $x$ cancel. The radial coefficient simplifies by the same mechanism:
\begin{align}
 D(r)&=\alpha h\bigl[r_+(r-r_-)^2-r_-(r-r_+)^2\bigr]=2\alpha\sigma h(r^2-r_+r_-).
\end{align}
Using the individual polarization relations, the remaining coefficients become
\begin{align}
 B(x)&=\alpha\sigma h\bigl[(a+n)(1-x)^2-(n-a)(1+x)^2\bigr]=\alpha\lambda\bigl[a(1+x^2)-2nx\bigr],\\
 C(r)&=\sigma h\bigl[-(a+n)(1+\alpha r)^2+(n-a)(1-\alpha r)^2\bigr]=-\lambda\bigl[a(1+\alpha^2r^2)+2\alpha nr\bigr].
\end{align}
Since $r_+r_-=a^2-n^2$ and $\lambda=2\sigma h$, all four coefficients reduce to
\begin{align}
 A&=\lambda[1-\alpha^2(a^2-n^2)x^2],&
 B&=\alpha\lambda[a(1+x^2)-2nx],\\
 C&=-\lambda[a(1+\alpha^2r^2)+2\alpha nr],&
 D&=\alpha\lambda(r^2-a^2+n^2).
\end{align}
For $\alpha>0$, the finite change of the \emph{limiting} Killing coordinates
\begin{equation}
 t=\sqrt{\alpha}\,T,\qquad \phi=-\Phi/\sqrt{\alpha}
\end{equation}
then gives
\begin{equation}
 A\dif t+B\dif\phi=\lambda\sqrt{\alpha}\,\vartheta_r,
 \qquad C\dif t+D\dif\phi=-\lambda\sqrt{\alpha}\,\vartheta_x.
\end{equation}
Together with $W=\lambda^2\Sigma$, these relations give the stationary sector of the physical metric. The finite Killing-coordinate transformation here is distinct from $S_L$, which was required to obtain a nondegenerate large-companion limit.

The polarization product relation fixes the remaining coordinate freedom on the generic branch, without imposing a further physical constraint. The identification with Eq.~\eqref{ism:eq:targetAKN} is local; exceptional parameter branches and global angular identifications require separate treatment.

\subsubsection{Orbit-space metric and normalization}
Agreement of the Killing block fixes the Weyl radius through $\det G=-\rho^2$, but not the orbit-space conformal factor. To determine the latter, we use $\rho=\sqrt{\Delta P}/\Omega^2$, whose logarithmic derivatives are
\begin{align}
 \frac{\partial_r\rho}{\rho}&=\frac{\Delta'}{2\Delta}-\frac{2\alpha x}{\Omega}, &
 \frac{\partial_x\rho}{\rho}&=\frac{P'}{2P}-\frac{2\alpha r}{\Omega}.
\end{align}
For an arbitrary change of orbit-space coordinates the Euclidean Weyl line
element is
\begin{align}
 \dif\rho^2+\dif z^2={}&
 (\rho_r^2+z_r^2)\dif r^2
 +2(\rho_r\rho_x+z_rz_x)\dif r\,\dif x
 +(\rho_x^2+z_x^2)\dif x^2.
\end{align}
Differentiating $z(r,x)$ gives the conformality relations in the chosen exterior orientation:
\begin{equation}
 z_r=-\sqrt{\frac{P}{\Delta}}\,\rho_x,
 \qquad z_x=\sqrt{\frac{\Delta}{P}}\,\rho_r.
\end{equation}
They imply, by direct substitution,
\begin{align}
 \rho_r\rho_x+z_rz_x&=0,\\
 \rho_x^2+z_x^2&=\frac{\Delta}{P}(\rho_r^2+z_r^2),\\
 \dif\rho^2+\dif z^2&=(\rho_r^2+z_r^2)
 \left(\dif r^2+\frac{\Delta}{P}\dif x^2\right).
\end{align}
The required orbit-space normalization is therefore
\begin{equation}\label{ism:eq:orbitcriterion}
 f_C(\rho_r^2+z_r^2)=\frac{\Sigma}{\Omega^2\Delta}.
\end{equation}

For the static seed, the Weyl Jacobian gives
\begin{equation}\label{ism:eq:seedjac}
 F_{0C}(\rho_r^2+z_r^2)\Omega^2\Delta
 =\alpha(r-r_-)^2
 \frac{(1-\alpha r_+)^2}{(1-\alpha r_-)^2}
 (1+\alpha r_-x)^2.
\end{equation}
Multiplying Eqs.~\eqref{ism:eq:gramphysical} and \eqref{ism:eq:seedjac}, and using $f_C=k_C^2F_{0C}\mathcal R_C$, gives
\begin{equation}\label{ism:eq:dressedjac}
 f_C(\rho_r^2+z_r^2)\Omega^2\Delta
 =\frac{\alpha k_C^2(X^2+\sigma^2Y^2)}
 {(1-\alpha r_-)^2(1+\alpha r_+)^2}.
\end{equation}
The physical parameter relations derived above imply $X^2+\sigma^2Y^2=\lambda^2\Sigma$, with
\begin{equation}
 \lambda=\frac{2\sigma(1-\alpha^2r_+^2)}{r_+}.
\end{equation}
Thus Eq.~\eqref{ism:eq:orbitcriterion} fixes the remaining constant:
\begin{equation}\label{ism:eq:kCphysical}
 k_C^2=\frac{(1-\alpha r_-)^2(1+\alpha r_+)^2}{\alpha\lambda^2}.
\end{equation}
The resulting normalization holds on the generic branch where the displayed denominators are nonzero. Exceptional limits must be treated separately.

\subsubsection{Final metric and horizon generators}
The preceding reductions, together with the orbit-space normalization,
produce the following local metric in the chosen accelerating coordinates:
\begin{align}\label{ism:eq:targetAKN}
 \dif s^2=\frac{1}{\Omega^2}\bigg[&-\frac{\Delta}{\Sigma}\,\vartheta_r^2
 +\frac{P}{\Sigma}\,\vartheta_x^2
 +\Sigma\left(\frac{\dif r^2}{\Delta}+\frac{\dif x^2}{P}\right)\bigg],\\
 \vartheta_r&=[1-\alpha^2(a^2-n^2)x^2]\dif T
 -[a(1+x^2)-2nx]\dif\Phi,\nonumber\\
 \vartheta_x&=[a(1+\alpha^2r^2)+2\alpha nr]\dif T
 +(r^2-a^2+n^2)\dif\Phi.\nonumber
\end{align}
Here $\Delta=(r^2-2mr+a^2-n^2)(1-\alpha^2r^2)$ follows from
$r_+r_-=a^2-n^2$ and $r_++r_-=2m$.  Likewise
$P=(1-x^2)[1+2m\alpha x+\alpha^2(a^2-n^2)x^2]$.
The equality holds on the generic parameter branch specified above, with both the Killing and orbit-space sectors fixed.

At a simple zero $r=r_H$ of $\Delta$, the surviving Killing one-form
is $\vartheta_x$.  The generator $\chi_H=\partial_T+\Omega_H\partial_\Phi$
therefore satisfies $\vartheta_x(\chi_H)=0$, giving
\begin{equation}
 \Omega_H=-\frac{a(1+\alpha^2r_H^2)+2\alpha nr_H}
 {r_H^2-a^2+n^2}.
\end{equation}
In particular, for $r_H=r_+$ and $r_H=1/\alpha$,
\begin{align}
 \Omega_{\mathrm{BH}}&=-\frac{a(1+\alpha^2r_+^2)+2\alpha nr_+}
 {r_+^2-a^2+n^2},\\
 \Omega_{\mathrm{A}}&=-\frac{2\alpha^2(a+n)}
 {1-\alpha^2(a^2-n^2)}.
\end{align}
The angular velocities refer to the displayed Killing basis. Their agreement with the kernels of the dressed Killing matrix verifies the constant Killing-coordinate transformation.

We next compare these generators with the rod directions of the finite binary before taking the zero-Kerr limit.

\subsection{Limits of the two finite horizon directions}\label{ism:sec:finite-rod-directions}

\begin{figure}[htbp]
 \centering
 \includegraphics[width=\linewidth]{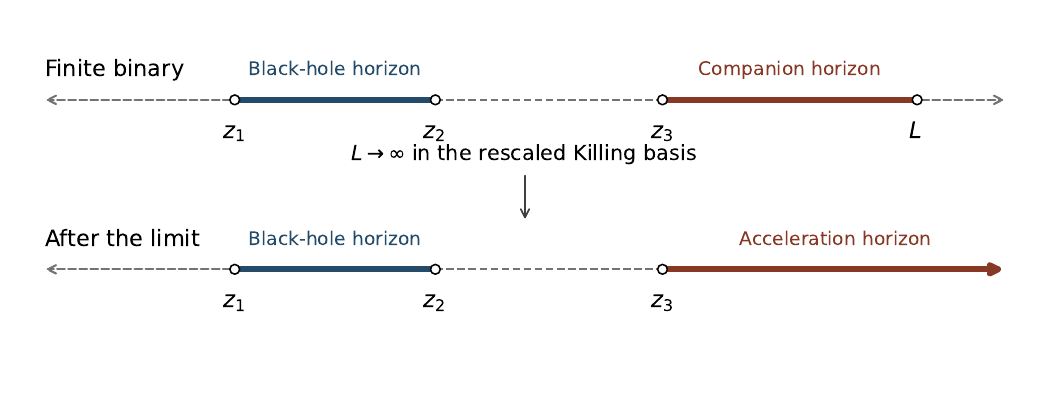}
 \caption{The two finite horizon rods and their limiting intervals. The finite kernels are evaluated on the open intervals $(z_1,z_2)$ and $(z_3,L)$ before taking $L\to\infty$. The left interval remains finite; the right becomes $(z_3,\infty)$. Dashed segments are axis intervals. The diagram gives interval incidence only; the directions are compared in the same rescaled Killing basis in the text. The qualitative rod limit is depicted in Fig.~2 of Ref.~\cite{AstorinoBoldi}. Not to scale.}
 \label{fig:finite-limit-horizon-rods}
\end{figure}

Convergence of the metric away from the axis does not determine the limiting horizon directions. We first take $\rho\to0$ at a fixed interior point of each finite horizon interval, and then let the companion recede. Write $B_i=C_i^B$ and specify a Killing direction by the
projective pair $[u:v]$ in the original finite $(t,\phi)$ basis.
The resulting directions are
\begin{align}
 [u_{\rm BH}:v_{\rm BH}]
 &=\left[z_2-z_1:
 \frac{B_1(z_3-z_1)}{4(L-z_1)^2}
 -\frac{B_2(z_3-z_2)}{4(L-z_2)^2}\right],\\
 [u_{\rm A}:v_{\rm A}]
 &=\left[4(L-z_1)^2(z_3-z_1)+B_1B_2(z_3-z_2):
                B_1(z_2-z_1)\right].
\label{ism:eq:finite-companion-direction}
\end{align}
In particular, neither direction depends on position along its rod. Figure~\ref{fig:finite-limit-horizon-rods} shows the open intervals used in the calculation.

To derive these expressions, set $s=\rho^2$ and rescale each row
in \eqref{ism:eq:binaryrows} so that its first entry is one. Its
second entry is then
\begin{equation*}
 \beta_i=\frac{B_i\mu_3(p_i-\mu_3)}
 {(s+\mu_3p_i)(\mu_4-p_i)^2}.
\end{equation*}
The rescaling leaves the reconstructed block unchanged. Let $M$
have rows $(1,\beta_i)$, and put $U=G_0 M^{\mathsf T}P_*^{-1}$,
where $G_0=G_{0B}$. With the Gram matrix of these rows the
numerator of the reconstructed block, before the factor $\mu_2/\mu_1$,
is
\begin{equation*}
 N=(\det\Gamma)G_0-U\,\operatorname{adj}(\Gamma)U^{\mathsf T}.
\end{equation*}
The leading coefficients follow from the $2\times2$ adjugate, with the pole and seed coefficients evaluated on each open interval.

On $z_1<z<z_2$, the poles tend to $p_i=2(z_i-z)$ and the seed has
the form $G_0=\operatorname{diag}(g,bs)$, with
$g=-\mu_3/\mu_4<0$ and $b=-1/g>0$. Put
$d=p_1-p_2$ and $e=p_1\beta_1-p_2\beta_2$. Then
\begin{align}
 \lim_{s\to0}\frac{\det\Gamma}{s}
 &=\frac{g[-gd^2+b(\beta_1-\beta_2)^2p_1^2p_2^2]}{p_1^4p_2^4},\\
 \lim_{s\to0}\frac{N}{s^2}
 &=-\frac{g^2b}{p_1^4p_2^4}
 \begin{pmatrix}e\\-d\end{pmatrix}
 \begin{pmatrix}e&-d\end{pmatrix}.
\end{align}
The determinant coefficient is nonzero since $d\ne0$. The factor
$\mu_2/\mu_1=-p_1p_2/s$ cancels the remaining power of $s$, so
the limiting block is nonzero and rank one. Its kernel is $[d:e]$.
Here $\mu_j\to2(z_j-z)$ for $j=3,4$, and
$p_i\beta_i=B_i(p_i-\mu_3)/(\mu_4-p_i)^2$ on the rod.
Substitution gives the first direction above.

On $z_3<z<L$, write $d_i=2(z-z_i)$ for $i=1,2,3$ and
$d_4=2(L-z)$. Now $\mu_i=s/d_i+O(s^2)$ for $i=1,2,3$.
Write the poles as $p_1=p$, $p_2=sq$, the seed as
$G_0=\operatorname{diag}(gs,b)$, and the normalized rows as
$(1,h)$ and $(1,sk)$. The leading values are
\begin{equation*}
\begin{gathered}
 p=-d_1,\quad q=d_2^{-1},\quad b=d_3d_4,\quad g=-b^{-1},\\
 h=\frac{B_1p}{(d_3+p)(d_4-p)^2},\qquad
 k=\frac{B_2(d_3-d_2)}{d_2d_3^2d_4^2}.
\end{gathered}
\end{equation*}
Put $j=(gpq-bhk)/(1+pq)$. Direct substitution gives
\begin{align}
 \lim_{s\to0}\det\Gamma
 &=-\frac{p^2(b^{-1}-bhk)^2+h^2(1+pq)^2}{p^2(1+pq)^2},\\
 \lim_{s\to0}N
 &=-\frac{b}{p^2q^2}
 \begin{pmatrix}gh\\-j\end{pmatrix}
 \begin{pmatrix}gh&-j\end{pmatrix}.
\end{align}
Since $1+pq\ne0$ and $b>0$, the determinant is nonzero;
when $h=0$, its first square is strictly positive. The
factor $\mu_2/\mu_1\to d_1/d_2$ is finite and positive, and the
outer product is nonzero. Thus the kernel is $[j:gh]$, which
reduces to \eqref{ism:eq:finite-companion-direction}. Keeping
projective pairs also covers a vanishing time component at finite $L$.

In the basis of \eqref{ism:eq:directscaling}, a direction
$[u:v]$ is represented by $[u:2Lv]$. With $B_i=2LC_i^C$, the
two rod-first limits are therefore
\begin{align}
 [z_2-z_1:&\,C_1^C(z_3-z_1)-C_2^C(z_3-z_2)], \\
 [(z_3-z_1)+C_1^CC_2^C(z_3-z_2):&\,C_1^C(z_2-z_1)] .
\end{align}
The rank-one limits are nonzero also in the rescaled basis:
the formulae above apply there with the seed's time coefficient
multiplied by $2L$ and its angular coefficient divided by $2L$,
and with $\beta_i$ multiplied by $2L$. Their leading coefficients
have finite limits with $gb=-1$; the same nonvanishing arguments
apply at fixed interior $z$.

For the physical comparison, the turning-point relation gives
\begin{equation*}
 2\alpha(z_3-z_1)=(1+\alpha r_+)(1-\alpha r_-),
 \qquad
 2\alpha(z_3-z_2)=(1-\alpha r_+)(1+\alpha r_-).
\end{equation*}
The subsequent constant change from $(t,\phi)$ to $(T,\Phi)$
multiplies the ratio $v/u$ in the normalized basis by $-\alpha$.
Inserting the polarization and product relations gives, respectively,
\begin{equation*}
 -\frac{a(1+\alpha^2r_+^2)+2\alpha nr_+}{r_+^2-a^2+n^2},
 \qquad
 -\frac{2\alpha^2(a+n)}{1-\alpha^2(a^2-n^2)}.
\end{equation*}
These agree with the black-hole and positive acceleration-horizon generators obtained from the physical Killing one-forms. The convergence holds on the two open rods in the stated limiting frame; turning points and global Killing identifications are not covered by this calculation.

\subsection{Zero-Kerr specialization and recovery of the physical \texorpdfstring{$A^+$}{A+} metric}
The accelerating Kerr--NUT metric obtained above uses the ISM angular
orientation, with conformal factor $1+\alpha r x_{\rm ISM}$.  Setting
$a=0$, $n=l$ and $x=-x_{\rm ISM}$ gives the physical zero-Kerr sector.
The Killing coordinates $T,\Phi$ are unchanged by this angular reversal.

The conformal, radial and angular functions reduce separately to
\begin{align}
 \Omega&=1-\alpha r x,\\
 \Delta&=(r^2-2mr-l^2)(1-\alpha^2r^2)=\Delta_r,\\
 P&=(1-x^2)(1-2\alpha mx-\alpha^2l^2x^2)=\Delta_x.
\end{align}
The quadratic denominator becomes
\begin{align}
 \Sigma&=(r+\alpha l^2x)^2+l^2(1+\alpha rx)^2\\
 &=(r^2+l^2)(1+\alpha^2l^2x^2)+4\alpha l^2rx
 =\rho^2.
\end{align}
The two separated Killing one-forms specialize to
\begin{align}
 \vartheta_r&=(1+\alpha^2l^2x^2)\dif T-2lx\,\dif\Phi,\\
 \vartheta_x&=2\alpha lr\,\dif T+(r^2+l^2)\dif\Phi.
\end{align}
It follows that the complete limiting metric, including its orbit-space
sector, is
\begin{equation}
\begin{aligned}
 \dif s^2=\frac{1}{(1-\alpha rx)^2}\bigg[&
 -\frac{\Delta_r}{\rho^2}\vartheta_r^2
 +\frac{\Delta_x}{\rho^2}\vartheta_x^2+\rho^2\left(\frac{\dif r^2}{\Delta_r}
 +\frac{\dif x^2}{\Delta_x}\right)\bigg].
\end{aligned}
\label{ism:zeroKerrFull}
\end{equation}
With $T=t$ and $\Phi=\phi$, this is Eq.~\eqref{Dmetric}, including
the orbit-space normalization.  The Killing-sector coefficient checks
are given in Appendix~\ref{ism:app:factorwise}, and the orbit-space
calculation in Sec.~\ref{ism:sec:physical}.  The local identification
does not fix the angular period or Misner-string identifications.

The PD reconstruction establishes the local type-D embedding, while the
ISM construction also traces the finite-binary horizon directions through
the large-companion limit.  Their limiting generators agree with those
computed directly from the physical metric.  We now examine the radial
and axial Killing directions, followed by the horizon geometry.

\section{Horizons, axes and projective Killing structure}
\label{killing}

Horizon angular velocities depend on the Killing basis. The coordinate
transformations relating the $A^+$ metric to older Pleba\'nski--Demia\'nski
forms mix the Killing fields~\cite{WuWu,OPA}. By contrast, equality or
inequality of two projective horizon directions is preserved by a common
invertible change of Killing basis.

\subsection{Radial horizons and parameter domain}

The four radial Killing horizons are located at
\begin{equation}
r=r_\pm,
\qquad
r=\pm\alpha^{-1}.
\label{horizons}
\end{equation}
We focus on the separated-horizon domain
\begin{equation}
-\alpha^{-1}<r_-<0<r_+<\alpha^{-1},
\label{ordering}
\end{equation}
with $\Delta_x>0$ for $-1<x<1$.  The positive exterior occupies $r_+<r<\alpha^{-1}$; the negative-$r$ region has its own radial blocks.

At a simple radial horizon $r=r_h$, the null Killing direction is
\begin{equation}
\chi_h=\partial_t+\Omega_h\partial_\phi,
\qquad
\Omega_h=-\frac{2\alpha lr_h}{r_h^2+l^2}.
\label{Omegah}
\end{equation}
Thus
\begin{equation}
\begin{aligned}
\chi_{b^+}&=\partial_t-\frac{\alpha l}{\sigma}\partial_\phi,
&\qquad
\chi_{b^-}&=\partial_t+\frac{\alpha l}{\sigma}\partial_\phi,\\
\chi_{a^+}&=\partial_t-\frac{2\alpha^2l}{1+\alpha^2l^2}\partial_\phi,
&
\chi_{a^-}&=\partial_t+\frac{2\alpha^2l}{1+\alpha^2l^2}\partial_\phi .
\end{aligned}
\label{fourgenerators}
\end{equation}
The angular velocities depend on the Killing basis, whereas equality of
projective horizon directions is preserved by a common invertible basis
change.  The rod interpretation follows the stationary-axisymmetric
framework~\cite{Harmark2004}.

\subsection{Axial directions}

At $x=\pm1$, $\Delta_x=0$ and the Killing block degenerates along
\begin{equation}
\ell_N\propto \frac{2l}{1+\alpha^2l^2}\partial_t+\partial_\phi,
\qquad
\ell_S\propto -\frac{2l}{1+\alpha^2l^2}\partial_t+\partial_\phi .
\label{axisvectors}
\end{equation}
The axial directions do not by themselves fix conical deficits.  In the
presence of NUT charge, the interpretation of conicity also depends on
the observer~\cite{OPA}.

\subsection{Projective horizon structure of type D}

Let $[v]$ denote the projective class of a nonzero Killing vector $v$.  In the regular type-D domain the positive black-hole and acceleration-horizon generators are distinct,
\begin{equation}
[\chi_{b^+}]\ne[\chi_{a^+}],
\label{distinctpositive}
\end{equation}
and similarly on the negative side.  Equality on the positive side would require
\begin{equation}
(1-\alpha r_+)(1+\alpha r_-)=0,
\label{degenerateN}
\end{equation}
which is excluded by the strict ordering in Eq.~\eqref{ordering}.
The negative-side generators have the opposite azimuthal components
of the corresponding positive-side generators, so their equality
has the same condition:
\begin{equation}
 \Omega_{b^-}=\Omega_{a^-}
 \quad\Longleftrightarrow\quad
 (1-\alpha r_+)(1+\alpha r_-)=0 .
 \label{degenerateS}
\end{equation}
At either boundary, equality of the corresponding projective generators
should be distinguished from the question of which two radial roots
actually coincide. Nor can an axis direction coincide with a regular
open horizon generator: if $s=\pm1$ and $\ell_s$ denotes the axial
vector of Eq.~\eqref{axisvectors} with unit azimuthal component, then
\begin{equation*}
 (C\,\dif t+D\,\dif\phi)(\ell_s)\big|_{x=s,r=r_h}
 =\frac{\rho^2(r_h,s)}{1+\alpha^2l^2}>0,
\end{equation*}
whereas the horizon generator annihilates that form. In the positive
exterior the two axial rods and the two horizon rods therefore carry
four distinct projective directions in the strict separated-horizon domain.  This statement concerns the distinguished Killing directions of the coordinate domain; it does not by itself classify a maximal analytic extension.


\section{Horizon mechanics and intrinsic geometry}
\label{horizongeom}

We calculate the surface gravities, areas and intrinsic Gaussian
curvature in the normalization of Eq.~\eqref{Dmetric}, using the general
type-D framework~\cite{PV2021,OPA}.  The polar curvatures enter the
comparison with type I in Sec.~\ref{comparison}.

\subsection{Area and surface gravity}

For a simple type-D radial horizon, the surface gravity in the normalization of Eq.~\eqref{Dmetric} is
\begin{equation}
\kappa_h=\frac{|\Delta_r'(r_h)|}{2(r_h^2+l^2)}.
\label{kappageneral}
\end{equation}
For the outer black-hole horizon,
\begin{equation}
\kappa_{b^+}^{\rm D}=\frac{1-\alpha^2r_+^2}{2\,r_+}.
\label{kappaBH}
\end{equation}
The positive acceleration horizon has
\begin{equation}
\kappa_{a^+}^{\rm D}=\frac{\alpha(1-2\alpha m-\alpha^2l^2)}{(1+\alpha^2l^2)}.
\label{kappaAH}
\end{equation}
The vanishing of Eq.~\eqref{kappaAH} corresponds to a degeneration of the positive acceleration-horizon root against a black-hole root on an exceptional parameter boundary. It must not be confused with the distinct question of equality of projective Killing directions in the open separated-horizon domain.

For an azimuthal period $\Delta\phi$, the outer black-hole horizon area is
\begin{equation}
A_{b^+}^{\rm D}=\frac{2(r_+^2+l^2)\Delta\phi}{1-\alpha^2r_+^2}.
\label{areaD}
\end{equation}
If $\Delta\phi=2\pi C$, then
\begin{equation}
A_{b^+}^{\rm D}=\frac{4\pi C(r_+^2+l^2)}{1-\alpha^2r_+^2}.
\end{equation}
The product simplifies to
\begin{equation}
\kappa_{b^+}^{\rm D}A_{b^+}^{\rm D}=4\pi C\sigma.
\label{kappaA}
\end{equation}
These expressions specialize the general type-D horizon formulas
~\cite{PV2021,OPA}.  Their dependence on Killing and angular
normalizations makes the intrinsic horizon geometry more suitable for
the comparison with type I.

\subsection{Intrinsic type-D horizon geometry}

On a spatial section of the outer black-hole horizon, the induced metric is
\begin{equation}
\dif s_H^2=\frac{1}{\Omega_+^2}\left[
\frac{\rho_+^2}{\Delta_x}\dif x^2
+\frac{(r_+^2+l^2)^2\Delta_x}{\rho_+^2}\dif\varphi_H^2
\right],
\label{horizon2metricD}
\end{equation}
where
\begin{equation}
\Omega_+=1-\alpha r_+x,
\end{equation}
\begin{equation}
\rho_+^2=(r_+^2+l^2)(1+\alpha^2l^2x^2)+4\alpha l^2r_+x.
\end{equation}
The Gaussian curvature can be written compactly as
\begin{equation}
K_D(x)=-\frac{\Omega_+^2}{2}\frac{\dif}{\dif x}
\left[\Omega_+^2\frac{\dif}{\dif x}
\left(\frac{\Delta_x}{\Omega_+^2\rho_+^2}\right)\right].
\label{KDcompact}
\end{equation}
Rather than using the lengthy expanded function, we characterize the intrinsic north--south distortion by
\begin{equation}
\mathcal R_D\equiv \frac{K_N^D}{K_S^D}.
\label{RDdef}
\end{equation}
An exact expression is given in Appendix~\ref{appcurvature}.

\section{Null geodesics and hidden symmetry}
\label{geodesics}

\subsection{Type-D separation}

The metric belongs to the conformal Carter class, for which null Hamilton--Jacobi separation is known~\cite{Carter1968CMP,Carter1968PR,FrolovKrtousKubiznak2017,Gray2026}. We give the separated equations in the parameters of Eq.~\eqref{Dmetric}. The conformal factor generally obstructs the corresponding separation for massive particles.

Write the two Killing one-forms in Eq.~\eqref{Dmetric} as
\begin{equation}
 \vartheta_r=(1+\alpha^2l^2x^2)\dif t-2lx\,\dif\phi,
 \qquad \vartheta_x=2\alpha lr\,\dif t+(r^2+l^2)\dif\phi.
\end{equation}
Their coefficient determinant is precisely $\rho^2$, as follows from Eq.~\eqref{rho2}.  Thus, for a covector with $p_t=-E$ and $p_\phi=L$, its components in the dual basis to $(\vartheta_r,\vartheta_x)$ are
\begin{equation}
 -\frac{E(r^2+l^2)+2\alpha lLr}{\rho^2},\qquad
 \frac{L(1+\alpha^2l^2x^2)-2lEx}{\rho^2}.
\end{equation}
For the separated principal function
\begin{equation}
 S=-Et+L\phi+S_r(r)+S_x(x),
\end{equation}
the null Hamilton--Jacobi equation $g^{\mu\nu}\partial_\mu S\partial_\nu S=0$, after multiplication by $\rho^2/\Omega^2$, becomes
\begin{equation}\label{eq:HJdirectD}
 \Delta_r(S_r')^2-\frac{[E(r^2+l^2)+2\alpha lLr]^2}{\Delta_r}
 +\Delta_x(S_x')^2
 +\frac{[L(1+\alpha^2l^2x^2)-2lEx]^2}{\Delta_x}=0.
\end{equation}
The first two terms depend only on $r$, and the remaining terms only on $x$.  Hence a separation constant $\calK$ exists, with
\begin{align}
 \Delta_r^2(S_r')^2&=\calR(r)
 =[E(r^2+l^2)+2\alpha lLr]^2-\calK\Delta_r,\label{RDgeo}\\
 \Delta_x^2(S_x')^2&=\calX(x)
 =\calK\Delta_x-[L(1+\alpha^2l^2x^2)-2lEx]^2.\label{XDgeo}
\end{align}
Real null trajectories require $\calR\geq0$ and $\calX\geq0$ in the relevant coordinate blocks.  With an affine parameter $\lambda$ and the rescaled parameter $\tau$ defined locally by $\dif\tau/\dif\lambda=\Omega^2/\rho^2$, Hamilton's equations give
\begin{equation}
 \left(\frac{\dif r}{\dif\tau}\right)^2=\calR(r),\qquad
 \left(\frac{\dif x}{\dif\tau}\right)^2=\calX(x).
\end{equation}
For massive geodesics, the Hamilton--Jacobi equation contains $\mu^2\rho^2/\Omega^2$, which is not generally a sum of separate radial and angular functions.

The radial null congruences used below follow from the separated equations. Fix an interior angular value $x=x_0$ and choose
\begin{equation}
 \calK=0,\qquad E=1+\alpha^2l^2x_0^2,\qquad L=2lx_0.
 \label{eq:principal-separation-choice}
\end{equation}
The angular square in Eq.~\eqref{XDgeo} then vanishes at $x=x_0$,
while the radial Killing combination obeys the exact identity
\begin{equation}
 E(r^2+l^2)+2\alpha lLr
 =(1+\alpha^2l^2x_0^2)(r^2+l^2)+4\alpha l^2rx_0
 =\rho^2(r,x_0).
 \label{eq:principal-rho-identity}
\end{equation}
Hence $\calR(r)=\rho^4(r,x_0)$ and the separated radial momentum is
$p_r=\varepsilon\rho^2(r,x_0)/\Delta_r$, with $\varepsilon=\pm1$.
The Hamilton equations also preserve $x=x_0$ and $p_x=0$ for this
choice, as verified directly in Sec.~\ref{sec:radial-geodesics}.
The Hamiltonian calculation in Sec.~\ref{sec:radial-geodesics} also determines the affine parametrization and azimuthal drift needed for horizon-penetrating coordinates.

At $\calK=0$, the angular potential is the negative of a square.  In the physical angular range, an admissible trajectory must satisfy
\begin{equation}
 L(1+\alpha^2l^2x^2)-2lEx=0.
\end{equation}
The corresponding first-order equations determine whether the root gives a constant-$x$ principal null trajectory.  For a causal geodesic to meet a locally regular axis, its momentum must annihilate the vanishing axial Killing field, $p\cdot\ell_{N,S}=0$.  In the Killing normalization of Eq.~\eqref{Dmetric}, this gives
\begin{equation}\label{axisratioD}
 \frac{L}{E}=\pm\frac{2l}{1+\alpha^2l^2},
\end{equation}
when $E\ne0$, with the sign fixed by the chosen axis.  This condition is local and does not determine global geodesic completeness.

\section{Global geometric and causal structure}
\label{causal}

The fixed-angle null congruences obtained from the separated equations, Eqs.~\eqref{eq:principal-separation-choice}--\eqref{eq:principal-rho-identity}, determine the null coordinates used to extend the radial geometry across its four simple horizons. We construct an unwrapped atlas of the resulting radial blocks at fixed angle, with the angular-dependent conformal boundary and axial identifications treated separately. The conformal-block method builds on the analysis of accelerating black holes by Griffiths and Podolsk\'y~\cite{GP2005}, the $C$-metric construction of Griffiths, Krtou\v{s} and Podolsk\'y~\cite{GKP}, the global study of the type-I accelerating NUT solution~\cite{PVtypeI}, and the general type-D horizon-block framework~\cite{PV2021}. The horizon charts, reciprocal crossing and five angular conformal-boundary cuts are calculated here for the physical $A^+$ metric of Eq.~\eqref{Dmetric}. The resulting atlas describes radial projections without fixing the global Killing-coordinate identifications.

\subsection{Radial blocks and regularity at the central surface}

For $l\ne0$ and the strict ordering $-\alpha^{-1}<r_-<0<r_+<\alpha^{-1}$, the sign of
\begin{equation}
 g^{rr}=\frac{\Omega^2\Delta_r}{\rho^2}
\end{equation}
changes at each of the four simple radial roots. It is positive on $(-\alpha^{-1},r_-)$ and $(r_+,\alpha^{-1})$, and negative on the three complementary radial intervals, wherever the chart is regular. Hence $r$ is spacelike in the former two intervals and timelike in the latter. This does not determine the sign of $g_{tt}$ or establish a global stationary Killing frame. Indeed, the five radial intervals must be intersected with the angular and conformal coordinate domains before they can be interpreted as connected causal blocks. The sign classification is summarized in Table~\ref{tab:radialblocks}.
\begin{table}[ht]
\centering
\begin{tabular}{llll}
\toprule
Block & Radial interval & Sign of $\Delta_r$ & Character of $r$ \\
\midrule
V & $r<-\alpha^{-1}$ & negative & timelike \\
IV & $-\alpha^{-1}<r<r_-$ & positive & spacelike \\
III & $r_-<r<r_+$ & negative & timelike \\
II & $r_+<r<\alpha^{-1}$ & positive & spacelike \\
I & $r>\alpha^{-1}$ & negative & timelike \\
\bottomrule
\end{tabular}
\caption{Local radial classification wherever the metric chart is regular and $\Omega\ne0$. This table does not assert that each interval is a single connected global region.}
\label{tab:radialblocks}
\end{table}
The distinction from stationarity can be seen explicitly from
\begin{equation}
 g_{tt}=\frac{-\Delta_r(1+\alpha^2l^2x^2)^2+4\alpha^2l^2r^2\Delta_x}{\Omega^2\rho^2}.
\end{equation}
Consequently the stationary-limit locus for $\partial_t$ obeys
\begin{equation}
 \Delta_r(1+\alpha^2l^2x^2)^2=4\alpha^2l^2r^2\Delta_x.
\end{equation}
When $\Delta_r>0$ and $\Delta_x>0$, the determinant of the Killing block is negative, so a timelike direction exists in the Killing two-plane at each point. This pointwise fact does not guarantee a single constant Killing combination timelike throughout an entire block.

The identity
\begin{equation}
 \rho^2=(r+\alpha l^2x)^2+l^2(1+\alpha rx)^2
 \label{rhosquares}
\end{equation}
shows that a zero of $\rho^2$ for real $l\ne0$ would require $\alpha^2l^2x^2=1$. This is impossible in the regular angular domain $|x|\le1$ since $\alpha^2l^2=(\alpha r_+)(-\alpha r_-)<1$ by
Eq.~\eqref{ordering}. At $r=0$ one has
\begin{equation}
 \Omega=1,\qquad \Delta_r=-l^2,\qquad
 \rho^2=l^2(1+\alpha^2l^2x^2),\qquad
 g^{rr}=-\frac{1}{1+\alpha^2l^2x^2}.
\end{equation}
The full determinant is $\det g=-\rho^4/\Omega^8$. Thus the metric and its inverse are smooth and nondegenerate across $r=0$ away from the axial coordinate degeneracies. In particular, polynomial curvature invariants remain finite there and $r=0$ is a spacelike hypersurface. This proves a local continuation to negative $r$, not global geodesic completeness. The limit $l\to0$ is nonuniform at this surface: the ordinary vacuum C-metric central singularity is recovered.

\subsection{Principal-null projection and tortoise coordinate}
\label{sec:radial-geodesics}

To construct the radial projection, we first find four-dimensional
null geodesics at fixed $x=x_0$. Write $\mathcal A=A\,\dif t-B\,\dif\phi$
and $\mathcal B=C\,\dif t+D\,\dif\phi$ as in Eq.~\eqref{Dmetric}.
Their coefficient determinant is \(AD+BC=\rho^2\). Their inverse Killing block gives the geodesic Hamiltonian
\begin{align}
 2\mathcal H=\frac{\Omega^2}{\rho^2}\bigg[
 &-\frac{[(r^2+l^2)p_t-2\alpha lr p_\phi]^2}{\Delta_r}
 +\frac{[2lx p_t+(1+\alpha^2l^2x^2)p_\phi]^2}{\Delta_x}
 +\Delta_r p_r^2+\Delta_x p_x^2
 \bigg].
 \label{eq:Hamiltonian}
\end{align}
At $x=x_0\in(-1,1)$, suppress the angular square by choosing the
conserved momenta $p_t=-(1+\alpha^2l^2x_0^2)$, $p_\phi=2lx_0$, and
take $p_x=0$. The first square then has argument $-\rho^2(r,x_0)$.
The null condition gives
\[
 p_r=\varepsilon\frac{\rho^2(r,x_0)}{\Delta_r},\qquad \varepsilon=\pm1 .
\]
These conditions are consistent with Hamiltonian evolution. In
$\partial_x\mathcal H$ at fixed momenta, the derivative of the prefactor
multiplies a vanishing bracket. Within the bracket the radial terms
have no $x$ dependence, and the two angular terms have vanishing
arguments at $x=x_0$. Thus $\dot p_x=-\partial_x\mathcal H=0$, and
$\dot x=\Omega^2\Delta_x p_x/\rho^2=0$. Similarly, the radial derivative
of the first square at the chosen conserved momenta is the derivative
of $\rho^4(r,x_0)$, while $p_r$ is held fixed in the term
$\Delta_r p_r^2$. This gives
\begin{equation}
 -\partial_r\mathcal H
 =\Omega^2\left[\frac{\partial_r\rho^2}{\Delta_r}
                  -\frac{\rho^2\Delta_r'}{\Delta_r^2}\right]_{x=x_0}
 =\dot r\,\frac{\dif}{\dif r}
       \left(\varepsilon\frac{\rho^2(r,x_0)}{\Delta_r}\right),
 \qquad \dot r=\varepsilon\Omega^2 .
\end{equation}
The dots denote derivatives with respect to an affine parameter. The
remaining Hamilton equations give the two null vector fields
\begin{equation}
 \ell_\varepsilon=\Omega^2\left[
   \frac{r^2+l^2}{\Delta_r}\partial_t+\varepsilon\partial_r
   -\frac{2\alpha lr}{\Delta_r}\partial_\phi
 \right].
 \label{eq:nullvectors}
\end{equation}

In particular $\dif t/\dif r=\varepsilon(r^2+l^2)/\Delta_r$ along either
family. Define the tortoise coordinate by
\begin{equation}
 \frac{\dif r_*}{\dif r}=\frac{r^2+l^2}{\Delta_r},\qquad
 r_*(r)=\sum_{r_h\in\{-1/\alpha,r_-,r_+,1/\alpha\}}
       \frac{r_h^2+l^2}{\Delta_r'(r_h)}\log|\alpha(r-r_h)| .
 \label{eq:tortoise}
\end{equation}
The residues sum to zero because $\dif r_*/\dif r=O(r^{-2})$ as
$|r|\to\infty$. The sum of logarithms is therefore independent of the
common length scale used in their arguments, and $r_*=0$ at $y=0$
from either side. In the reciprocal chart \(y=(\alpha r)^{-1}\), with
\(G(z)=(1-z^2)(1-2\alpha mz-\alpha^2l^2z^2)\),
\begin{equation}
 \frac{\dif r_*}{\dif y}=\frac{1+\alpha^2l^2 y^2}{\alpha G(y)} .
 \label{eq:tortoise-reciprocal}
\end{equation}
This derivative is finite at $y=0$ and the chosen normalization joins the
two branches there. The residues in Eq.~\eqref{eq:tortoise} have signs
$+,-,+,-$ in increasing $r_h$; hence $r_*$ diverges at the successive
roots as $-\infty,+\infty,-\infty,+\infty$. The functions
$u=t-r_*$ and $v=t+r_*$ are respectively constant along $\ell_+$ and
$\ell_-$; they need not define null hypersurfaces of the four-dimensional
metric.

The resulting two-dimensional conformal representative is
\begin{equation}
 -F(r)\,\dif u\,\dif v,\qquad F(r)=\frac{\Delta_r}{r^2+l^2}.
 \label{eq:gkp-radial-projection}
\end{equation}
It represents the projected null directions; it is not the
metric obtained by setting \(\dif\phi=0\). In fact the rays found above
obey \(\mathcal B=0\), so their azimuth changes. The distinction also
matters for the character of conformal infinity below.

\subsection{Analytic extensions at the four simple horizons}
\label{sec:horizon-extensions}

\subsubsection{Penetrating coordinates}

The radial geodesics~\eqref{eq:nullvectors} have an azimuthal drift as well
as a time drift. To obtain coordinates regular across a horizon branch, we
therefore transform both Killing coordinates. With $\eta=\pm1$, set
\begin{equation}
 \dif t_\eta=\dif t+\eta\frac{r^2+l^2}{\Delta_r}\dif r,
 \qquad
 \dif\varphi_\eta=\dif\phi-\eta\frac{2\alpha lr}{\Delta_r}\dif r.
 \label{EFchange}
\end{equation}
Here $t_+=v$ is the advanced time and $t_-=u$ the retarded time. Write
$\mathcal A_\eta,\mathcal B_\eta$ for the two forms in
Eq.~\eqref{Dmetric} with $(\dif t,\dif\phi)$ replaced by
$(\dif t_\eta,\dif\varphi_\eta)$. The determinant identity
\begin{equation}
 (r^2+l^2)(1+\alpha^2l^2x^2)+4\alpha l^2rx=\rho^2
\end{equation}
gives $\mathcal A=\mathcal A_\eta-\eta\rho^2\dif r/\Delta_r$ and
$\mathcal B=\mathcal B_\eta$. The radial square in the first Killing
term cancels the original $\dif r^2$ term, giving
\begin{equation}
 \dif s^2=\frac{1}{\Omega^2}\left[
 -\frac{\Delta_r}{\rho^2}\mathcal A_\eta^2
 +2\eta\mathcal A_\eta\dif r
 +\frac{\Delta_x}{\rho^2}\mathcal B_\eta^2
 +\frac{\rho^2}{\Delta_x}\dif x^2\right].
 \label{EFmetric}
\end{equation}
The metric is analytic at a simple root where $\Omega\ne0$ and
$\Delta_x>0$. In the $(\mathcal A_\eta,\dif r)$ part of the coframe
the determinant is $-1$ before the conformal rescaling. The other
two coframe coefficients have product one, and the coframe
determinant has magnitude $\rho^2$. Thus
$\det g=-\rho^4/\Omega^8$ in the penetrating coordinates. The two
signs cover complementary horizon branches.

\subsubsection{Bifurcation charts}

At a simple root $r_h$, introduce the corotating angle and signed horizon
parameter
\begin{equation}
 \phi_h=\phi-\Omega_ht,\qquad
 \Omega_h=-\frac{2\alpha lr_h}{r_h^2+l^2},\qquad
 \beta_h=\frac{\Delta_r'(r_h)}{2(r_h^2+l^2)}.
\end{equation}
At the four roots the signed parameters simplify to
\begin{equation}
 \beta_{b^\pm}=\frac{1-\alpha^2r_\pm^2}{2r_\pm},\qquad
 \beta_{a^\pm}=\mp\alpha
 \frac{1\mp2\alpha m-\alpha^2l^2}{1+\alpha^2l^2}.
 \label{eq:signed-gravities}
\end{equation}
The generator is $\chi_h=\partial_t+\Omega_h\partial_\phi$.
Let $D=r^2+l^2$ and write the Killing forms as
\begin{equation}
\begin{aligned}
 \mathcal A&=A_h\dif t-2lx\dif\phi_h,&
 A_h&=1+\alpha^2l^2x^2-2lx\Omega_h,\\
 \mathcal B&=B_h\dif t+D\dif\phi_h,&
 B_h&=2\alpha lr+\Omega_hD.
\end{aligned}
\end{equation}
Their horizon values satisfy
\begin{equation}
 B_h(r_h)=0,\qquad A_h(r_h,x)=\frac{\rho^2(r_h,x)}{r_h^2+l^2}.
 \label{eq:kruskalkey}
\end{equation}
These identities are the two cancellations required by the four-dimensional
metric: the first removes the leading time term in $\mathcal B$ and the
second relates the time and radial coefficients in the other square.

In a neighborhood of $r_h$, the common tortoise primitive is
$r_*=(2\beta_h)^{-1}\log|r-r_h|+f_h(r)$, with $f_h$ analytic there.
Use $u=t-r_*$, $v=t+r_*$ and define
\begin{equation}
 U_h=\epsilon_{Uh}e^{-\beta_hu},\qquad
 V_h=\epsilon_{Vh}e^{\beta_hv},\qquad
 \epsilon_{Uh},\epsilon_{Vh}\in\{+1,-1\}.
\end{equation}
On each open quadrant choose the product sign so that
\begin{equation}
 U_hV_h=-(r-r_h)e^{2\beta_h f_h(r)}.
 \label{eq:kruskalproduct}
\end{equation}
The derivative of the right-hand side at $r_h$ is nonzero, so $r$ is
an analytic function of $U_hV_h$ near the horizon.

To check all components, substitute $\dif t=(\dif u+\dif v)/2$ and
$\dif r=\Delta_r(\dif v-\dif u)/(2D)$ into Eq.~\eqref{Dmetric}. In
particular,
\begin{equation}
 g_{uu}=g_{vv}=
 \frac{-\Delta_r A_h^2+\Delta_x B_h^2+\rho^4\Delta_r/D^2}
 {4\Omega^2\rho^2}.
\end{equation}
By Eq.~\eqref{eq:kruskalkey}, $B_h$ and $\rho^4/D^2-A_h^2$ are
divisible by $\Delta_r$ with analytic quotients. The remaining
components that involve $u$ or $v$ give, with the same analytic
interpretation of the orders,
\begin{equation}
 g_{uu}=g_{vv}=O(\Delta_r^2),\qquad
 g_{uv}=-\frac{\Delta_r\rho^2}{2\Omega^2 D^2}+O(\Delta_r^2),\qquad
 g_{u\phi_h}=g_{v\phi_h}=O(\Delta_r).
\end{equation}
Since $\Delta_r/(U_hV_h)$ is analytic and nonzero at the root,
$\dif u=-\dif U_h/(\beta_hU_h)$ and
$\dif v=\dif V_h/(\beta_hV_h)$ leave analytic coefficients. In particular,
\begin{equation}
 g_{U_hU_h}=O(V_h^2),\quad g_{V_hV_h}=O(U_h^2),\quad
 g_{U_h\phi_h}=O(V_h),\quad g_{V_h\phi_h}=O(U_h).
\end{equation}
At the bifurcation surface the $U_hV_h$ component is nonzero, the
pure null and null-angular components vanish, and the angular
diagonal components are
\[
 g_{xx}=\frac{\rho^2}{\Omega^2\Delta_x},\qquad
 g_{\phi_h\phi_h}=\frac{\Delta_x D^2}{\Omega^2\rho^2}.
\]
The metric is therefore nondegenerate there. This establishes a local analytic extension with four
quadrants at each regular simple horizon, away from the axes and
conformal infinity.

\subsubsection{The four local Kruskal blocks}
\label{sec:local-nets}

Before attaching different horizons, we display the four quadrants of each
individual extension. We use the product and ratio construction of
Podolsk\'y and Vr\'atn\'y~\cite[Sec.~5.4, Eqs.~(108)--(119)]{PV2021}.
Their horizon residue $k_h$ plays the role of $(2\beta_h)^{-1}$ here.
We use the $A^+$ radial function and sign convention of
Eq.~\eqref{eq:kruskalproduct}, and compactify the two null coordinates
separately. On a neighborhood of $r_h$, the sign of $F$ is
$\operatorname{sgn}(\beta_h)\operatorname{sgn}(r-r_h)$, and
Eq.~\eqref{eq:kruskalproduct} fixes the sign of $U_hV_h$. Define
\begin{equation}
 P_h=\frac{\operatorname{sgn}(\beta_h)U_h}{\sqrt{M_h}},\qquad
 Q_h=\frac{V_h}{\sqrt{M_h}},\qquad M_h>0 .
 \label{eq:local-display}
\end{equation}
The radial conformal metric $-F\dif u\dif v$ has a negative coefficient
of $\dif P_h\dif Q_h$ on all four quadrants:
$F/\bigl[\beta_h^2P_hQ_h\bigr]<0$. We orient the local drawing so that
increasing $P_h$ and $Q_h$ is future. The two spacelike-\(r\) quadrants in this projection are
then $P_hQ_h<0$; the other two lie to the future and past of the
bifurcation. The sign reversal in Eq.~\eqref{eq:local-display} for
$\beta_h<0$ changes the plotting convention, not the signed chart
of Eq.~\eqref{eq:kruskalproduct}.

The exact coordinate curves in each open quadrant are given by
\begin{equation}
\begin{aligned}
 \log|P_hQ_h| &= 2\beta_h r_*(r)-\log M_h
       &&(r\ \text{fixed}),\
 \log|Q_h/P_h| &= 2\beta_h t
       &&(t\ \text{fixed}).
 \end{aligned}
\label{eq:local-levels}
\end{equation}
The first relation gives a hyperbola in the uncompactified null
coordinates; the second gives a ray of fixed ratio. In plotting them
we apply the same compactification $2\arctan(z)/\pi$ to $P_h$ and $Q_h$.
Writing the results as $\bar P_h,\bar Q_h$,
we draw $\mathcal T=(\bar P_h+\bar Q_h)/2$ and
$\mathcal X=(\bar P_h-\bar Q_h)/2$. The horizon branches thus
retain slopes $\pm1$. Every level is computed separately on the
appropriate side of $r_h$; a fixed value of $|P_hQ_h|$ need not
give the same radius in different radial intervals.

\begin{figure}[p]
 \centering
 \includegraphics[width=.94\linewidth]{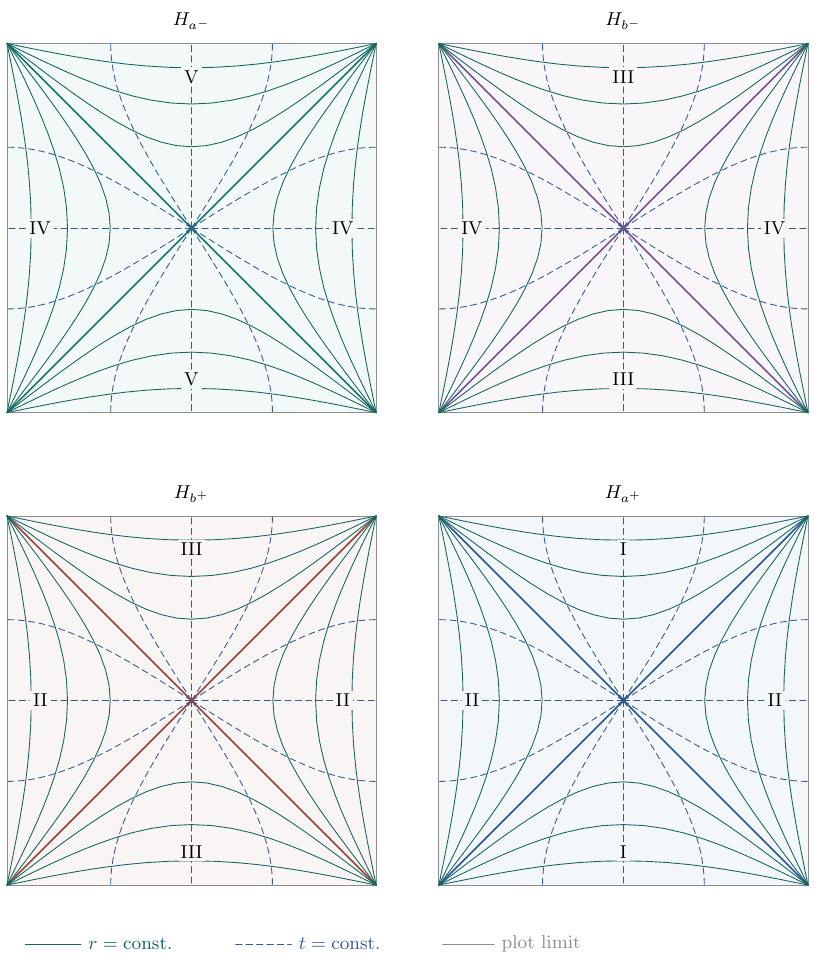}
 \caption{Four-quadrant neighborhoods of the four simple horizons.
 The diagonals are the branches of the indicated horizon; their
 intersection is its bifurcation. The future direction is upwards. Each quadrant is labelled by its
 radial block. Solid green curves are constant $r$, and dashed blue
 curves are constant $t$, calculated from Eq.~\eqref{eq:local-levels}.}
 \label{fig:local-four-horizons}
\end{figure}

Figure~\ref{fig:local-four-horizons} uses $(m,l,\alpha)=(1,0.6,0.3)$
and $x=0$. We take $M_h$ to be the smaller of
$\exp[2\beta_h r_*(r)]$ evaluated at two reference radii: the midpoint
between $r_h$ and the next root on each bounded side, and the point
a distance $\alpha^{-1}$ from $r_h$ on an unbounded side.
This common scale on both sides permits a local plotting window
$|P_hQ_h|\leq1$ without reaching either neighboring horizon or
conformal infinity. The radius levels have $|P_hQ_h|=0.13,0.34,0.66$,
and the time levels have $2\beta_h t=-1.5,0,1.5$.
The gray limits are plot boundaries only. The signed transitions
below specify how the quadrants are attached beyond these local
drawings.

\subsubsection{Signed transitions on adjacent open blocks}

Let $h$ and $k$ be the two distinct horizons of a common open block.
Using the same $r_*$ and Killing coordinates in both charts, eliminate
$u,v$ to obtain
\begin{align}
 U_k&=\epsilon_{Uk}
       \left(\frac{U_h}{\epsilon_{Uh}}\right)^{\beta_k/\beta_h},
 &V_k&=\epsilon_{Vk}
       \left(\frac{V_h}{\epsilon_{Vh}}\right)^{\beta_k/\beta_h},
 \label{eq:signed-overlap}\\
 \phi_k&=\phi_h+\frac{\Omega_h-\Omega_k}{2\beta_h}
       \log\frac{V_h/\epsilon_{Vh}}{U_h/\epsilon_{Uh}}.
 \label{eq:signed-angular-overlap}
\end{align}
The bases of the powers are positive, and the angular equation follows
from $t=(u+v)/2$. With $x$ unchanged, the determinant of the transition is
\begin{equation}
 \det\frac{\partial(U_k,V_k,\phi_k,x)}
 {\partial(U_h,V_h,\phi_h,x)}
 =\left(\frac{\beta_k}{\beta_h}\right)^2
   \frac{U_kV_k}{U_hV_h}\ne0.
 \label{eq:adjacent-chart-jacobian}
\end{equation}
For example, in region II, $r_+<r<\alpha^{-1}$, the overlap is on the
right of $h=b^+$ and on the left of $k=a^+$. One may take
$(\epsilon_{Uh},\epsilon_{Vh})=(-1,+1)$ and
$(\epsilon_{Uk},\epsilon_{Vk})=(+1,+1)$. The powers are then
$U_k=(-U_h)^{\beta_k/\beta_h}$ and $V_k=V_h^{\beta_k/\beta_h}$.
Although the exponent ratio is negative, the signed Jacobian is finite
and negative at every point of the open overlap. A divergence or zero
at one chart's unrelated endpoint is not an obstruction: that endpoint
is covered by its own horizon chart.

In increasing order of $r$, the left--right pairs inferred from
Eq.~\eqref{eq:kruskalproduct} are
$(\mathrm V,\mathrm{IV})$, $(\mathrm{IV},\mathrm{III})$,
$(\mathrm{III},\mathrm{II})$ and $(\mathrm{II},\mathrm I)$.
Each product sign occurs in two sign pairs, accounting for the four
local quadrants. On crossing either branch in the chart adapted to that
horizon, the vanishing Kruskal coordinate changes sign and the other
coordinate stays fixed on the edge. No global periodicity of $t$ or
$\phi$ is needed.

\subsection{Reciprocal crossing and conformal infinity}
\label{sec:reciprocal}\label{sec:gkp-integration}

The inverse radial coordinate brings the asymptotic regions into a finite
chart. Use the same \(y=(\alpha r)^{-1}\) as in
Eq.~\eqref{eq:tortoise-reciprocal}, and write \(T=\alpha t\) and
\[
 H(x,y)=(1+\alpha^2l^2xy)^2+\alpha^2l^2(x+y)^2 .
\]
On the overlap $r\ne0$, $y\ne0$,
\begin{equation}
 \Delta_r=-\frac{G(y)}{\alpha^2y^4},\qquad
 \Delta_x=G(x),\qquad
 \rho^2=\frac{H(x,y)}{\alpha^2y^2},\qquad
 \Omega=\frac{y-x}{y}.
\end{equation}
The radial Jacobian is $\dif r=-\dif y/(\alpha y^2)$, while
\[
 \mathcal B=\frac{2\alpha ly\,\dif T+(1+\alpha^2l^2y^2)\dif\phi}
                  {\alpha^2y^2}.
\]
Substitution into Eq.~\eqref{Dmetric} gives
\begin{align}
 \dif s^2=\frac{1}{\alpha^2(x-y)^2}\bigg\{
 &\frac{G(y)}{H(x,y)}[(1+\alpha^2l^2x^2)\dif T-2\alpha lx\,\dif\phi]^2
 \nonumber\\
 &+\frac{G(x)}{H(x,y)}[2\alpha ly\,\dif T+(1+\alpha^2l^2y^2)\dif\phi]^2
 \nonumber\\
 &+H(x,y)\left(-\frac{\dif y^2}{G(y)}+\frac{\dif x^2}{G(x)}\right)
 \bigg\}.
 \label{eq:factor}
\end{align}
The radial roots in this chart are ordered as
\[
 \frac{1}{\alpha r_-}<-1<1<\frac{1}{\alpha r_+}.
\]
The new defining function for conformal infinity is proportional to
\(x-y\). The locus \(y=0\) is of a different kind and need not be a
physical boundary. Conversely, the original \(r\) chart covers \(r=0\),
where \(y\) is undefined.

In the reciprocal chart the four radial Killing horizons are surfaces of
constant $y$, whereas conformal infinity is given by $y=x$. Its location
therefore varies with the angular coordinate and cuts different radial
intervals. The locus $y=0$, corresponding to the limits $r\to\pm\infty$
of the original chart, is distinct from conformal infinity except at
$x=0$. Away from that equatorial intersection it admits the regular
reciprocal crossing described below. These distinctions determine the
truncations of the five angular conformal diagrams.

\subsubsection{The reciprocal crossing}

The two acceleration regions lie in the reciprocal interval
$-1<y<1$, where $G(y)>0$. At $y=0$ we have
$G(0)=1$ and $H(x,0)=1+\alpha^2l^2x^2$. For $0<|x|<1$ the metric
coefficients in \eqref{eq:factor} are smooth there, and
both the Killing block and the $x,y$ block are nondegenerate: the
two Killing one-forms have coefficient determinant $H(x,y)$ and
$G(0)G(x)>0$.
In particular, the restriction at fixed $x$ and $\phi$ is
\begin{equation}
 \left.\dif s^2_{x,\phi}\right|_{y=0}
 =\frac{1+\alpha^2l^2x^2}{x^2}
 \left(\dif t^2-\frac{\dif y^2}{\alpha^2}\right).
\label{eq:reciprocal-limit}
\end{equation}
This restriction is a check on the crossing, not the metric used to draw
the radial conformal diagrams.

The null congruences of Eq.~\eqref{eq:nullvectors} give a second check.
For the normalization used there, $\dif r/\dif\lambda=\varepsilon\Omega^2$
at fixed $x=x_0$, with $\varepsilon=\pm1$. The reciprocal equation is
\begin{equation}
 \frac{\dif y}{\dif\lambda}=-\varepsilon\alpha(y-x_0)^2,
 \qquad
 \frac{1}{y-x_0}=\varepsilon\alpha\lambda+C .
 \label{eq:affine-reciprocal}
\end{equation}
Thus for $x_0\ne0$ the crossing at $y=0$ occurs at finite affine parameter,
with $\dif y/\dif\lambda=-\varepsilon\alpha x_0^2\ne0$ there.
In contrast, $y=x_0$ is at infinite affine parameter along these null
rays; in particular, there is no equatorial crossing at $y=0$.
These affine statements concern the displayed null family, not arbitrary
causal curves.

\subsubsection{The interior angular sections}

Let $\omega=\alpha(x-y)$ and $\widehat g=\omega^2g$. In the interior
angular range, the metric $\widehat g$ obtained from \eqref{eq:factor}
is nondegenerate at $y=x$. The inverse radial
and angular coefficients give
\begin{equation}
 \widehat g^{ab}\partial_a\omega\partial_b\omega
   =\frac{\alpha^2}{H(x,y)}\big[G(x)-G(y)\big].
\label{eq:reciprocal-boundary-norm}
\end{equation}
The four-dimensional boundary is therefore null. In particular, the
reciprocal representative is regular at \(x=y=0\): there
\(G(x)=G(y)=H(x,y)=1\) and
\[
 \widehat g=\dif T^2+\dif\phi^2-\dif y^2+\dif x^2 .
\]
This point is on the local null conformal boundary, not an interior
crossing of the physical metric. The old representative
\(\Omega^2g=\widehat g/(\alpha^2y^2)\) fails at \(y=0\);
its divergence is due to the choice of defining function rather
than a failure of the reciprocal conformal chart. At $y=x$ its null
generator is proportional to $G(x)(\partial_x+\partial_y)$, which is
not tangent to a fixed-$x$ section. In particular, on $-1<y<1$ the
Killing block of $\widehat g$ is positive definite for $|x|<1$. Its
restriction to a nonzero fixed-angle tangent in the Killing directions
is spacelike, also at the equatorial cut.

The radial diagrams represent the null directions selected in
Eq.~\eqref{eq:nullvectors}, not the induced metric at constant
$(x,\phi)$. Indeed, imposing $\dif x=0$ and
$2\alpha lr\dif t+(r^2+l^2)\dif\phi=0$ on tangent vectors reduces
the metric to a positive multiple of
\begin{equation}
 -\frac{\Delta_r}{r^2+l^2}\dif t^2
       +\frac{r^2+l^2}{\Delta_r}\dif r^2 .
\end{equation}
The second condition is in general nonintegrable:
the required coefficient of $\dif t$ in $\dif\phi$ depends on $r$.
In the acceleration regions $\Delta_r<0$, so a constant-radius
boundary is spacelike in this radial model. This is compatible with
the null character of the full boundary in
Eq.~\eqref{eq:reciprocal-boundary-norm}.

At fixed $x=x_0$ with $|x_0|<1$, the two acceleration components are
$-1<y<x_0$ and $x_0<y<1$. The first reaches the negative acceleration
horizon, the second the positive one. For $x_0>0$ the regular crossing
at $y=0$ is on the first component; for $x_0<0$ it is on the second.
For $x_0=0$ their common coordinate endpoint belongs to conformal infinity.
Both components end at $y=x_0$, not at a further Killing horizon.
Their coincident coordinate images do not identify the two boundary limits.

With the normalization in Eq.~\eqref{eq:tortoise}, the primitive can
also be written as
\begin{equation}
 r_*(y)=\sum_h\frac{r_h^2+l^2}{\Delta_r'(r_h)}
 \log|1-\alpha r_h y|.
 \label{eq:reciprocal-tortoise}
\end{equation}
The terms proportional to $\log|r|$ cancel in the sum. This expression
gives $r_*=0$ at $y=0$; Eq.~\eqref{eq:tortoise-reciprocal} shows that
it is increasing on $-1<y<1$. The boundary has the single level
$r_*\big|_{y=x_0}$ in the common normalization, although its two banks
are distinct.

\subsubsection{The axial sections}
\label{sec:axial}

The two axes are $x=s$ with $s=\pm1$. The time coordinate appropriate
to the axis is, locally,
\begin{equation}
 \tau_s=t-\frac{2ls}{1+\alpha^2l^2}\phi .
\end{equation}
It is constant along the degenerating axial Killing vector
$\partial_\phi+[2ls/(1+\alpha^2l^2)]\partial_t$. On the axis the
surviving Killing one-form is $(1+\alpha^2l^2)\dif\tau_s$, and
the induced axial two-metric is
\begin{equation}
 \dif s^2_{\mathrm{axis}}=\frac{1}{(1-\alpha rs)^2}
 \left[-\frac{(1+\alpha^2l^2)^2\Delta_r}{\rho^2(r,s)}\dif\tau_s^2
 +\frac{\rho^2(r,s)}{\Delta_r}\dif r^2\right].
\label{eq:axis-metric}
\end{equation}
The axial null coordinates are $\tau_s\mp r_{*,s}$, where
\begin{equation}
 \frac{\dif r_{*,s}}{\dif r}
 =\frac{\rho^2(r,s)}{(1+\alpha^2l^2)\Delta_r},\qquad
 k_h^{(s)}=\frac{\rho^2(r_h,s)}
                 {(1+\alpha^2l^2)\Delta_r'(r_h)}.
\label{eq:axial-tortoise}
\end{equation}
The axial normalization also follows from the four-dimensional
horizon generator. On the axis
\begin{equation}
 \chi_h(\tau_s)
 =\frac{\rho^2(r_h,s)}{(1+\alpha^2l^2)(r_h^2+l^2)},\qquad
 \frac{1}{2k_h^{(s)}}=\frac{\beta_h}{\chi_h(\tau_s)}.
\end{equation}
Thus the axial time changes the logarithmic residues without shifting
a radial root. The $k_h^{(s)}$ are the logarithmic residues, and their sum is
zero, so the reciprocal primitive normalized at $y=0$ is
$\sum_h k_h^{(s)}\log|1-\alpha r_h y|$. Its derivative is
$H(s,y)/[\alpha(1+\alpha^2l^2)G(y)]$, positive for $-1<y<1$.
In particular, the reciprocal crossing remains regular on the
opposite acceleration component of each axial section.

The matching acceleration root $r=s/\alpha$ also satisfies
$1-\alpha rs=0$. This coincidence does not give a regular physical
horizon crossing on the axis. For the null geodesics of
Eq.~\eqref{eq:axis-metric}, conservation of the Killing energy
$E=-g(\partial_{\tau_s},\dot\gamma)$ and the null equation give
\begin{equation}
 \frac{\dif r}{\dif\lambda}
   =\pm\frac{E}{1+\alpha^2l^2}(1-\alpha rs)^2.
\end{equation}
The coincident surface is at infinite affine parameter. The axial
two-metric rescaled by $(1-\alpha rs)^2$ has a regular null surface
there: with $v_s=\tau_s+r_{*,s}$ its radial cross term is
$2(1+\alpha^2l^2)\dif v_s\dif r$ and its $\dif v_s^2$
coefficient has a simple zero. Accordingly the matching acceleration
edges on the axial diagram are null conformal ends, rather than
attachments across a regular horizon. The other three simple roots
are crossed using the corresponding residues in
Eq.~\eqref{eq:axial-tortoise}. These local adapted times do not impose
a Misner time period or select a globally regular axial identification.

\subsection{Repeated radial blocks and the five angular diagrams}
\label{sec:global-atlas}

\subsubsection{Block incidence and horizon attachments}
For a fixed interior angle, let $u=t-r_*$ and $v=t+r_*$. In each radial block the null directions are represented by the conformal metric $-F\dif u\,\dif v$, where $F=\Delta_r/(r^2+l^2)$. The signs of the residues in Eq.~\eqref{eq:tortoise} give the horizon limits of $r_*$ in Table~\ref{tab:block-ends}. The labels I and V refer to the acceleration blocks attached across $H_{a^+}$ and $H_{a^-}$, respectively, including any reciprocal continuation within the same block.

\begin{table}[htbp]
\centering
\begin{tabular}{@{}ccccc@{}}
\toprule
Block & Index class & Sign of $F$ & Horizon at $r_*\to-\infty$ & Horizon at $r_*\to+\infty$\\
\midrule
V   & $0$ & $-$ & $H_{a^-}$ & ---\\
IV  & $1$ & $+$ & $H_{a^-}$ & $H_{b^-}$\\
III & $2$ & $-$ & $H_{b^+}$ & $H_{b^-}$\\
II  & $3$ & $+$ & $H_{b^+}$ & $H_{a^+}$\\
I   & $0$ & $-$ & --- & $H_{a^+}$\\
\bottomrule
\end{tabular}
\caption{Horizon endpoints of the radial blocks for $|x|<1$. The two components of class $0$ end at conformal infinity on their non-horizon sides.}
\label{tab:block-ends}
\end{table}

To display the repeated attachments, place each block in the unit null square with lower corner $(i,j)$, where $i,j\in\mathbb Z$. Choose
\begin{equation}
 p=i+L\big(-(-1)^iu\big),\qquad
 q=j+L\big((-1)^jv\big),\qquad
 L(z)=\frac{1}{1+\exp(-\alpha z)}.
 \label{eq:display-map}
\end{equation}
The alternating signs account for the reversal of the orientation of one null coordinate at each horizon. The future direction is increasing $p$ and $q$: the product $(\partial u/\partial p)(\partial v/\partial q)$ has sign $-(-1)^{i+j}$, the sign of $F$ in that square. In the plane
\begin{equation}
 \mathcal T=\frac{p+q-1}{2},\qquad \mathcal X=\frac{p-q}{2},
 \label{eq:display-plane}
\end{equation}
the two null families have slopes $\pm1$. The particular monotone function $L$ fixes the appearance of the drawing without changing the null branches used for gluing.

Number the radial blocks from V to I by $n=0,1,2,3,4$ in increasing order of the original patch radius. Where the lower radial horizon exists, its divergence has sign $(-1)^n$; the upper radial horizon, where present, has the opposite sign. From Eq.~\eqref{eq:display-map}, $r_*$ on the right edge has divergence of sign $(-1)^i$, and on the top edge of sign $(-1)^j$. At the left and bottom edges the signs are reversed. Since $n$ and $i+j$ have the same parity, crossing the right edge changes $n$ by $-(-1)^j$, and crossing the top edge changes it by $-(-1)^i$. These changes apply to the physical edge on the appropriate component. Taking the reference block $(0,0)$ to be III gives
\begin{equation}
 n(i,j)\equiv 2-i-j+2ij\pmod 4.
 \label{eq:block-index}
\end{equation}
For example, in the reference block III the left and bottom edges have
$r_*\to-\infty$ and carry $H_{b^+}$, adjoining II at $(-1,0)$ and
$(0,-1)$. The right and top edges have $r_*\to+\infty$ and carry
$H_{b^-}$, adjoining IV at $(1,0)$ and $(0,1)$. Thus the two
branches of each black-hole horizon form adjacent edges of III,
as also seen in the local quadrants of Fig.~\ref{fig:local-four-horizons}.
For $i+j$ odd, these are II when $i+j\equiv3\pmod4$ and IV
when $i+j\equiv1\pmod4$. For $i+j$ even, they are III when
$j-i\equiv0\pmod4$ and class $0$ when $j-i\equiv2\pmod4$.
These are the index classes of Fig.~4 of Ref.~\cite{PV2021}.
That figure gives a comparison for the incidence of the four horizons.
The cut of conformal infinity is determined here from the $A^+$ metric,
not from the drawing in the earlier parametrization. The two components in class $0$ are distinguished
by their horizon attachments. The periodicity of Eq.~\eqref{eq:block-index} is a periodicity of the labels, not an identification of the blocks.

On a common vertical edge the value of $q$ fixes the same finite $v$ on either side, and the coordinate $U_h$ vanishes. On a horizontal edge $p$ fixes $u$, and $V_h$ vanishes. In a neighborhood of a retained bifurcation vertex $(I,J)$ take
\begin{equation}
 U_h=e_U\operatorname{sgn}(p-I)e^{-\beta_hu},\qquad
 V_h=e_V\operatorname{sgn}(q-J)e^{\beta_hv},\qquad e_U,e_V=\pm1 .
 \label{eq:vertex-signs}
\end{equation}
The constants are fixed in one quadrant by Eq.~\eqref{eq:kruskalproduct} and by the choice of signed chart; they are then held fixed at that vertex. A single edge crossing changes the radial side and reverses the sign of the vanishing coordinate. The nonvanishing coordinate and the corotating azimuth $\phi_h$ are matched across that edge. Two horizon charts on the same open block are related by Eq.~\eqref{eq:signed-overlap}; analyticity is supplied by those charts, not by the display coordinates.

At the upper-right corner of square $(i,j)$, the two incident edges have the same divergent sign of $r_*$ if $i+j$ is even. All four edges there then belong to the same horizon, and the corner is its bifurcation if that horizon is retained. When $i+j$ is odd the two signs are opposite and the edges belong to different horizons. Such a corner is not added as an analytic point. Along either incident horizon its surviving Kruskal coordinate tends to infinite magnitude there. These checks distinguish the repeated bifurcations from the apparent contact of different horizon endpoints in the drawing.

\subsubsection{The five angular diagrams}
Within a class-$0$ square put $w=L^{-1}$, so that
\begin{equation}
 r_* = \frac{(-1)^i w(p-i)+(-1)^j w(q-j)}{2},
 \qquad
 w(z)=\frac{1}{\alpha}\log\frac{z}{1-z}.
 \label{eq:display-tortoise}
\end{equation}
In the reciprocal interval $-1<y<1$, the common $r_*$ increases from $-\infty$ to $+\infty$. Thus, at $|x_0|<1$, Eq.~\eqref{eq:display-tortoise} with $r_* = r_*\big|_{y=x_0}$ gives conformal infinity. The V component has $r_* < r_*\big|_{y=x_0}$ and the I component has $r_*>r_*\big|_{y=x_0}$. For even $i$ the future direction in class $0$ increases $r_*$, so this is future infinity for V and past infinity for I. For odd $i$ the assignments reverse. The cut equation can be solved explicitly. Put $z=p-i$ and
$c=\exp\bigl(2\alpha(-1)^i r_*\big|_{y=x_0}\bigr)$. Then
\begin{equation}
 q-j=\frac{c(1-z)}{z+c(1-z)},\qquad 0<z<1 .
 \label{eq:explicit-cut}
\end{equation}
With the same tortoise normalization the two boundary banks have
this same image in the display plane. We use a double line for that
image, without identifying the banks.

\begin{figure}[p]
\centering
\includegraphics[width=.97\linewidth]{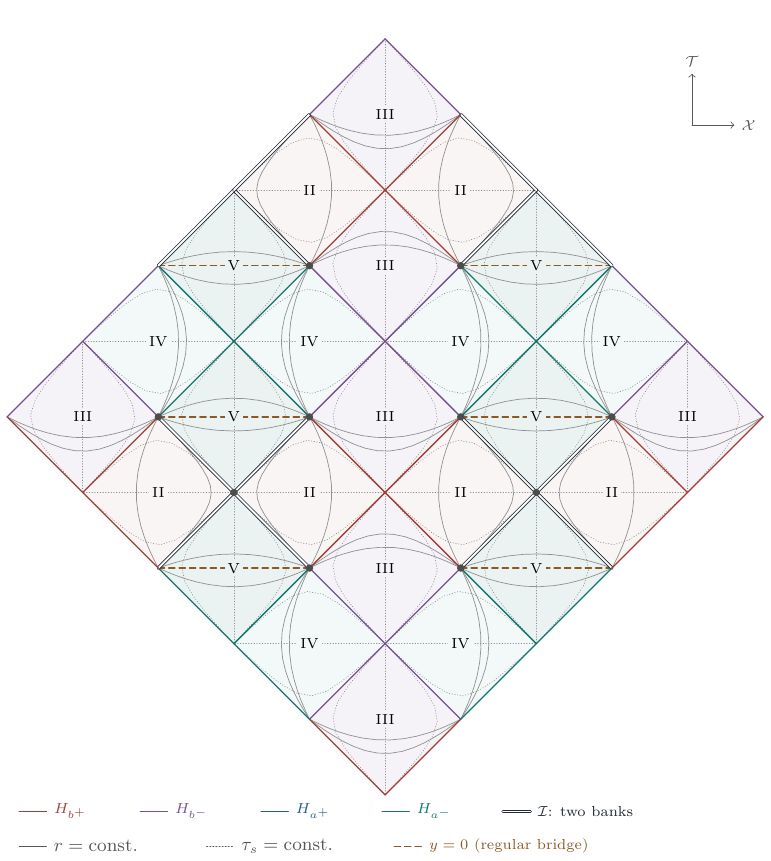}
\caption{Northern axis ($x=1$). The positive acceleration edges are null conformal ends. The reciprocal crossing $y=0$ is contained in V.}
\label{fig:north-axis}
\end{figure}

\begin{figure}[p]
\centering
\includegraphics[width=.97\linewidth]{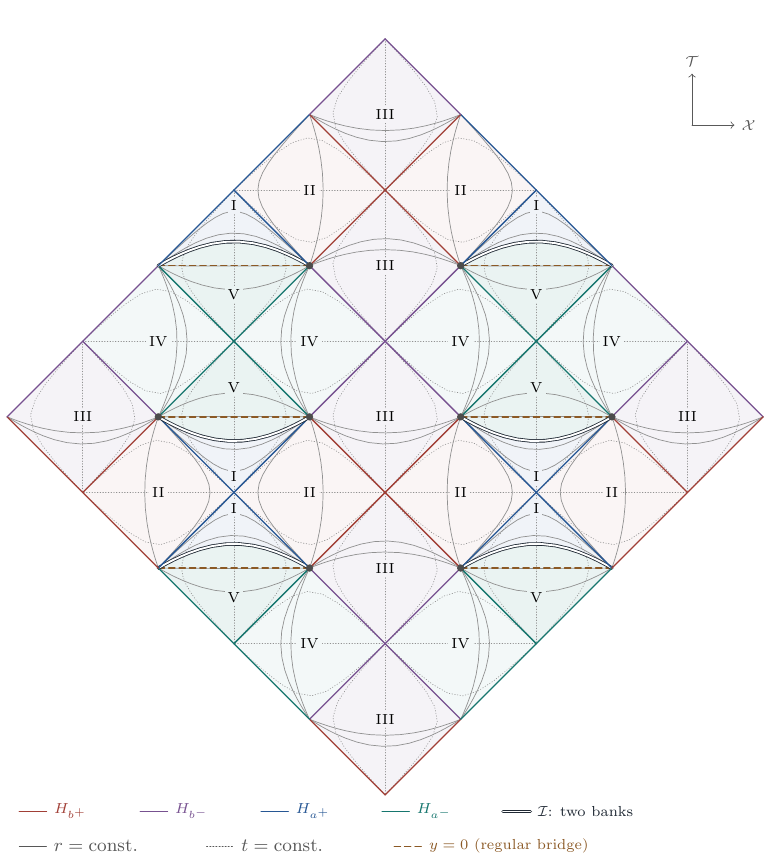}
\caption{Northern interior section ($x=1/2$). The regular crossing at $y=0$ lies in V. The double line at $y=x$ represents the two distinct banks of conformal infinity.}
\label{fig:north}
\end{figure}

\begin{figure}[p]
\centering
\includegraphics[width=.97\linewidth]{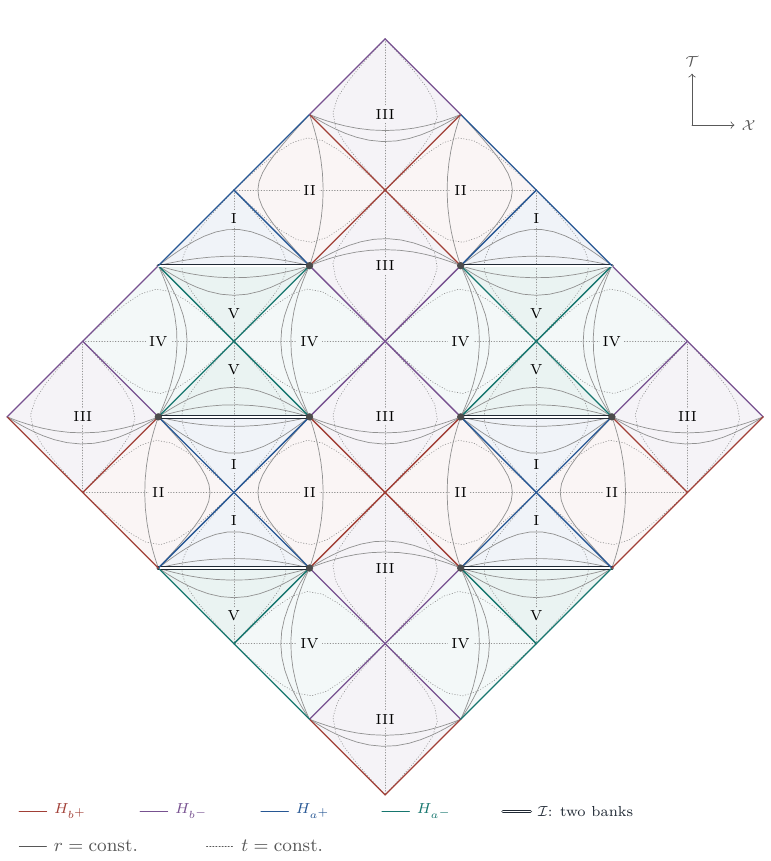}
\caption{Equatorial section ($x=0$). Both acceleration components end at $y=0$. There is no reciprocal crossing.}
\label{fig:equator}
\end{figure}

\begin{figure}[p]
\centering
\includegraphics[width=.97\linewidth]{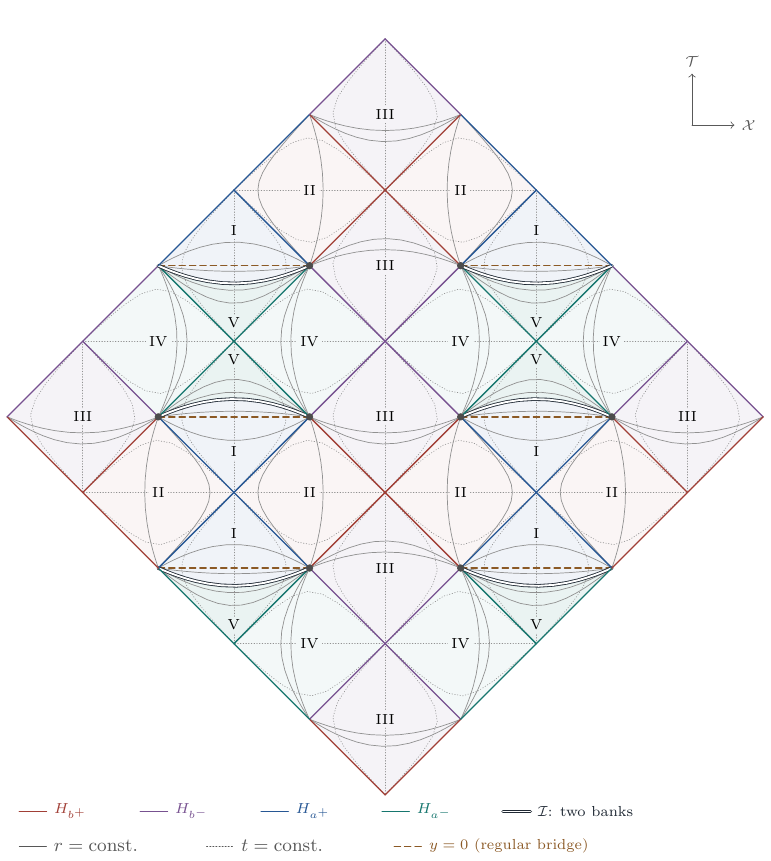}
\caption{Southern interior section ($x=-1/2$). The reciprocal crossing now lies in I. The double line marks the distinct banks of $\mathcal I$.}
\label{fig:south}
\end{figure}

\begin{figure}[p]
\centering
\includegraphics[width=.97\linewidth]{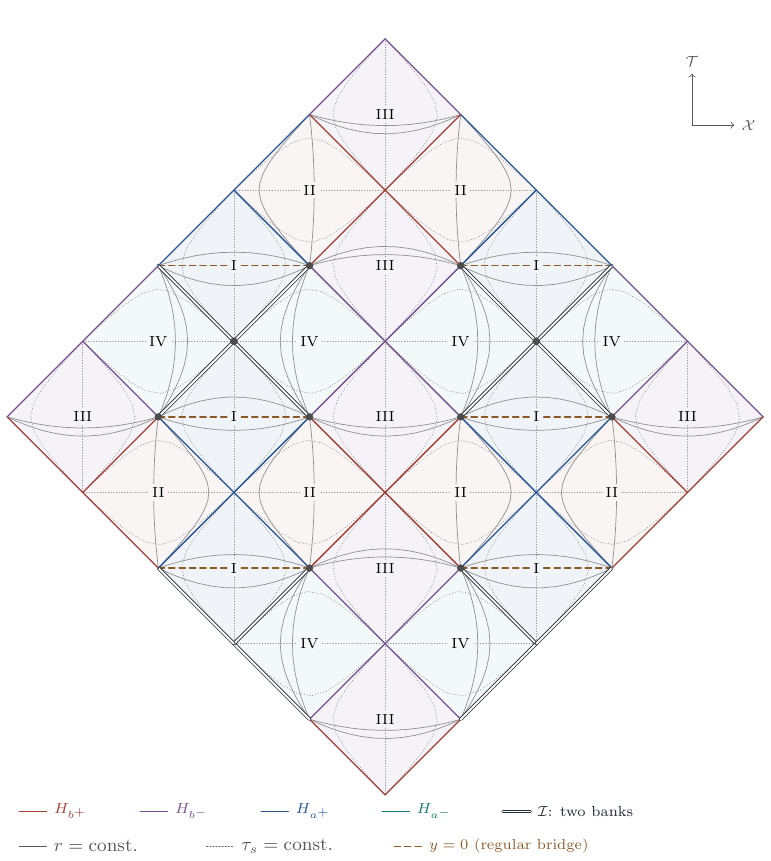}
\caption{Southern axis ($x=-1$). The negative acceleration edges are null conformal ends, and I contains the reciprocal crossing.}
\label{fig:south-axis}
\end{figure}

The arrangement of the black-hole blocks is the same in
Figs.~\ref{fig:north-axis}--\ref{fig:south-axis}. The differences occur
in the acceleration blocks and at their attachments to II and IV.
The negative acceleration block V has $-1<y<x_0$, while the positive
acceleration block I has $x_0<y<1$. Since $r_*$ increases with $y$ in
this interval, the relative positions of $y=0$ and $y=x_0$ determine
which block includes the reciprocal crossing.

\paragraph{Northern and southern interior sections.}
Both acceleration horizons are regular crossings for $0<|x_0|<1$.
In Fig.~\ref{fig:north}, $-1<0<x_0<1$: the crossing $y=0$ lies in
V, between its acceleration horizon and its bank of conformal infinity.
In Fig.~\ref{fig:south}, $-1<x_0<0<1$, and the crossing lies instead
in I. The dashed brown line is this regular level $r_*=0$, given by
$q-j=1-(p-i)$ in Eq.~\eqref{eq:explicit-cut}; the double line is
$y=x_0$. A ray may cross the dashed line within one acceleration
block, but cannot pass from I to V through the double line.

\paragraph{Equatorial section.}
In Fig.~\ref{fig:equator} the boundary itself is $y=0$. Thus $c=1$
and the double line lies on the level that marks a regular crossing
in the non-equatorial diagrams. There is no separate dashed line:
neither acceleration block continues through $y=0$.

\paragraph{Axial sections.}
For $x=s=\pm1$ the display uses $\tau_s$ and $r_{*,s}$ of
Sec.~\ref{sec:axial}. As $x_0$ tends to either axis,
$c\to0$ or $c\to\infty$ in Eq.~\eqref{eq:explicit-cut}, according
to the parity of $i$ and the sign of the divergent tortoise value.
The cut tends to the two null edges on that side of the acceleration
square. In Fig.~\ref{fig:north-axis}, $y=x=1$ coincides with
$H_{a^+}$. I has zero radial extent and the positive acceleration
attachment is replaced by distinct null ends of II and V. The
reciprocal crossing in V survives. In Fig.~\ref{fig:south-axis},
$y=x=-1$ coincides with $H_{a^-}$. V has zero radial extent, the
negative acceleration attachment is replaced by distinct null
ends of IV and I, and the crossing survives in I. In these two
figures the double null edges therefore denote conformal infinity,
not horizons across which the adjacent blocks are glued.

Figures~\ref{fig:north-axis}--\ref{fig:south-axis} show $-2\leq i,j\leq2$ for $(m,l,\alpha)=(1,0.6,0.3)$. The thin curves are obtained from $v-u=2r_*(r)$ at fixed radius and from $u+v=2t$ at fixed time, with the corresponding axial substitutions. We use $\alpha t\in\{-1.7,0,1.7\}$, or the same values of $\alpha\tau_s$. The two radius levels in each bounded radial interval are at fractions $0.30$ and $0.68$ from its lower $r$ endpoint. In each acceleration component they are at fractions $0.23$ and $0.66$ from the lower endpoint in $y$. The same map gives the boundary curves and the regular reciprocal level. The constant-time and constant-radius curves are level sets computed on each open block separately. The two banks drawn on the same limiting curve are not identified. The outer diamond only bounds the plotted index window; the incidence rule~\eqref{eq:block-index} applies outside it. Open circles denote excluded vertices in the interior of the window.

\subsection{Misner patches and axial identifications}

The two axial kernels from Eq.~\eqref{axisvectors} are proportional to
\begin{equation}
 \ell_N=\partial_\phi+\frac{2l}{1+\alpha^2l^2}\partial_t,
 \qquad
 \ell_S=\partial_\phi-\frac{2l}{1+\alpha^2l^2}\partial_t.
\end{equation}
Locally adapted time coordinates $t_N=t-2l\phi/(1+\alpha^2l^2)$ and $t_S=t+2l\phi/(1+\alpha^2l^2)$ satisfy
\begin{equation}
 t_S-t_N=\frac{4l}{1+\alpha^2l^2}\phi.
\end{equation}
If $\phi$ has period $\Delta\phi$ at fixed $t$, the transition $t_S-t_N$ changes by $4l\Delta\phi/(1+\alpha^2l^2)$ after one circuit. Thus the transition cannot be single-valued on the ordinary angular quotient for nonzero $l$ while both adapted times remain globally nonperiodic. Thus the north and south adapted times cannot both be single-valued under an ordinary periodic angular identification with nonperiodic time. The distinction between periodic-time and string-retaining descriptions dates to the classical interpretations of Taub--NUT geometry~\cite{Misner1963,Bonnor1969,MankoRuiz2005}. The torsion singularities considered for the rotating $C$-metric by Hong and Teo~\cite{HongTeo2005} are a related axial issue, but their conclusions do not replace the present analysis of the NUT-dependent axial Killing directions. It is distinct from conical regularity, whose local coefficients for the axial generators normalized above are
\begin{equation}
 \mathcal C_N=\frac{1-2\alpha m-\alpha^2l^2}{1+\alpha^2l^2},
 \qquad
 \mathcal C_S=\frac{1+2\alpha m-\alpha^2l^2}{1+\alpha^2l^2}.
\end{equation}
For $\alpha m\ne0$ these coefficients differ. A common angular period cannot eliminate both local conical defects, while imposing periodic time changes the global spacetime. We retain the nonperiodic-time description and state chronology conclusions relative to the chosen periodic $\phi$ identification.

\subsection{Closed timelike coordinate circles}

For periodic $\phi$ at fixed $t$, the axial-coordinate circles are timelike precisely where
\begin{equation}
 g_{\phi\phi}=\frac{(r^2+l^2)^2\Delta_x-4l^2x^2\Delta_r}{\Omega^2\rho^2}<0.
 \label{gphiphiD}
\end{equation}
Equivalently,
\begin{equation}
 (r^2+l^2)^2\Delta_x<4l^2x^2\Delta_r.
 \label{CTCD}
\end{equation}
For $\Delta_x>0$ this is impossible in blocks with $\Delta_r<0$. For $\Delta_r>0$ it holds sufficiently close to either axis when $l\ne0$, away from the horizons. Near $x=1-\epsilon$ and $x=-1+\epsilon$, the leading small-width estimates are
\begin{equation}
 \epsilon_N\simeq\frac{2l^2\Delta_r}{(1-2\alpha m-\alpha^2l^2)(r^2+l^2)^2},
 \qquad
 \epsilon_S\simeq\frac{2l^2\Delta_r}{(1+2\alpha m-\alpha^2l^2)(r^2+l^2)^2}.
 \label{CTCwidths}
\end{equation}
These estimates apply to the specified coordinate circles. Comparing the corresponding regions in different accelerating NUT families would require compatible global Killing identifications.

\subsection{Causal geodesics and the CGG viewpoint}\label{cgg}

Cl\'ement, Gal'tsov and Guenouche argued that Misner strings and chronology-violating regions do not, by themselves, rule out ordinary Lorentzian Taub--NUT with nonperiodic time~\cite{CGG}.  Their analysis distinguishes the existence of CTCs from the stronger question of whether causal geodesics themselves close and emphasizes the local geodesic transparency of the Misner strings. The distinction between alternative analytic extensions is also historically important: the ordinary Taub--NUT metric admits inequivalent extension prescriptions~\cite{MillerKruskalGodfrey1971}, so the present radial atlas is not by itself a uniqueness statement for the four-dimensional extension.

For the type-D accelerating solution, the local Misner structure survives.  The north and south axes are described by the adapted Killing directions~\eqref{axisvectors}, and null axis crossing obeys the regularity condition~\eqref{axisratioD}.  The black-hole and acceleration horizons are regular simple Killing horizons, and the nonzero NUT parameter removes the finite central curvature singularity.  These local properties are compatible with the geometrical setting considered by CGG.

The global CGG no-closed-causal-geodesic argument, however, does not transfer directly.  Acceleration breaks the $SO(3)$ structure of ordinary Taub--NUT.  The type-D solution retains a conformal Carter structure and separable null dynamics, but these local properties do not establish global causal-geodesic completeness.  The absence of closed causal geodesics in the accelerating geometry remains a separate global question.

\subsection{Scope of the global conclusions}
The signed horizon charts and the reciprocal metric determine the
unwrapped radial conformal atlas shown in
Figs.~\ref{fig:north-axis}--\ref{fig:south-axis}. These fixed-angle
projections do not determine the number of connected components of
four-dimensional null infinity, the topology of a maximal extension,
or geodesic completeness. In particular, the smooth equatorial
reciprocal representative is a local conformal statement; it does not
identify the two physical banks of the boundary. Likewise, a local choice of angular period removing one conical defect is not a globally consistent prescription for removing both Misner strings. These distinctions are essential when interpreting the causal regions of an accelerating NUT spacetime.

\begin{figure}[htbp]
\centering
\includegraphics[width=.82\linewidth]{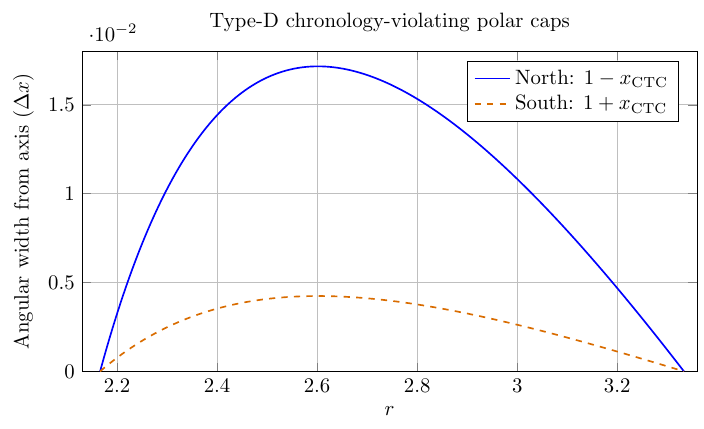}
\caption{Chronology boundary for the type-D metric in the regular positive exterior, for $m=1$, $l=0.6$ and $\alpha=0.3$. The two curves give the angular widths $1-x_{\rm CTC}$ and $1+x_{\rm CTC}$ of the northern and southern polar caps bounded by the exact zeros of $g_{\phi\phi}$, assuming a periodic $\phi$ identification. The coordinate circles at fixed $t,r,x$ are timelike within these caps. The radial plotting interval lies between the outer black-hole and positive acceleration horizons, which are its limiting endpoints. The plot does not classify other closed causal curves or impose a global Misner identification.}
\label{fig:chronologyD}
\end{figure}

\section{Type D versus type I}
\label{comparison}

We finally compare the type-D metric with the earlier type-I accelerating Taub--NUT solution. Their common Taub--NUT and vacuum $C$-metric limits do not establish equivalence for nonzero acceleration and NUT charge.

\subsection{The type-I metric and comparison conventions}
\label{sec:typeI-definition}
The type-I metric is written in the physical coordinates used in Ref.~\cite{PVtypeI}. The Ehlers--Harrison extension and its type-I character were further developed by Barrientos, Cisterna and Pallikaris~\cite{BarrientosCisternaPallikaris2024}; we use the vacuum metric of Ref.~\cite{PVtypeI} for the comparison below. The following functions refer only to type I. Using the same parameter symbols does not imply an invariant identification of parameters between the two families.

For the type-I metric of Ref.~\cite{PVtypeI}, write
\begin{equation}
\dif s^2=\frac{1}{\Omega^2}\left[
-\frac{Q}{R^2}(\dif t-W\dif\phi)^2
+\frac{R^2}{Q}\dif r^2
+R^2\left(\frac{\dif\theta^2}{P}+P\sin^2\theta\dif\phi^2\right)
\right],
\label{ImetricHJ}
\end{equation}
where
\begin{equation}
W=2l\left(\cos\theta-\alpha T\sin^2\theta\right),
\qquad
T=\frac{(r-r_-)^2P}{(r_+-r_-)\Omega^2},
\end{equation}
and
\begin{equation}
\Omega=1-\alpha(r-r_-)\cos\theta.
\end{equation}

The remaining functions are given explicitly by Eqs.~(44)--(45) of Ref.~\cite{PVtypeI}:
\begin{align}
 P(\theta)&=1-\alpha(r_+-r_-)\cos\theta,\nonumber\\
 Q(r)&=(r-r_+)(r-r_-)\big[1-\alpha^2(r-r_-)^2\big],\nonumber\\
 R^2(r,\theta)&=\frac{1}{r_+^2+l^2}\left\{
 r_+^2(r-r_-)^2+\frac{l^2(r-r_+)^2
 [1-\alpha^2(r-r_-)^2]^2}{\Omega^4}\right\}.
 \label{eq:typeI-functions}
\end{align}
Here $r_\pm=m\pm\sqrt{m^2+l^2}$ are the parameters of the type-I representation, not a declaration that the two accelerating metrics have an invariantly matched radial coordinate. The function $P$ depends only on $\theta$, $Q$ only on $r$, whereas $R^2$ depends on both. This last dependence is important for the subsequent Hamilton--Jacobi comparison. The type-I separated-horizon angular range requires $\alpha(r_+-r_-)<1$ so that $P>0$ for $0\leq\theta\leq\pi$.

Two limiting checks help fix the normalization. At $\alpha=0$, the identity $r_+r_-=-l^2$ gives $R^2=r^2+l^2$, $Q=r^2-2mr-l^2$, and $P=\Omega=1$, recovering the ordinary Taub--NUT metric. At $l=0$ (for $m>0$), $r_-=0$, $r_+=2m$ and $R^2=r^2$, recovering the standard vacuum $C$-metric in this coordinate convention. These common limits do not establish an isometry between the two accelerating NUT geometries for generic nonzero $\alpha l$.

\subsection{Comparison of radial Killing directions}
In the physical coordinates of Podolsk\'y and Vr\'atn\'y, all four type-I radial horizons have a common projective Killing generator~\cite{PVtypeI}:
\begin{equation}
[\chi_{b^+}^{\rm I}]=[\chi_{b^-}^{\rm I}]=[\chi_{a^+}^{\rm I}]=[\chi_{a^-}^{\rm I}].\label{Iparallel}
\end{equation}
The type-D horizon directions were determined in Sec.~\ref{killing}.  Their noncoincidence in the positive exterior is preserved by any common invertible change of Killing basis.  In that exterior the four boundary rods therefore have four distinguished projective directions for type D, compared with three for type I.

On each regular open horizon the Killing block is positive semidefinite and its kernel is the horizon generator. By continuity, any constant Killing combination timelike throughout the adjacent exterior must lie in the kernel on both bounding horizons. In type D the two kernels have trivial intersection, so no such Killing field exists in the stationary-axisymmetric algebra. In type I the common generator has norm $-Q/(\Omega^2 R^2)<0$ in the open exterior, in the conventions of Eq.~\eqref{ImetricHJ}. Thus the distinction concerns a constant Killing combination that is timelike throughout the open exterior and extends continuously to both regular bounding horizons. It is not a statement about arbitrary timelike vector fields or the stationary-limit surface of a particular coordinate field.

\subsection{Differential frame dragging at the outer-horizon merger}
\label{comparison:merger}

The preceding comparisons assume separated horizons. In type D the positive pair merges at $\alpha r_+=1$. In type I the acceleration radii are $r_-\pm\alpha^{-1}$, so the positive pair merges at $\alpha(r_+-r_-)=1$~\cite{PVtypeI}. At finite acceleration and nonzero $l$ there is no triple radial root in either family: in type D each black-hole root can meet only the acceleration root on its own side of zero, while in type I $r_-$ lies strictly between the two acceleration roots and cannot coincide with either. At $m=0$ type D can have two simultaneous double roots at $r=\pm |l|$ when $\alpha|l|=1$; the corresponding type-I inner root remains simple.

For $m>0$ and fixed $l\ne0$, approach the positive merger from the separated side. Write $r_{\rm A}-r_+=\epsilon r_+$ and set $r=r_+(1+\epsilon \xi)$, $t=\tau/\epsilon$, keeping $0<\xi<1$ fixed. Each radial function has two simple factors of order $\epsilon$, so its order $\epsilon^2$ compensates both the radial and time scalings. In type I the off-diagonal one-form appears only in the square multiplied by $Q$ in Eq.~\eqref{ImetricHJ}; its azimuthal coefficient is finite at fixed interior angle. The time--azimuth term consequently vanishes in the limit.

For type D, the angular square in Eq.~\eqref{Dmetric} contains the one-form
\begin{equation}
    \dif\phi+\frac{2\alpha lr}{r^2+l^2}\,\dif t .
\end{equation}
Its coefficient at the black-hole horizon must first be removed by a corotating azimuth, otherwise the angular square diverges as $\epsilon^{-2}$. The radial variation of that coefficient survives: in the corotating azimuth $\varphi$ the one-form tends to
\begin{equation}
    \dif\varphi+k\xi\,\dif\tau,\qquad
    k=\left.r_+\frac{\partial}{\partial r}\frac{2\alpha lr}{r^2+l^2}
        \right|_{r=r_+,\,\alpha=1/r_+}
      =\frac{2l(l^2-r_+^2)}{(r_+^2+l^2)^2}.
\end{equation}
Since $r_+^2-l^2=2mr_+$, $k$ is nonzero on the stated branch. The first Killing square contributes no time--azimuth mixing in the limit, since its prefactor is of order $\epsilon^2$. A trajectory of zero axial angular momentum therefore has $\dif\varphi/\dif\tau=-k\xi$, whereas in the type-I limit it has zero angular velocity in the static frame. A rigid change of rotating frame cannot remove the difference of angular velocities at two radii in type D. This is a local result at fixed interior angle. The angular factor degenerates at the north pole at the merger in both families; the calculation does not classify the angular endpoints or the global limiting spacetimes.

\subsection{Comparison of intrinsic horizon geometry}

For the type-I metric of Ref.~\cite{PVtypeI}, the induced outer-horizon metric simplifies considerably and its Gaussian curvature is
\begin{equation}
K_I(x)=\frac{\left(1-2\alpha\sigma x\right)^3}{r_+^2+l^2}.
\label{KI}
\end{equation}
Consequently
\begin{equation}
\mathcal R_I\equiv\frac{K_N^I}{K_S^I}
=\left(\frac{1-2\alpha\sigma}{1+2\alpha\sigma}\right)^3.
\label{RI}
\end{equation}
The two intrinsic distortions coincide in either common limit $\alpha\to0$ or $l\to0$, but differ generically when both parameters are nonzero.  At weak acceleration,
\begin{equation}
\mathcal R_D-\mathcal R_I
=-\frac{24\alpha m l^2}{r_+^2+l^2}+O(\alpha^2).
\label{Rdifference}
\end{equation}
This first-order difference compares equal coordinate parameters in the two metric representations. It is not by itself an intrinsic test, since the acceleration parameters need not have the same normalization.

A comparison that does not require this identification uses the cone angles at the poles. Let $2\pi\beta_N$ and $2\pi\beta_S$ be the angles obtained from the limiting ratios of circumference to proper distance in the respective horizon metrics. A common change of azimuthal period multiplies both factors by the same constant. For type I the horizon angular function is $(1-x^2)(1-2\alpha\sigma x)$ and the remaining conformal factors cancel out of the cone-angle calculation. Hence $\beta_N^I:\beta_S^I=(1-2\alpha\sigma):(1+2\alpha\sigma)$. Equation~\eqref{KI} then gives the exact relation
\begin{equation}
 K_N^I(\beta_S^I)^3=K_S^I(\beta_N^I)^3 .
 \label{eq:typeI-cone-curvature}
\end{equation}
The curvatures here are limits from the smooth part of the section, not distributional curvatures at the conical points.

For type D, introduce $u=\alpha r_+$ and $\lambda=l/r_+$ for the following exact expression. The corresponding difference is
\begin{align}
&K_N^D(\beta_S^D)^3-K_S^D(\beta_N^D)^3=-\frac{16\lambda^4u^3(1-\lambda^2)^3(1-\lambda^2u^2)
(1-u^2)^3(1+\lambda^2)^3}
{r_+^2\mathcal Q_N^3\mathcal Q_S^3},
\label{eq:typeD-cone-curvature}
\end{align}
where
\begin{equation}
\mathcal Q_N=1+\lambda^2+4\lambda^2u+\lambda^2(1+\lambda^2)u^2,
\qquad
\mathcal Q_S=1+\lambda^2-4\lambda^2u+\lambda^2(1+\lambda^2)u^2.
\end{equation}
The polar curvature and cone-angle formulas from which this identity follows are collected in Appendix~\ref{appcurvature}. For $m>0$, $l\ne0$ and $0<u<1$, one has $0<\lambda^2<1$ and both $\mathcal Q_N$ and $\mathcal Q_S$ are positive. The difference is therefore strictly negative. It begins at cubic order in the acceleration because the lower-order asymmetry cancels from this combination. A common change of azimuthal period rescales the difference without affecting its vanishing, while exchanging the poles reverses its sign. Thus the polar curvature limits and cone angles distinguish the two axisymmetric horizon geometries without matching the mass, NUT or acceleration parameters. The comparison concerns the intrinsic sections equipped with their axial circle action, not a choice of four-dimensional angular or Misner quotient.

\begin{figure}[htbp]
\centering
\includegraphics[width=.82\linewidth]{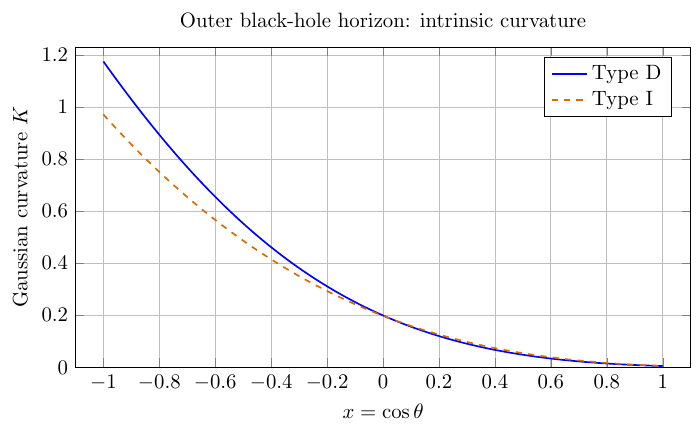}
\caption{Gaussian curvature of the outer black-hole horizon sections for type D (solid) and type I (dashed), evaluated at the illustrative equal coordinate parameters $m=1$, $l=0.6$, $\alpha=0.3$. The plot demonstrates distinct intrinsic profiles for this parameter choice, but equal numerical acceleration parameters do not provide an invariant matching between the families. The curvature--cone relations in Eqs.~\eqref{eq:typeI-cone-curvature} and \eqref{eq:typeD-cone-curvature} supply the parameter-independent diagnostic.}
\label{fig:curvatureDI}
\end{figure}

\subsection{Comparison of null Hamilton--Jacobi dynamics}

The Petrov types have a direct optical interpretation. In vacuum the repeated principal null directions are tangent to shear-free null geodesic congruences~\cite{GoldbergSachs}. Hence type D has two such families of rays, while a generic type-I region has none. The four simple principal directions of type I can have vanishing instantaneous Weyl shear source at a point; this does not give a shear-free congruence on an open set. We now distinguish this local optical fact from the question of additive separability.

Using the type-I line element and conventions defined at the beginning of Sec.~\ref{comparison}, the natural ansatz $S=-Et+L\phi+S_r(r)+S_\theta(\theta)$, the null HJ equation becomes
\begin{equation}
Q(S_r')^2-\frac{R^4}{Q}E^2
+P(S_\theta')^2
+\frac{(L-WE)^2}{P\sin^2\theta}=0.
\label{IHJ}
\end{equation}
The $EL$ coefficient contains the mixed term
\begin{equation}
\frac{4\alpha l(r-r_-)^2}{(r_+-r_-)\Omega^2}.
\label{mixedEL}
\end{equation}

The origin of this term is seen by expanding the angular square,
\begin{equation}
\frac{(L-WE)^2}{P\sin^2\theta}
=\frac{L^2}{P\sin^2\theta}
-\frac{2ELW}{P\sin^2\theta}
+\frac{E^2W^2}{P\sin^2\theta}.
\end{equation}
The coefficient of $EL$ is
\begin{equation}
-\frac{4l\cos\theta}{P\sin^2\theta}
+\frac{4\alpha l(r-r_-)^2}{(r_+-r_-)\Omega^2}.
\end{equation}
The first term depends only on $\theta$, whereas the second mixes
$r$ and $\theta$. Since $E$ and $L$ are independent constants,
terms proportional to $E^2$ or $L^2$ cannot cancel this obstruction
within the stated additive ansatz.
Setting $y=r-r_-$ and $x=\cos\theta$,
\begin{equation}
\frac{\partial^2}{\partial y\,\partial x}
\left[\frac{y^2}{(1-\alpha yx)^2}\right]
=\frac{6\alpha y^2}{(1-\alpha yx)^4},
\label{mixedderivative}
\end{equation}
which is nonzero generically.  Thus the $EL$ coefficient in Eq.~\eqref{IHJ} cannot be written as a sum of a purely radial and a purely angular function.  This obstructs additive separation in the displayed PV coordinates after the indicated common multiplier.  A fully coordinate-independent nonseparability theorem would additionally require excluding other admissible St\"ackel multipliers and nonstandard separable coordinate systems.  We do not claim such a theorem, nor complete dynamical nonintegrability.


\begin{table}[H]
\centering
\caption{Structural comparison of the type-D and type-I accelerating Taub--NUT spacetimes.}
\label{tab:comparison}
\begin{tabular}{p{0.31\textwidth}p{0.29\textwidth}p{0.29\textwidth}}
\toprule
Property & Type D & Type I \\
\midrule
Petrov class & D & I generically \\
Radial horizon directions & four distinct directions generically & common Killing direction \\
Positive-exterior boundary Killing directions & four distinct generically & three distinct (BH/AH coincide) \\
Constant timelike Killing combination across the exterior & None under the stated regularity assumptions & Common horizon generator \\
Outer merger, local scaled geometry & Differential frame dragging & Static at fixed interior angle \\
Intrinsic horizon curvature and cone angles & Cubic relation fails for $m>0$ & Cubic relation holds \\
Shear-free null geodesic families & Two & None generically \\
Null HJ dynamics & separated in $A^+$ coordinates & additive separation obstructed in the displayed PV form \\
CTC condition & factorized in $\Delta_r,\Delta_x$ & mixed through $T(r,\theta)$ \\
Local CGG interpretation & survives & survives \\
$\alpha\to0$ & Taub--NUT & Taub--NUT \\
$l\to0$ & C-metric & C-metric \\
\bottomrule
\end{tabular}
\end{table}

In the context of the established solution-generating distinction~\cite{Astorino,WuWu}, the projective horizon structure distinguishes the two families: the type-D black-hole and acceleration horizons have different Killing directions, whereas the type-I radial horizons share one. The curvature--cone relation and the null-geodesic structures provide independent distinctions, without comparing coordinate coefficients or angular velocities across different Killing normalizations.

Neither family is singled out as the uniquely physical accelerating Taub--NUT spacetime. Both retain the local features relevant to the CGG interpretation, but they combine acceleration and NUT charge in geometrically inequivalent ways.

\section{Conclusions}
\label{conclusion}

We have studied the vacuum type-D accelerating Taub--NUT spacetime in the nondegenerate four-horizon regime, using the $A^+$ form of the metric and a fixed normalization of its Killing coordinates. The aim has been to determine its horizon, null-geodesic and radial conformal properties, and to identify the geometrical distinctions from the type-I accelerating Taub--NUT solution. The type-D solution itself is known, as is its origin in an accelerating binary limit. The calculations here concern its explicit reconstruction and the properties of the resulting spacetime.

The direct reduction from the Pleba\'nski--Demia\'nski metric establishes the coordinate and Killing transformations needed to obtain the $A^+$ form, including the normalization of the orbit-space metric. The finite-binary inverse-scattering calculation provides a complementary description. In that construction the kernels of the dressed Killing block are evaluated on both open horizon rods before the large-companion limit is taken. Their projective directions can then be followed into the accelerating metric. This gives a direct connection between the finite-binary rod data and the horizon generators of the limiting spacetime. It does not determine the global periodic identifications of the Killing coordinates, which must be specified separately.

The horizon analysis shows that the black-hole and acceleration horizons generally have distinct projective Killing generators. Although each simple radial root admits a regular local horizon-penetrating chart, the different generators matter when comparing horizon frames and constructing their overlapping extensions. Under the assumptions stated in the text, they also preclude a single constant Killing combination that remains timelike throughout the regular exterior. The surface gravities and horizon quantities must therefore be interpreted with their Killing normalization made explicit; the rod directions alone do not fix that normalization.

The intrinsic geometry of the black-hole horizon gives a distinction that is less dependent on such choices. Its polar Gaussian curvatures and cone-angle coefficients satisfy an exact relation which differs from the corresponding identity for the type-I solution. For axisymmetric horizon sections equipped with their axial circle action, the type-I combination vanishes, whereas the type-D expression is nonzero throughout the nondegenerate parameter range considered here. The discrepancy first appears at cubic order in weak acceleration. Since the relation does not require the acceleration parameters of the two families to be identified, it provides a direct comparison of their marked horizon geometries. Its scope is the intrinsic horizon geometry with the specified axial action, rather than an unrestricted invariant of the full four-dimensional spacetime.

The null Hamilton--Jacobi equation separates in the conformal Carter geometry of the type-D metric. The fixed-angle null congruences used for the radial conformal construction follow from a particular sector of these separated equations; they are not an independent assumption about all null geodesics. This relation supplies the connection between the local null dynamics and the horizon extensions. It also makes clear why the conformal separability of null motion should be distinguished from a claim of complete integrability for massive geodesics. In the type-I comparison, the obstruction found concerns the natural additive Carter ansatz in the displayed coordinates and does not exclude every possible separation in other coordinates.

Using the distinguished null congruences, we constructed horizon-adapted coordinates at each of the four simple radial roots and assembled an unwrapped repeated radial conformal atlas. The reciprocal coordinate continues regularly across $r=\pm\infty$ away from the equatorial conformal endpoint, so those coordinate limits do not generally represent physical boundaries. The conformal boundary instead depends on the angular coordinate. Its position relative to the two acceleration horizons leads to five angular cases, including the distinct northern and southern axial limits. The resulting diagrams describe the incidence of the radial blocks and their conformal truncations. They do not establish a unique maximal four-dimensional analytic extension or settle the possible global identifications of its Killing coordinates.

These qualifications are particularly relevant in the presence of NUT charge. The local conical coefficients of the two axes, the transition between Misner-adapted coordinates, and the periodicity chosen for the angular generator are separate pieces of information. For the specified periodic angular identification, the metric admits regions in which the corresponding coordinate circles are timelike. This observation should not be conflated with a classification of all closed causal curves or geodesics. In particular, conclusions about causal geodesics in ordinary Taub--NUT spacetime cannot be transferred to the accelerating case without further analysis.

The comparison with the type-I accelerating Taub--NUT metric brings these results together. Although the two families share the Taub--NUT and $C$-metric limits, they have different horizon Killing directions and marked intrinsic horizon geometries for generic nonzero acceleration and NUT charge. The natural null-separation structures also differ, while the outer-horizon merger exhibits a local distinction in differential frame dragging. Thus the common limits do not imply equivalence of the two families.

The distinct horizon Killing directions, the curvature--cone relation and the angular dependence of conformal infinity together characterize the type-D accelerating Taub--NUT geometry. In particular, its horizon structure and marked intrinsic horizon geometry distinguish it from the type-I family despite their common nonaccelerating and zero-NUT limits.

\section*{Acknowledgments}
The author thanks Liang Zhengxin, James Winston Tumbokon, Hong Dezheng, Chan Hong He Kevin and Tan Hao Yu for their work on related undergraduate research projects under his supervision during 2022--2026. These projects contributed to the author's broader understanding of accelerating Taub--NUT spacetimes, the Pleba\'nski--Demia\'nski family, type-D black holes and the binary-system interpretation of acceleration.

\appendix
\section{Detailed algebraic verification of the limiting Killing sector}\label{ism:app:factorwise}
The reduction of the limiting Weyl metric to accelerating coordinates is most efficiently checked before expanding the full rational matrix. This appendix gives a system-independent coefficient test. Write $s=\rho^2$, $R_{ij}=s+\mu_i\mu_j$, and let $r_\pm=m\pm\sigma$. For the exterior chart use
\begin{align}
 \Omega&=1+\alpha rx,\qquad \Delta=(r-r_+)(r-r_-)(1-\alpha^2r^2),\\
 P&=(1-x^2)(1+\alpha r_+x)(1+\alpha r_-x),\qquad s=\Delta P/\Omega^4,\\
 \mu_1&=\frac{(1-x)(1-\alpha r)(r-r_+)(1+\alpha r_-x)}{\Omega^2},\\
 \mu_2&=\frac{(1-x)(1-\alpha r)(r-r_-)(1+\alpha r_+x)}{\Omega^2},\\
 \mu_3&=\frac{(1-\alpha^2r^2)(1+\alpha r_+x)(1+\alpha r_-x)}{\alpha\Omega^2}.
\end{align}
For example, these substitutions imply the elementary factorizations
\begin{align}
 \mu_1-\mu_2&=-\frac{2\sigma(1-x)(1-\alpha r)}{\Omega},\\
 \mu_1-\mu_3&=-\frac{(1-\alpha r)(1+\alpha r_+)(1+\alpha r_-x)}{\alpha\Omega},\\
 R_{12}&=\frac{2(r-r_+)(r-r_-)(1-x)(1-\alpha r)(1+\alpha r_+x)(1+\alpha r_-x)}{\Omega^3},\\
 R_{13}&=\frac{(r-r_+)(1-x)(1-\alpha^2r^2)(1-\alpha r_-)(1+\alpha r_+x)(1+\alpha r_-x)}{\alpha\Omega^3}.
\end{align}
These identities can be established by substituting the three rational $\mu_i$ and collecting common factors. Their role is to expose cancellations that would otherwise be hidden in a fully expanded stationary metric.

The mixed Killing numerator of the limiting two-soliton solution is cubic in the polarizations:
\begin{equation}
 N_{12}=C_1N_1+C_2N_2+C_1^2C_2N_{112}+C_1C_2^2N_{122}.
\end{equation}
Its four coefficients, written directly in the Weyl variables, are
\begin{align}
 N_1={}&-(\mu_1-\mu_2)\mu_2 R_{11}(\mu_1-\mu_3)R_{13}R_{12}R_{23}^2,\\
 N_2={}&(\mu_1-\mu_2)\mu_1sR_{22}(\mu_2-\mu_3)R_{23}(\mu_1-\mu_3)^2R_{12},\\
 N_{112}={}&-(\mu_1-\mu_2)\mu_1R_{22}(\mu_2-\mu_3)R_{23}R_{13}^2R_{12},\\
 N_{122}={}&(\mu_1-\mu_2)\mu_2sR_{11}(\mu_1-\mu_3)R_{13}R_{12}(\mu_2-\mu_3)^2.
\end{align}
To state the comparison, let $A=A_0+C_1C_2A_{12}$, $B=C_1B_1+C_2B_2$, $C=C_1C_1^{(r)}+C_2C_2^{(r)}$, and $D=D_0+C_1C_2D_{12}$, where
\begin{align}
 A_0&=(1-\alpha^2r_+^2)(1+\alpha r_-x)^2,\\
 A_{12}&=(1-\alpha^2r_-^2)(1+\alpha r_+x)^2,\\
 B_1&=\alpha\sigma(1-\alpha r_+)(1-\alpha r_-)(1-x)^2,\\
 B_2&=-\alpha\sigma(1+\alpha r_+)(1+\alpha r_-)(1+x)^2,\\
 C_1^{(r)}&=-\sigma(1+\alpha r)^2(1-\alpha r_+)(1-\alpha r_-),\\
 C_2^{(r)}&=\sigma(1-\alpha r)^2(1+\alpha r_+)(1+\alpha r_-),\\
 D_0&=\alpha(1-\alpha^2r_+^2)(r-r_-)^2,\\
 D_{12}&=\alpha(1-\alpha^2r_-^2)(r-r_+)^2.
\end{align}
The common factor in the denominator reduction is
\begin{equation}\label{ism:eq:Kfactor}
 K=\frac{2(1-x)^2(1-\alpha r)^3(1+\alpha r)(r-r_+)(r-r_-)
 (1+\alpha r_+x)^2(1+\alpha r_-x)^2}{\alpha^2\Omega^7}.
\end{equation}
The four independent component tests are the rational identities
\begin{align}
 \alpha\Omega^2N_i&=\mu_1\mu_2K^2[-\Delta A_0B_i+PC_i^{(r)}D_0],\quad i=1,2,\\
 \alpha\Omega^2(N_{112},N_{122})&=\mu_1\mu_2K^2\bigl(-\Delta A_{12}(B_1,B_2)+P(C_1^{(r)},C_2^{(r)})D_{12}\bigr),
\end{align}
The ordered-pair notation in the second line represents two separate scalar equations. A reader can reproduce the check by substituting the displayed rational expressions, clearing the common denominators, and factoring each of the four differences. No physical Kerr--NUT parameter relations are used in this coefficientwise test. The identity for the canonical determinant is a separate consistency check and is not a substitute for these four comparisons.

\subsection{Common polynomial normalization of the time--time component}\label{ism:app:timepoly}
We supply the detailed coefficientwise algebra supporting Eqs.~\eqref{ism:eq:mainpolydet} and \eqref{ism:eq:mainpolytt}, without using the physical accelerating-coordinate parametrization.  The following calculation also fixes the relative normalization of its numerator and denominator.  Set $v=\rho^2$ and retain $R_{ij}=v+\mu_i\mu_j$ and the poles $p_1=-v/\mu_1$, $p_2=\mu_2$.  The polarization-independent and polarization-dependent Gram entries are
\begin{align}
 \Gamma_{ij}^{(0)}&=-\frac{\mu_3}{(p_i-\mu_3)(p_j-\mu_3)(v+p_ip_j)},\\
 \Gamma_{ij}^{(2)}&=\frac{v\mu_3}{(v+\mu_3p_i)(v+\mu_3p_j)(v+p_ip_j)},\qquad
 \Gamma_{ij}=\Gamma_{ij}^{(0)}+C_iC_j\Gamma_{ij}^{(2)}.
\end{align}
Introduce the polynomial factors, used only in this calculation,
\begin{align}
 F_0&=R_{12}R_{23}(\mu_1-\mu_3),&F_{12}&=R_{12}R_{13}(\mu_2-\mu_3),\\
 G_1&=(\mu_1-\mu_2)R_{23}R_{13},&G_2&=v(\mu_1-\mu_2)(\mu_1-\mu_3)(\mu_2-\mu_3).
\end{align}
A common factor, independent of both polarizations, is
\begin{align}
 Q={}&-\frac{\mu_2v^2}{\mu_1^3\mu_3^2}
 (\mu_1-\mu_2)^2(\mu_1-\mu_3)^2(\mu_2-\mu_3)^2\\
 &\quad\times R_{11}R_{22}R_{13}^2R_{23}^2.
\end{align}
Expanding the $2\times2$ determinant in $1,C_1^2,C_2^2,C_1C_2,C_1^2C_2^2$ and multiplying every coefficient by this same $Q$ gives
\begin{equation}
 Q\det\Gamma=\mu_1\mu_2\bigl[(F_0+C_1C_2F_{12})^2+(C_1G_1+C_2G_2)^2\bigr].
\end{equation}
To obtain the time--time numerator, one must use the same $Q$, not a separately chosen clearing factor.  Writing $\ell_i=L_t^{(i)}/p_i=-\mu_3/[p_i(p_i-\mu_3)]$, the original reconstruction yields
\begin{align}
 g_{tt}&=\frac{\mathcal N_{11}}{Q\det\Gamma},\\
 \mathcal N_{11}&=-\frac{\mu_2}{\mu_1}Q\bigl[\mu_3\det\Gamma+\Gamma_{22}\ell_1^2-2\Gamma_{12}\ell_1\ell_2+\Gamma_{11}\ell_2^2\bigr].
\end{align}
The five polarization coefficients reduce to the single identity
\begin{align}
 \mathcal N_{11}=\mu_3\biggl\{&v\left(C_1G_1-\frac{\mu_1\mu_2}{v}C_2G_2\right)^2-\left(\mu_1F_0+C_1C_2\mu_2F_{12}\right)^2\biggr\}.
\end{align}
The overall factor $\mu_3$ is essential.  At zero polarization the ratio gives
$g_{tt}=-\mu_1\mu_3/\mu_2$, whereas omitting it would incorrectly give
$-\mu_1/\mu_2$.  This unpolarized limit is an independent normalization check.
The coefficientwise verification uses only rational arithmetic in
$v,\mu_1,\mu_2,\mu_3$; no Weyl quadratic identities or physical-parameter relations are needed.

To connect this Weyl factorization with the accelerating-coordinate
one-forms, one must make a second, logically distinct comparison.  In the
exterior chart, substitute the rational expressions for $v$ and $\mu_i$
from Appendix~\ref{ism:app:factorwise}.  The denominator has the form
\begin{equation}
 Q\det\Gamma=\mu_1\mu_2 K^2W,
\end{equation}
where $K$ is given in Eq.~\eqref{ism:eq:Kfactor} and
$W=X^2+\sigma^2Y^2$ is defined in the main text.  The corrected numerator
must then obey
\begin{equation}\label{ism:eq:gttfullcheck}
 \alpha\Omega^2\mathcal N_{11}
 =\mu_1\mu_2K^2(-\Delta A^2+PC^2).
\end{equation}
This identity is checked \emph{before} the physical parameter dictionary
is imposed.  Its verification is particularly economical because the two
sides have only five polarization monomials.  On the left their coefficients,
ordered as $1,C_1^2,C_2^2,C_1C_2,C_1^2C_2^2$, are
\begin{align}
 &-\mu_3\mu_1^2F_0^2,\quad
 \mu_3vG_1^2,\quad
 \frac{\mu_3\mu_1^2\mu_2^2}{v}G_2^2, \quad
 -2\mu_3\mu_1\mu_2(F_0F_{12}+G_1G_2),\quad
 -\mu_3\mu_2^2F_{12}^2.
\end{align}
On the right, after removing the common multiplier $\mu_1\mu_2K^2$,
the corresponding coefficients are
\begin{equation}
 -\Delta A_0^2,\quad PC_1^{(r)2},\quad PC_2^{(r)2},\quad
 -2\Delta A_0A_{12}+2PC_1^{(r)}C_2^{(r)},\quad -\Delta A_{12}^2.
\end{equation}
Here $A_0,A_{12},C_1^{(r)},C_2^{(r)}$ are given in
Appendix~\ref{ism:app:factorwise}.  Equality of each pair after multiplying by
its respective common prefactor is a rational-function identity in
$r,x,\alpha,r_+,r_-$.  This provides five explicit checks that can be
implemented independently in any exact computer-algebra system.  The
corrected factor $\mu_3$ is required in every one of them; checking the
unpolarized coefficient first is an especially simple safeguard.

The analogous mixed-component comparison is recorded in
Appendix~\ref{ism:app:factorwise}.  These two component identities also suffice to determine the third component, because the canonical determinant of the original ISM metric has been proved independently in Eq.~\eqref{ism:eq:canonicalproof}.  Indeed, if $G$ is the ISM Killing matrix and $\widehat G$ the proposed two-square matrix, then the coefficient identity $AD-BC=\alpha W$ implies $\det\widehat G=-\rho^2=\det G$.  Having independently established $g_{tt}=\widehat g_{tt}$ and $g_{t\phi}=\widehat g_{t\phi}$, subtraction of the determinant equations gives
\begin{equation}\label{ism:eq:thirdcomponent}
 g_{tt}\bigl(g_{\phi\phi}-\widehat g_{\phi\phi}\bigr)=0.
\end{equation}
Thus $g_{\phi\phi}=\widehat g_{\phi\phi}$ wherever $g_{tt}\ne0$.  Because the entries are rational functions, the equality extends as a rational identity across regular portions of the locus $g_{tt}=0$.  This argument does not replace either of the two independent component comparisons; it completes them using the separately derived determinant normalization.

\section{Exact type-D polar curvatures}
\label{appcurvature}

We use the parameters $u$ and $\lambda$ defined in the comparison section.
The exact north and south polar curvatures can be written as
\begin{equation}
K_N^D=\frac{(1-u)^3}{r_+^2}
\frac{\mathcal P_N(u,\lambda)}{\mathcal Q_N(u,\lambda)^2},
\label{KNDapp}
\end{equation}
\begin{equation}
K_S^D=\frac{(1+u)^3}{r_+^2}
\frac{\mathcal P_S(u,\lambda)}{\mathcal Q_S(u,\lambda)^2},
\label{KSDapp}
\end{equation}
where
\begin{equation}
\begin{aligned}
\mathcal P_N={}&1+\lambda^2-\lambda^2(1-3\lambda^2)u
-\lambda^2(3-\lambda^2)u^2
-\lambda^4(1+\lambda^2)u^3,\\
\mathcal P_S={}&1+\lambda^2+\lambda^2(1-3\lambda^2)u
-\lambda^2(3-\lambda^2)u^2
+\lambda^4(1+\lambda^2)u^3,
\end{aligned}
\end{equation}
with $\mathcal Q_N$ and $\mathcal Q_S$ as defined in the comparison section.
Hence
\begin{equation}
\mathcal R_D=
\left(\frac{1-u}{1+u}\right)^3
\frac{\mathcal P_N}{\mathcal P_S}
\left(\frac{\mathcal Q_S}{\mathcal Q_N}\right)^2.
\label{RDexactapp}
\end{equation}
The reflection identity $K_S^D(u,\lambda)=K_N^D(-u,\lambda)$ provides a useful algebraic check.

For azimuthal period $2\pi$, the cone-angle factors of
\eqref{horizon2metricD} are
\begin{equation}
 \beta_N^D=\frac{(1+\lambda^2)(1-u)(1+\lambda^2u)}{\mathcal Q_N},
 \qquad
 \beta_S^D=\frac{(1+\lambda^2)(1+u)(1-\lambda^2u)}{\mathcal Q_S}.
\end{equation}
Indeed, for $\dif s_H^2=E(x)\dif x^2+F(x)\dif\varphi_H^2$,
the cone factor is the absolute limiting value of
$F'(x)/(2\sqrt{E(x)F(x)})$. In the present case this is
$(r_+^2+l^2)|\Delta_x'|/(2\rho_+^2)$ at either pole.
Substituting these factors and Eqs.~\eqref{KNDapp} and \eqref{KSDapp}
gives Eq.~\eqref{eq:typeD-cone-curvature} in the main text.
The factorization follows from the polynomial identity
$\mathcal P_N\mathcal Q_N(1-\lambda^2u)^3-
 \mathcal P_S\mathcal Q_S(1+\lambda^2u)^3$.

\section*{AI Use Disclosure}
During the preparation of this work, the author used OpenAI ChatGPT (GPT-6) to assist with analytical calculations, preparation of figure-plotting code, preliminary literature searches, and language editing. All mathematical derivations were independently verified by the author, who also checked the cited literature against primary sources and reviewed and edited the manuscript. The author takes full responsibility for the scientific content and conclusions of this work.

\end{document}